\pdfoutput=1
\newcommand*{\ATLASLATEXPATH}{}
\documentclass[PAPER, atlasdraft=true, texlive=2016, cernpreprint, UKenglish, subfigure, LANGEDIT=true]{\ATLASLATEXPATH atlasdoc}
 
\usepackage{\ATLASLATEXPATH atlaspackage}
\usepackage{\ATLASLATEXPATH atlasbiblatex}
 
\usepackage{\ATLASLATEXPATH atlasphysics}
 
\graphicspath{{logos/}{figures/}}
 
\usepackage{ANA-FTAG-2018-01-PAPER-defs}

\AtlasTitle{ATLAS $b$-jet identification performance and efficiency measurement with $t\bar{t}$ events in $pp$ collisions at $\sqrt{s}=13$~TeV}
 
\AtlasAbstract{
The algorithms used by the ATLAS Collaboration during Run 2 of the Large Hadron Collider to identify jets containing $b$-hadrons are presented. The performance of the algorithms is evaluated in the simulation and the efficiency with which these algorithms identify jets containing $b$-hadrons is measured in collision data. The measurement uses a likelihood-based method in a sample highly enriched in $t\bar{t}$ events. The topology of the $t \to W b$ decays is exploited to simultaneously measure both the jet flavour composition of the sample and the efficiency in a transverse momentum range from 20~GeV to 600~GeV. The efficiency measurement is subsequently compared with that predicted by the simulation. The data used in this measurement, corresponding to a total integrated luminosity of 80.5~fb$^{-1}$, were collected in proton--proton collisions during the years 2015 to 2017 at a centre-of-mass energy $\sqrt{s}=$~13~TeV. By simultaneously extracting both the efficiency and jet flavour composition, this measurement significantly improves the precision compared to previous results, with uncertainties ranging from 1\% to 8\% depending on the jet transverse momentum.
}
 
\author{The ATLAS Collaboration}
 
\AtlasRefCode{FTAG-2018-01}

\AtlasJournal{EPJC}
 
\AtlasJournalRef{Eur. Phys. J. C 79 (2019) 970}
\AtlasDOI{10.1140/epjc/s10052-019-7450-8}
 
\PreprintIdNumber{CERN-EP-2019-132}

\AtlasCoverSupportingNote{Algorithm supporting documentation}{https://cds.cern.ch/record/2639796}
\AtlasCoverSupportingNote{Calibration supporting documentation}{https://cds.cern.ch/record/2631547}
\AtlasCoverSupportingNote{High \pt\ extrapolation supporting documentation}{https://cds.cern.ch/record/2658194}
 
\AtlasCoverCommentsDeadline{20 June 2019}
 
\AtlasCoverAnalysisTeam{Marco~Battaglia, Nihal~Brahimi, Alessandro~Calandri,
Andrea~Coccaro, Rachael~Ann Creager, Markus~Cristinziani,
Francesco~Armando~Di~Bello, Binbin~Dong, Rima~El~Kosseifi, James~Ferrando,
Maria~Giovanna~Foti, Sanmay~Ganguly, Jannik~Geisen, Mazuza~Ghneimat,
Geoffrey~Gilles, Tobias~Golling, Daniel~Guest, Sebastian~Heer,
Todd~Brian~Huffman, Zihao~Jiang, Michael~Kagan, Vadim~Kostyukhin, Joe~Kroll,
Marie~Christine~Lanfermann, Emmanuel~Le~Guirriec, Kazuya~Mochizuki,
Emery~Nibigira, Michaela~Paganini, Giacinto~Piacquadio,
Christopher~Samuel~Pollard, Yulia~Rodina, Matthias~Saimpert,
Julian~Constantin~Schmoeckel, Ariel~Schwartzman, Andrea~Sciandra,
Jeffrey~David~Shahinian,
Jonathan~Shlomi, Weimin~Song, Beojan~Stanislaus, Paul~Tipton, Lauren~Tompkins,
Michael~Ughetto, Laurent~Vacavant, Christian~Weiser, Daniel~Whiteson, Qi~Zeng,
Ning~Zhou}
 
\AtlasCoverEdBoardMember{Andrew Chisholm, Giuseppe Salamanna, Timothy Paul Scanlon~(chair)}

\AtlasCoverEgroupEditors{atlas-FTAG-2018-01-editors@cern.ch}
 
\AtlasCoverEgroupEdBoard{atlas-FTAG-2018-01-editorial-board@cern.ch}
\hypersetup{pdftitle={ATLAS document},pdfauthor={The ATLAS Collaboration}}
 
\begin{document}
 
\LEcontact{Pippa Wells pippa.wells@cern.ch}
 
\maketitle
 
\tableofcontents
 
\section{Introduction}
\label{sec:intro}
 
The identification of jets containing $b$-hadrons ($b$-jets) against the large jet background containing $c$-hadrons but no $b$-hadron ($c$-jets) or containing neither $b$- or $c$-hadrons (light-flavour jets) is of major importance in many areas of the physics programme of the ATLAS experiment~\cite{PERF-2007-01} at the Large Hadron Collider (LHC)~\cite{Evans:2008zzb}. It has been decisive in the recent observations of the Higgs boson decay into bottom quarks~\cite{HIGG-2018-04} and of its production in association with a top-quark pair~\cite{HIGG-2018-13}, and plays a crucial role in a large number of Standard Model (SM) precision measurements, studies of the Higgs boson properties, and searches for new phenomena.
 
The ATLAS Collaboration uses various algorithms to identify $b$-jets~\cite{PERF-2012-04}, referred to as \btag algorithms, when analysing data recorded during Run 2 of the LHC (2015--2018).
These algorithms exploit the long lifetime, high mass and high decay multiplicity of $b$-hadrons as well as the properties of the $b$-quark fragmentation. Given a lifetime of the order of 1.5 ps ($<c\tau> \approx 450~\mu m$), measurable $b$-hadrons have a significant mean flight length, $<l>=\beta \gamma c \tau$, in the detector before decaying, generally leading to at least one vertex displaced from the hard-scatter collision point.
The strategy developed by the ATLAS Collaboration is based on a two-stage approach. Firstly, low-level algorithms reconstruct the characteristic features of the $b$-jets via two complementary approaches, one that uses the individual properties of charged-particle tracks, later referred to as tracks, associated with a hadronic jet, and a second which combines the tracks to explicitly reconstruct displaced vertices. These algorithms, first introduced during Run 1~\cite{PERF-2012-04}, have been improved and retuned for Run~2. Secondly, in order to maximise the \btag performance, the results of the low-level \btag algorithms are combined in high-level algorithms consisting of multivariate classifiers. The performance of a \btag algorithm is characterised by the probability of tagging a $b$-jet ($b$-jet tagging efficiency, \beff) and the probability of mistakenly identifying a $c$-jet or a light-flavour jet as a $b$-jet, labelled \ceff\ (\leff). In this paper, the performance of the algorithms is quantified in terms of $c$-jet and light-flavour jet rejections, defined as $1/\ceff$ and $1/\leff$, respectively.

The imperfect description of the detector response and physics modelling effects in Monte Carlo (MC) simulations necessitates the measurement of the performance of the \btag algorithms with collision data~\cite{PERF-2016-05,ATLAS-CONF-2018-045,ATLAS-CONF-2018-006}.
In this paper, the measurement of the $b$-jet tagging efficiency of the high-level \btag algorithms used in proton--proton ($pp$) collision data recorded during Run 2 of the LHC at $\sqrt{s}=13~\TeV$ is presented. The corresponding measurements for $c$-jets and light-flavour jets, used in the measurement of the $b$-jet tagging efficiency to correct the simulation such that the overall tagging
efficiency of $c$-jets and light-flavour jets match that of the data, are described elsewhere~\cite{ATLAS-CONF-2018-045,ATLAS-CONF-2018-006}.
The production of $t\bar{t}$ pairs at the LHC provides an abundant source of $b$-jets by virtue of the high cross-section and the $t\rightarrow Wb$ branching ratio being close to 100\%.
A very pure sample of $t\bar{t}$ events is selected by requiring that both $W$ bosons decay leptonically, referred to as dileptonic $t\bar{t}$ decays in the following.
A combinatorial likelihood approach is used to measure the $b$-jet tagging efficiency of the high-level $b$-tagging algorithms as a function of the jet transverse momentum (\pt). This version of the analysis builds upon the approach used previously by the ATLAS Collaboration~\cite{PERF-2016-05}, extending the method to derive additional constraints on the flavour composition of the sample, which reduces the uncertainties by up to a factor of two relative to previous publication.
 
The paper is organised as follows. In Section~\ref{sec:detector}, the ATLAS detector is described. Section~\ref{sec:objects} contains a description of the objects reconstructed in the detector which are key ingredients for $b$-tagging algorithms, while Section~\ref{sec:algorithms} describes the $b$-tagging algorithms and the evaluation of their performance in the simulation. The second part of the paper focuses on the $b$-jet tagging efficiency measurement carried out in collision data and the application of these results in ATLAS analyses. The data and simulated samples used in this work are described in Section~\ref{sec:samples}. The event selection and classification performed for the measurement of the $b$-jet tagging efficiency are summarised in Section~\ref{sec:selection}. The measurement technique is presented in Section~\ref{sec:efficiency} and the sources of uncertainties are described in Section~\ref{sec:uncertainties}. The results and their usage within the ATLAS Collaboration are discussed in Sections~\ref{sec:result} and~\ref{sec:usage}, respectively.


\section{ATLAS detector}
\label{sec:detector}
 
The ATLAS detector~\cite{PERF-2007-01} at the LHC covers nearly the entire solid angle around the collision point.
It consists of an inner tracking detector (ID) surrounded by a superconducting solenoid, electromagnetic and hadronic calorimeters and a muon spectrometer incorporating three large superconducting toroid magnets.
 
The ID consists in a high-granularity silicon pixel detector which covers the vertex region and typically provides four measurements per track. The innermost layer, known as the insertable B-layer (IBL)~\cite{ATLAS-TDR-2010-19}, was added in 2014 and provides high-resolution hits at small radius to improve the tracking performance. For a fixed $b$-jet efficiency, the incorporation of the IBL improves the light-flavour jet rejection of the $b$-tagging algorithms by up to a factor of four~\cite{PIX-2018-001}.
The silicon pixel detector is followed by a silicon
microstrip tracker (SCT) that
typically provides eight measurements from four strip double layers.
These silicon detectors are complemented by a transition radiation tracker (TRT),
which enables radially extended track reconstruction up to the pseudorapidity\footnote{ATLAS uses a right-handed coordinate system with its origin at the nominal interaction point (IP)
in the centre of the detector and the $z$-axis along the beam pipe.
The $x$-axis points from the IP to the centre of the LHC ring,
and the $y$-axis points upwards.
Cylindrical coordinates $(r,\phi)$ are used in the transverse plane,
$\phi$ being the azimuthal angle around the $z$-axis.
The pseudorapidity is defined in terms of the polar angle $\theta$ as $\eta = -\ln \tan(\theta/2)$.
Angular distance is measured in units of $\Delta R \equiv \sqrt{(\Delta\eta)^{2} + (\Delta\phi)^{2}}$.}
$|\eta| = 2.0$. The TRT also provides electron identification information
based on the fraction of hits (typically 33 in the barrel and up to an average of 38 in the endcaps) above a higher energy-deposit threshold corresponding to transition radiation.
The ID is immersed in a \SI{2}{\tesla} axial magnetic field
and provides charged-particle tracking in the pseudorapidity range $|\eta| < 2.5$.
 
The calorimeter system covers the pseudorapidity range $|\eta| < 4.9$. Within the region $|\eta|< 3.2$, electromagnetic calorimetry is provided by barrel and
endcap high-granularity lead/liquid-argon (LAr) sampling calorimeters,
with an additional thin LAr presampler covering $|\eta| < 1.8$
to correct for energy loss in material upstream of the calorimeters.
Hadronic calorimetry is provided by a steel/scintillator-tile calorimeter,
segmented into three barrel structures within $|\eta| < 1.7$, and two copper/LAr hadronic endcap calorimeters.
The solid angle coverage is completed with forward copper/LAr and tungsten/LAr calorimeter modules
optimised for electromagnetic and hadronic measurements, respectively.
 
The muon spectrometer  comprises separate trigger and
high-precision tracking chambers measuring the deflection of muons in a magnetic field generated by superconducting air-core toroids.
The precision chamber system covers the region $|\eta| < 2.7$ with three layers of monitored drift tubes,
complemented by cathode-strip chambers in the forward region.
The muon trigger system covers the range $|\eta| < 2.4$ with resistive-plate chambers in the barrel and thin-gap chambers in the endcap regions.
 
A two-level trigger system~\cite{TRIG-2016-01} is used to select interesting events.
The first level of the trigger is implemented in hardware and uses a subset of detector information
to reduce the event rate to a design value of at most \SI{100}{\kHz}.
It is followed by a software-based trigger that reduces the event rate to a maximum of around \SI{1}{\kHz} for offline storage.

 
\section{Key ingredients for $b$-jet identification}
\label{sec:objects}

The identification of $b$-jets is based on several objects reconstructed in the ATLAS detector which are described here.
 
\begin{description}
 
\item {\textbf{Tracks}} are reconstructed in the ID~\cite{PERF-2015-08}. 
The $b$-tagging algorithms only consider tracks with \pt larger than 500~\MeV, with
further selection criteria applied to reject fake and poorly measured tracks~\cite{ATL-PHYS-PUB-2015-051}. The combined efficiency of the track reconstruction and selection criteria, evaluated by using minimum-bias simulated events (in which more than 98\% of charged-particles are pions), ranges from 91\% in the central (${|\eta| < 0.1}$) region to 73\% in the forward (${2.3 \le |\eta| < 2.5}$) region of the detector. Additional selections on reconstructed tracks are applied in the low-level \btag algorithms described in Section~\ref{sec:LowLevelAlgorithm} 
to specifically select $b$- and $c$-hadron decay track candidates and improve the rejection of tracks originating from pile-up.\footnote{Pile-up interactions correspond to additional $pp$ collisions accompanying the hard-scatter $pp$ interaction in proton bunch collisions at the LHC.}

\item {\textbf{Primary vertex}} (PV) reconstruction~\cite{PERF-2015-01,ATL-PHYS-PUB-2015-026} on an event-by-event basis is particularly important for \btag\ since it defines the reference point from which track and vertex displacements are computed. 
A longitudinal vertex position resolution of about $30~\mu$m is achieved for events with a high multiplicity of reconstructed tracks, while the transverse resolution ranges from 10 to 12~$\mu$m, depending on the LHC running conditions that determine the beam spot size.
At least one PV is required in each event, with the PV that has the highest sum of squared transverse momenta of contributing tracks selected as the primary interaction point.
Displaced charged-particle tracks originating from $b$-hadron decays can then be selected by requiring large transverse and longitudinal impact parameters, $\mathrm{IP}_{r\phi}=|d_0|$ and $\mathrm{IP}_{z}=|z_0 \sin \theta|$, respectively, where $d_0$ ($z_0$) represents the transverse (longitudinal) perigee parameter defined at the point of the closest approach of the trajectory to the $z$-axis~\cite{Billoir:245288}.  
Upper limits on the values of these parameters are used to reduce the contamination from pile-up, secondary and fake tracks.

\item{\textbf{Hadronic jets}} are built from topological clusters of energy in the calorimeter~\cite{PERF-2014-07}, calibrated to the
electromagnetic scale, using the anti-$k_t$ algorithm with radius parameter
$R=0.4$~\cite{Cacciari:2008gp}.
Jet transverse momenta are further corrected to the corresponding
particle-level jet \pt, based on the simulation~\cite{PERF-2016-04}. Remaining differences between data and simulated events are evaluated and corrected for using in situ techniques
, which exploit the transverse momentum balance between a jet and a reference object such as a photon, $Z$ boson, or multi-jet system in data.
After these calibrations, all jets in the event with $\pt \ge 20~\GeV$ and $|\eta| < 4.5$ must satisfy a set of loose jet-quality requirements~\cite{ATLAS-CONF-2015-029}, else the event is discarded. These requirements are designed to reject fake jets originating from sporadic bursts, large coherent noise or isolated pathological cells in the calorimeter system, hardware issues, beam-induced background or cosmic-ray muons.
Jets with $\pt < 20~\GeV$ or $|\eta| \ge 2.5$ are
then discarded.
In order to reduce the number of jets with large energy fractions from pile-up
collision vertices, the Jet Vertex Tagger (JVT) algorithm is
used~\cite{ATLAS-CONF-2014-018}.  The JVT procedure builds a multivariate discriminant for each jet within
$|\eta| < 2.5$ based on the
ID tracks ghost-associated with the jet~\cite{Cacciari:2008gn}; in particular, jets with a large fraction of high-momentum tracks from pile-up vertices are
less likely to pass the JVT requirement.
The rate of pile-up jets with $\pt \ge 120~\GeV$ is
sufficiently small that the JVT requirement is removed above this threshold.
The JVT efficiency for jets originating from the hard  $pp$ scattering is 92\% in the simulation. Scale factors of order unity are applied to account for efficiency differences relative to collision data.
 
\item{\textbf{Track--jet matching}} is performed using the angular separation $\Delta R$ between the track momenta, defined at the perigee, 
and the jet axis, defined as the vectorial sum of the four-momenta of the clusters associated with the jet. Given that the decay products from higher-\pt\ $b$-hadrons are more collimated, the $\Delta R$ requirement varies as a function of jet \pt, being wider for low-\pt\ jets (0.45 for jet $\pt=20~\GeV$) and narrower for high-\pt\ jets (0.26 for jet $\pt=150~\GeV$). If more than one jet fulfils the matching criteria, the closest jet is preferred. The jet axis is also used to assign signed impact parameters to tracks, where the sign is defined as positive if the track intersects the jet axis in the transverse plane in front of the primary vertex, and as negative if the intersection lies behind the primary vertex.
 
\item{\textbf{Jet flavour labels}} are attributed to the jets in the simulation. Jets are labelled as $b$-jets if they are matched to at least one weakly decaying
$b$-hadron having $\pt \ge 5~\GeV$ within a cone of size $\Delta R=0.3$ around the jet axis. If no
$b$-hadrons are found, $c$-hadrons and then $\tau$-leptons are searched for, based on the same selection criteria. The jets matched to a $c$-hadron ($\tau$-lepton) are labelled as $c$-jets ($\tau$-jets). The remaining jets are labelled as light-flavour jets.

\end{description}

 
\section{Algorithms for $b$-jet identification}
\label{sec:algorithms}
 
This section describes the different algorithms used for $b$-jet identification and the evaluation of their performance in simulation. Low-level \btag algorithms fall into two broad categories. A first approach, implemented in the \textsc{IP2D} and \textsc{IP3D} algorithms~\cite{ATL-PHYS-PUB-2017-013} and described in Section~\ref{sec:IPTag}, is inclusive and based on exploiting the large impact parameters of the tracks originating from the $b$-hadron decay. The second approach explicitly reconstructs displaced vertices. The \textsc{SV1} algorithm~\cite{ATL-PHYS-PUB-2017-011}, discussed in Section~\ref{sec:SV1}, attempts to reconstruct an inclusive secondary vertex, while the \textsc{JetFitter} algorithm~\cite{ATL-PHYS-PUB-2018-025}, presented in Section~\ref{sec:JetFitter}, aims to reconstruct the full $b$- to $c$-hadron decay chain. These algorithms, first introduced during Run 1~\cite{PERF-2012-04}, benefit from improvements and a new tuning for Run 2. To maximise the $b$-tagging performance, low-level algorithm results are combined using multivariate classifiers. To this end, two high-level tagging algorithms have been developed. The first one, \textsc{MV2}~\cite{ATL-PHYS-PUB-2017-013}, is based on a boosted decision tree (BDT) discriminant, while the second one, \textsc{DL1}~\cite{ATL-PHYS-PUB-2017-013}, is based on a deep feed-forward neural network (NN). These two algorithms are presented in Sections~\ref{sec:MV2} and~\ref{sec:DL1}, respectively.

\subsection{Training and tuning samples}

The new tuning and training strategies of the \btag algorithms for Run 2 are based on the use of a hybrid sample composed of \ttbar\ and $Z'$ simulated events. Only $\ttbar$ decays with at least one lepton from a subsequent $W$-boson decay are considered in order to ensure a sufficiently large fraction of $c$-jets in the event whilst maintaining a jet multiplicity profile similar to that in most analyses. A dedicated sample of $Z^\prime$ decaying into hadronic jet pairs is included to optimise the $b$-tagging performance at high jet \pt. 
The cross-section of the hard-scattering process is modified by applying an event-by-event weighting factor to broaden the natural width of the resonance and widen the transverse momentum distribution of the jets produced in the hadronic decays up to a jet $p_\mathrm{T}$ of 1.5~\TeV.
The branching fractions of the decays are set to be one-third each for the $bb$, $cc$ and light-flavour quark pairs to give a $p_\mathrm{T}$ spectrum uniformly populated by jets of all flavours.
The hybrid sample is obtained by selecting $b$-jets from \ttbar\ events if the corresponding $b$-hadron \pt\ is below $250~\GeV$ and from the $Z^\prime$ sample if above, with a similar strategy applied for $c$-jets and light-flavour jets. More details about the production of the \ttbar\ simulated sample, referred to as the baseline \ttbar\ sample in the following, and the $Z^\prime$ simulated sample are given in Section~\ref{sec:samples}. Events with at least one jet are selected, excluding the jets overlapping with a generator-level electron originating from a $W$- or $Z$-boson decay.

\subsection{Low-level \btag algorithms}
\label{sec:LowLevelAlgorithm}

\subsubsection{Algorithms based on impact parameters}
\label{sec:IPTag}
 
There are two complementary impact parameter-based algorithms, \textsc{IP2D} and \textsc{IP3D}~\cite{ATL-PHYS-PUB-2017-013}. The \textsc{IP2D} tagger makes use of the signed transverse impact parameter significance of tracks to construct a discriminating variable, whereas \textsc{IP3D} uses both the track signed transverse and the longitudinal impact parameter significance in a two-dimensional template to account for their correlation.
Probability density functions (pdf) obtained from reference histograms
of the signed transverse and longitudinal impact parameter significances of tracks associated with $b$-jets, $c$-jets and light-flavour jets are derived from MC simulation. The pdfs are computed in exclusive categories that depend on the hit pattern of the tracks to increase the discriminating power. The pdfs are used to calculate ratios of the $b$-jet, $c$-jet and light-flavour jet probabilities on a per-track basis. 
Log-likelihood ratio (LLR) discriminants are then defined as the sum of the per-track probability ratios for each jet-flavour hypothesis, e.g. $\sum_{i=1}^{N} \log \left({p_b}/{p_u}\right)$ for the $b$-jet and light-flavour jet hypotheses, where $N$ is the number of tracks associated with the jet and $p_b$ ($p_u$) is the template pdf for the $b$-jet (light-flavour jet) hypothesis. 
The flavour probabilities of the different tracks contributing to the sum are assumed to be independent of each other.
In addition to the LLR separating $b$-jets from light-flavour jets, two extra LLR functions are defined to separate $b$-jets from $c$-jets and $c$-jets from light-flavour jets, respectively. These three likelihood discriminants for both the \textsc{IP2D} and \textsc{IP3D} algorithms are used as inputs to the high-level taggers.

Both the \textsc{IP2D} and \textsc{IP3D} algorithms benefited from a complete retuning prior to the 2017--2018 ATLAS data taking period~\cite{ATL-PHYS-PUB-2017-013}. In particular, a reoptimisation of the track category definitions was performed, allowing the IBL hit pattern expectations and the next innermost layer information to be fully exploited. The rejection of tracks originating from photon conversions, long-lived particles decays ($K_\text{s}$, $\Lambda$) and interactions with detector material by the secondary vertex algorithms has also been improved.
An additional set of new template pdfs was also produced using a 50\%/50\% mixture of \ttbar and $Z^\prime$ simulated events for extra track-categories with no hits in the first two layers, which are populated by long-lived $b$-hadrons traversing the first layers before they decay. The \ttbar\ sample is used to populate all remaining categories.
 
\subsubsection{Secondary vertex finding algorithm}
\label{sec:SV1}

The secondary vertex tagging algorithm, \textsc{SV1}~\cite{ATL-PHYS-PUB-2017-011}, reconstructs a single displaced secondary vertex in a jet. The reconstruction starts from identifying the possible two-track vertices built with all tracks associated with the jet, while rejecting tracks that are compatible with the decay of long-lived particles ($K_\text{s}$ or $\Lambda$), photon conversions or hadronic interactions with the detector material. The \textsc{SV1} algorithm runs iteratively on all tracks contributing to the cleaned two-tack vertices, trying to fit one secondary vertex. In each iteration, the track-to-vertex association is evaluated using a $\chi^2$ test. The track with the largest $\chi^2$ is removed and the vertex fit is repeated until an acceptable vertex $\chi^2$ and a vertex invariant mass less than 6 GeV are obtained. With this approach, the decay products from $b$- and $c$-hadrons are assigned to a single common secondary vertex.
 
Several refinements in the track and vertex selection were made prior to the 2016--2017 ATLAS data taking period to improve the performance of the algorithm, resulting in an increased pile-up rejection and an overall enhancement of the performance at high jet \pt~\cite{ATL-PHYS-PUB-2017-011}. Among the various algorithm improvements, additional track-cleaning requirements are applied for jets in the high-pseudorapidity region ($|\eta |\ge1.5$) to mitigate the negative influence of the increasing amount of detector material on the secondary vertex finding efficiency. The fake-vertex rate is also better controlled by limiting the algorithm to only consider the 25 highest-\pt tracks in the jets, which preserves all reconstructed tracks from $b$-hadron decays, whilst limiting the influence of additional tracks in the jet. The selection of two-track vertex candidates prior to the $\chi^2$--fit was also reoptimised. Extra candidate-cleaning requirements were introduced to further reduce the number of fake vertices and material interactions, such as the rejection of two-track vertex candidates with an invariant mass greater than $6~\GeV$, which are not likely to originate from $b$- and $c$-hadron decays. Eight discriminating variables, including the number of tracks associated with the \textsc{SV1} vertex, the invariant mass of the secondary vertex, its energy fraction (defined as the total energy of all the tracks associated with the secondary vertex divided by the energy of all the tracks associated with the jet), and the three-dimensional decay length significance are used as inputs to the high-level taggers. The \btag performance of the \textsc{SV1} algorithm is evaluated using a LLR discriminant based on pdfs for the $b$-jet, $c$-jet and light-flavour jet hypotheses computed from three-dimensional histograms built from three \textsc{SV1} output variables: the vertex mass, the energy fraction and the number of two-track vertices.
 
\subsubsection{Topological multi-vertex finding algorithm}
\label{sec:JetFitter}
 
The topological multi-vertex algorithm, \textsc{JetFitter}~\cite{ATL-PHYS-PUB-2018-025}, exploits the topological structure of weak $b$- and $c$-hadron decays inside the jet and tries to reconstruct the full $b$-hadron decay chain. A modified Kalman filter~\cite{Fruhwirth:1987fm} is used to find a common line on which the primary, bottom and charm vertices lie, approximating the $b$-hadron flight path as well as the vertex positions. With this approach, it is possible to resolve the $b$- and $c$-hadron vertices even when a single track is attached to them.
 
Several improvements~\cite{ATL-PHYS-PUB-2018-025}, prior to the 2017-2018 ATLAS data taking period, have been introduced in the current version of the \textsc{JetFitter} algorithm. These include, a reoptimisation of the track selection to better mitigate the effect of pile-up tracks, an improvement in the rejection of material interactions, and the introduction of a vertex-mass dependent selection during the decay chain fit to increase the efficiency for tertiary vertex reconstruction. Eight discriminating variables, including the track multiplicity at the \textsc{JetFitter} displaced vertices, the invariant mass of tracks associated with these vertices, their energy fraction and their average three-dimensional decay length significance, are used as inputs to the high-level taggers. The \btag performance of the \textsc{JetFitter} algorithm is evaluated using a LLR discriminant based on likelihood functions combining pdfs extracted from some of the \textsc{JetFitter} output variables (vertex mass, energy fraction and decay length significance) and parameterised for each of the three jet flavours.

\subsection{High-level tagging algorithms}

\subsubsection{MV2}
\label{sec:MV2}
 
The \textsc{MV2} algorithm~\cite{ATL-PHYS-PUB-2017-013} consists of a boosted decision tree (BDT) algorithm that combines the outputs of the low-level tagging algorithms described in Section~\ref{sec:LowLevelAlgorithm} and listed in Table~\ref{tab:HLTaggerInputs}. The BDT algorithm is trained using the ROOT Toolkit for Multivariate Data Analysis (TMVA)~\cite{TMVA} on the hybrid $\ttbar+Z^\prime$ sample. 
The kinematic properties of the jets, namely \pt\ and $|\eta|$, are included in the training in order to take advantage of the correlations with the other input variables. However, to avoid differences in the kinematic distributions of signal ($b$-jets) and background ($c$-jets and light-flavour jets) being used to discriminate between the different jet flavours, the $b$-jets and $c$-jets are reweighted in \pt\ and $|\eta|$ to match the spectrum of the light-flavour jets. No kinematic reweighting is applied at the evaluation stage of the multivariate classifier. For training, the $c$-jet fraction in the background sample is set to 7\%, with the remainder composed of light-flavour jets. This allows the charm rejection to be enhanced whilst preserving a high light-flavour jet rejection.
The BDT training hyperparameters of the \textsc{MV2} tagging algorithm are listed in Table~\ref{tab:MV2HyperPar}. They have been optimised to provide the best separation power between the signal and the background. The output discriminant of the \textsc{MV2} algorithm for $b$-jets, $c$-jets and light-flavour jets evaluated with the baseline \ttbar\ simulated events 
are shown in Figure~\ref{fig:MV2}.
 
\begin{table}[htbp]
\caption{Input variables used by the \textsc{MV2} and the \textsc{DL1} algorithms. The \textsc{JetFitter} $c$-tagging variables are only used by the \textsc{DL1} algorithm.}
\label{tab:HLTaggerInputs}
\begin{center}
\small
\begin{tabular}{|l|c|p{0.4\textwidth}|}
\hline
Input & Variable & Description \\
\hline
\multirow{2}{*}{Kinematics}
& \pt & Jet \pt \\
& $\eta$ & Jet $|\eta|$ \\
\hline
\multirow{3}{*}{IP2D/IP3D}
& $\log(P_{b}/P_{\mathrm{light}})$  & Likelihood ratio between the $b$-jet and light-flavour jet hypotheses  \\
& $\log(P_{b}/P_{\mathrm{c}})$  & Likelihood ratio between the $b$- and $c$-jet hypotheses  \\
& $\log(P_{c}/P_{\mathrm{light}})$  & Likelihood ratio between the $c$-jet and light-flavour jet hypotheses  \\
\hline
\multirow{8}{*}{SV1}
& $m$(SV) & Invariant mass of tracks at the secondary vertex assuming pion mass \\
& $f_E$(SV) & Energy fraction of the tracks associated with the secondary vertex \\
& $N_{{\mathrm{TrkAtVtx}}}$(SV) & Number of tracks used in the secondary vertex \\
& $N_{{\mathrm{2TrkVtx}}}$(SV) & Number of two-track vertex candidates \\
& $L_{xy}$(SV) & Transverse distance between the primary and secondary vertex \\
& $L_{xyz}$(SV)& Distance between the primary and the secondary vertex  \\
& $S_{xyz}$(SV)& Distance between the primary and the secondary vertex divided by its uncertainty \\
& $\Delta R(\vec p_{\mathrm{jet}}, \vec p_{\mathrm{vtx}})$(SV) & $\Delta R$ between the jet axis and the direction of the secondary vertex relative to the primary vertex. \\
\hline
\multirow{8}{*}{\textsc{JetFitter}}
& $m$(JF) & Invariant mass of tracks from displaced vertices \\
& $f_E$(JF) &  Energy fraction of the tracks associated with the displaced vertices \\
& $\Delta R(\vec p_{\mathrm{jet}}, \vec p_{\mathrm{vtx}})$(JF) & $\Delta R$ between the jet axis and the vectorial sum of momenta of all tracks attached to displaced vertices \\
& $S_{xyz}$(JF) & Significance of the average distance between PV and displaced vertices \\
& $N_{{\mathrm{TrkAtVtx}}}$(JF) & Number of tracks from multi-prong displaced vertices \\
& $N_{{\mathrm{2TrkVtx}}}$(JF) & Number of two-track vertex candidates (prior to decay chain fit) \\
& $N_{{\mathrm {1\mbox{-}trk\ vertices}}}$(JF) & Number of single-prong displaced vertices \\
& $N_{{\geq \mathrm{2\mbox{-}trk\ vertices}}}$(JF) & Number of multi-prong displaced vertices \\
\hline
\multirow{6}{*}{\textsc{JetFitter} $c$-tagging}
& $L_{xyz}(2^{{\mathrm{nd}}}/3^{{\mathrm{rd}}}{\mathrm{vtx}})$(JF) & Distance of $2^{{\mathrm{nd}}}$ or $3^{{\mathrm{rd}}}$ vertex from PV \\
& $L_{xy}(2^{{\mathrm{nd}}}/3^{{\mathrm{rd}}}{\mathrm{vtx}})$(JF) & Transverse displacement of the $2^{{\mathrm{nd}}}$ or $3^{{\mathrm{rd}}}$ vertex \\
& $m_{\mathrm{Trk}}(2^{{\mathrm{nd}}}/3^{{\mathrm{rd}}}{\mathrm{vtx}})$(JF) & Invariant mass of tracks associated with $2^{{\mathrm{nd}}}$ or $3^{{\mathrm{rd}}}$ vertex \\
& $E_{\mathrm{Trk}}(2^{{\mathrm{nd}}}/3^{{\mathrm{rd}}}{\mathrm{vtx}})$(JF) &  Energy fraction of the tracks associated with $2^{{\mathrm{nd}}}$ or $3^{{\mathrm{rd}}}$ vertex   \\
& $f_E (2^{{\mathrm{nd}}}/3^{{\mathrm{rd}}}{\mathrm{vtx}})$(JF) & Fraction of charged jet energy in $2^{{\mathrm{nd}}}$ or $3^{{\mathrm{rd}}}$ vertex\\
& $N_{{\mathrm{TrkAtVtx}}}(2^{{\mathrm{nd}}}/3^{{\mathrm{rd}}}{\mathrm{vtx}}) $(JF) & Number of tracks associated with $2^{{\mathrm{nd}}}$ or $3^{{\mathrm{rd}}}$ vertex \\
& $ Y_{\mathrm{trk}}^{\mathrm{min}}, Y_{\mathrm{trk}}^{\mathrm{max}}, Y_{\mathrm{trk}}^{\mathrm{avg}} (2^{{\mathrm{nd}}}/3^{{\mathrm{rd}}}{\mathrm{vtx}}) $(JF) & Min., max. and avg. track rapidity of tracks at $2^{{\mathrm{nd}}}$ or $3^{{\mathrm{rd}}}$ vertex \\
\hline
\end{tabular}
\end{center}
\end{table}

\begin{table}[htbp]
\caption{List of optimised hyperparameters used in the \textsc{MV2} tagging algorithm.}
\label{tab:MV2HyperPar}
\begin{center}
\small
\begin{tabular}{|l|c|}
\hline
Hyperparameter & Value \\
\hline
Number of trees & 1000 \\
Depth & 30 \\
Minimum node size & 0.05\% \\ 
Cuts & 200 \\
Boosting type &  Gradient boost \\
Shrinkage &  0.1 \\
Bagged sample fraction &  0.5 \\
\hline
\end{tabular}
\end{center}
\end{table}
 
\subsubsection{DL1}
\label{sec:DL1}
 
The second high-level \btag algorithm, \textsc{DL1}~\cite{ATL-PHYS-PUB-2017-013}, is based on a deep feed-forward neural network (NN) trained using {\textsc Keras}~\cite{chollet2015keras} with the {\textsc Theano}~\cite{2016arXiv160502688short} backend and the {\textsc Adam} optimiser~\cite{Kingma:2014aa}. The \textsc{DL1} NN has a multidimensional output corresponding to the probabilities for a jet to be a $b$-jet, a $c$-jet or a light-flavour jet. The topology of the output consists of a mixture of fully connected hidden and Maxout layers~\cite{Goodfellow:2013aa}. The input variables to \textsc{DL1} consist of those used for the \textsc{MV2} algorithm with the addition of the \textsc{JetFitter} $c$-tagging variables listed in Table~\ref{tab:HLTaggerInputs}. The latter relate to the dedicated properties of the secondary and tertiary vertices (distance to the primary vertex, invariant mass and number of tracks, energy, energy fraction, and rapidity of the tracks associated with the secondary and tertiary vertices). A jet \pt and  $|\eta|$ reweighting similar to the one used for \textsc{MV2} is performed. The \textsc{DL1} algorithm parameters, listed in Table~\ref{tab:DL1HyperPar}, include the architecture of the NN, the number of training epochs, the learning rates and training batch size. All of these are optimised in order to maximise the \btag\ performance. Batch normalisation~\cite{Ioffe:2015aa} is added by default since it is found to improve the performance.

\begin{table}[htbp]
\caption{List of optimised hyperparameters used in the \textsc{DL1} tagging algorithm.}
\label{tab:DL1HyperPar}
\begin{center}
\small
\begin{tabular}{|l|c|}
\hline
Hyperparameter & Value  \\
\hline
Number of input variables & 28 \\
Number of hidden layers & 8 \\
Number of nodes [per layer] & [78, 66, 57, 48, 36, 24, 12, 6] \\
Number of Maxout layers [position] & 3 [1, 2, 6] \\
Number of parallel layers per Maxout layer & 25 \\
Number of training epochs & 240 \\
Learning rate & 0.0005 \\
Training minibatch size & 500 \\
\hline
\end{tabular}
\end{center}
\end{table}

Training with multiple output nodes offers additional flexibility when constructing the final output discriminant by combining the $b$-jet, $c$-jet and light-flavour jet probabilities. Since all flavours are treated equally during training, the trained network can be used for both $b$-jet and $c$-jet tagging. In addition, the use of a multi-class network architecture provides the \textsc{DL1} algorithm with a smaller memory footprint than BDT-based algorithms.
The final \textsc{DL1} \btag discriminant is defined as:
\begin{equation*}
D_{\mathrm{DL1}}= \ln \left( \frac{p_{b}}{f_{c} \cdot p_{c} + (1-f_{c}) \cdot p_{\mathrm{light}}}\right),
\end{equation*}
where $p_{b}$, $p_{c}$, $p_{\mathrm{light}}$ and $f_{c}$ represent respectively the $b$-jet, $c$-jet and light-flavour jet probabilities, and the effective $c$-jet fraction in the background training sample. Using this approach, the  $c$-jet fraction in the background can be chosen a posteriori in order to optimise the performance of the algorithm. An optimised $c$-jet fraction of 8\% is used to evaluate the performance of the \textsc{DL1} \btag algorithm in this paper.

\begin{figure}[htbp]
\centering
\subfigure[]{\includegraphics[width=0.45\textwidth]{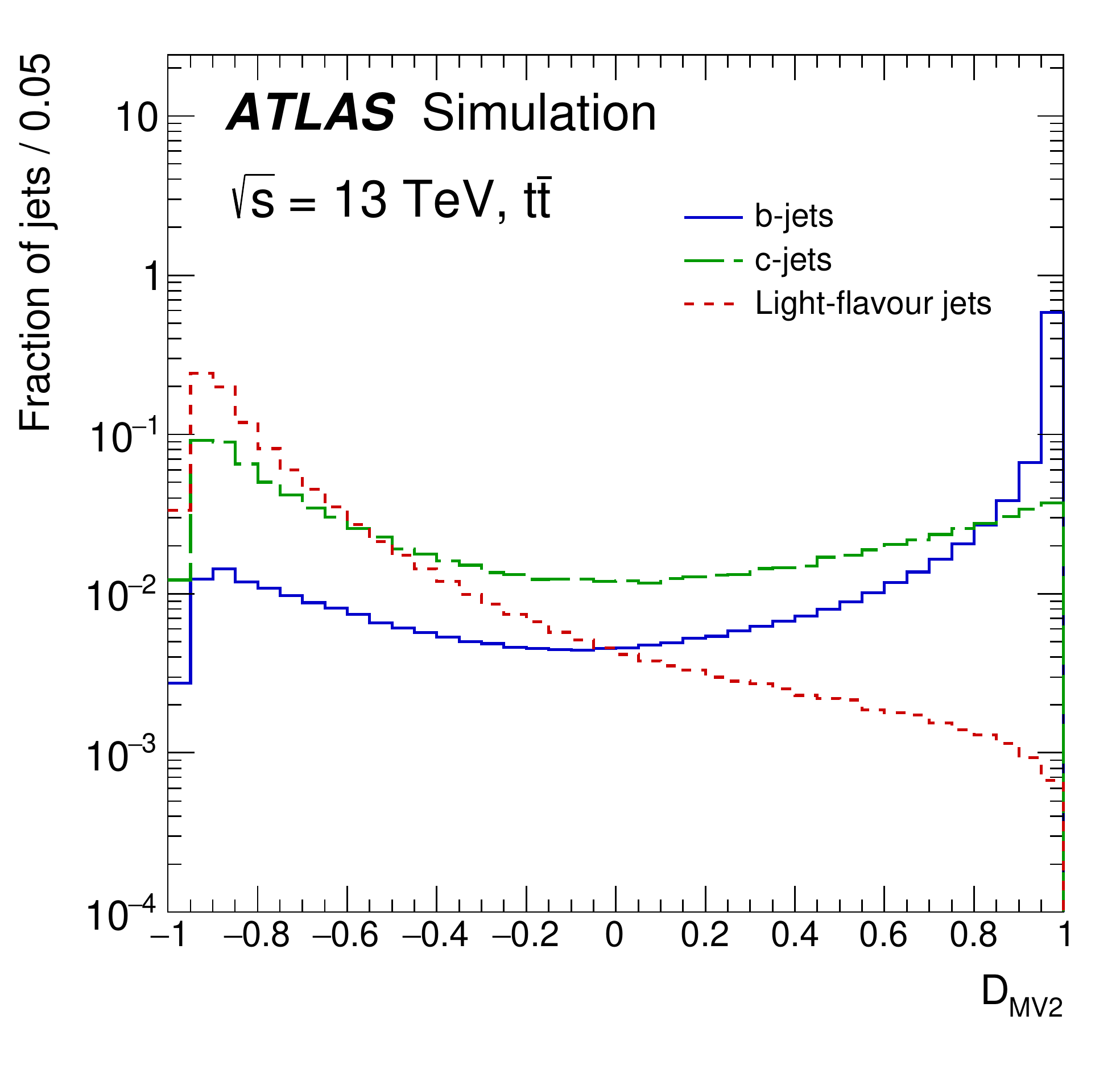} \label{fig:MV2}}
\subfigure[]{\includegraphics[width=0.45\textwidth]{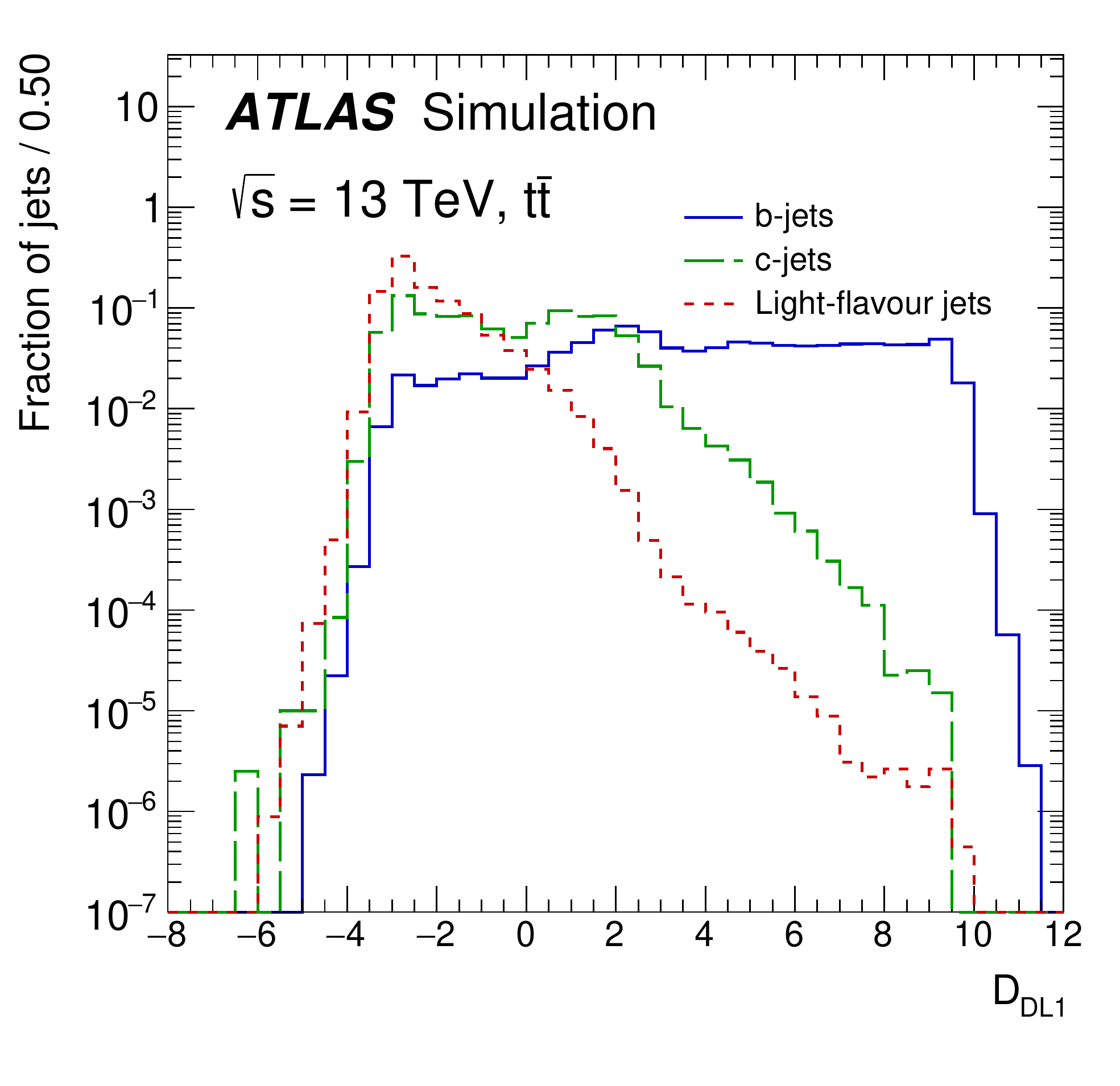} \label{fig:DL1}}
\caption{Distribution of the output discriminant of the (a) \textsc{MV2} and (b) \textsc{DL1} \btag algorithms for $b$-jets, $c$-jets and light-flavour jets in the baseline \ttbar\ simulated events. 
}
\end{figure}

The output discriminants of the \textsc{DL1} $b$-tagging algorithms for $b$-jets, $c$-jets and light-flavour jets in the baseline \ttbar\ simulated events 
are shown in Figure~\ref{fig:DL1}.

\subsection{Algorithm performance}
\label{sec:TaggerPerformance}

The evaluation of the performance of the algorithms is carried out using $b$-jet tagging single-cut operating points (OPs). These are based on a fixed selection requirement on the \btag algorithm discriminant distribution ensuring a specific $b$-jet tagging efficiency, \beff, for the $b$-jets present in the baseline \ttbar\ simulated sample. 
The selections used to define the single-cut OPs of the \textsc{MV2} and the \textsc{DL1} algorithms, as well as the corresponding $c$-jet, $\tau$-jet and light-flavour jet rejections, are shown in Table~\ref{tab:EffCuts}. The \textsc{MV2} and the \textsc{DL1} discriminant distributions are also divided into five `pseudo-continuous' bins, $(O^k)_{k=1..5}$, delimited by the selections used to define the $b$-jet tagging single-cut OPs for 85\%, 77\%, 70\% and 60\% efficiency, and bounded by the trivial 100\% and 0\% selections. The value of the pdf in each bin is called the $b$-jet tagging probability and labelled $\Pro_b$ in the following. The $b$-jet tagging efficiency of the $\beff=X\%$ single-cut OP can then be defined as the sum of the $b$-jet tagging probabilities in the range [$X$\%,0\%].
 
\begin{figure}[htbp]
\begin{center}
\subfigure[]{\includegraphics[width=0.45\textwidth]{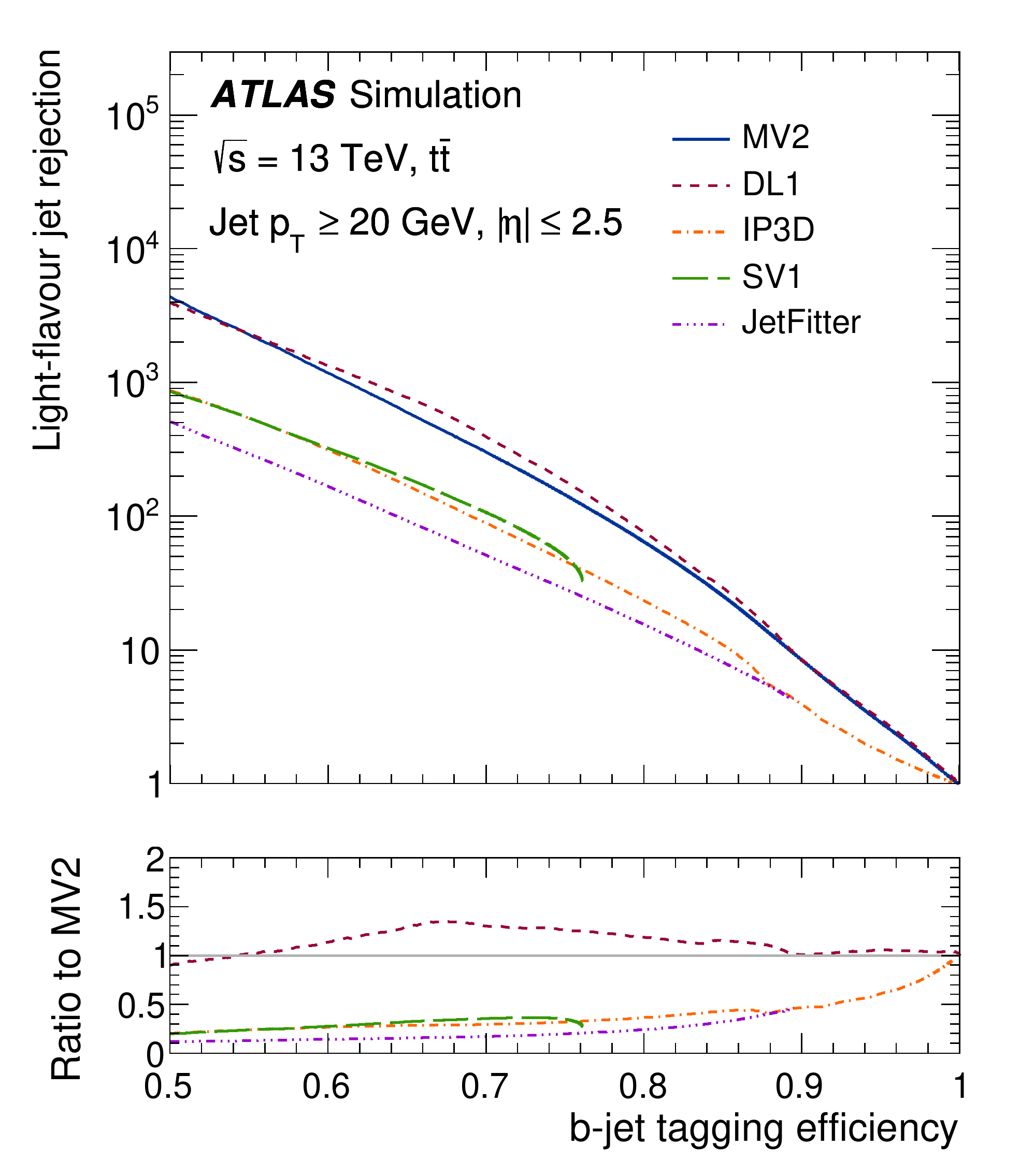}}
\subfigure[]{\includegraphics[width=0.45\textwidth]{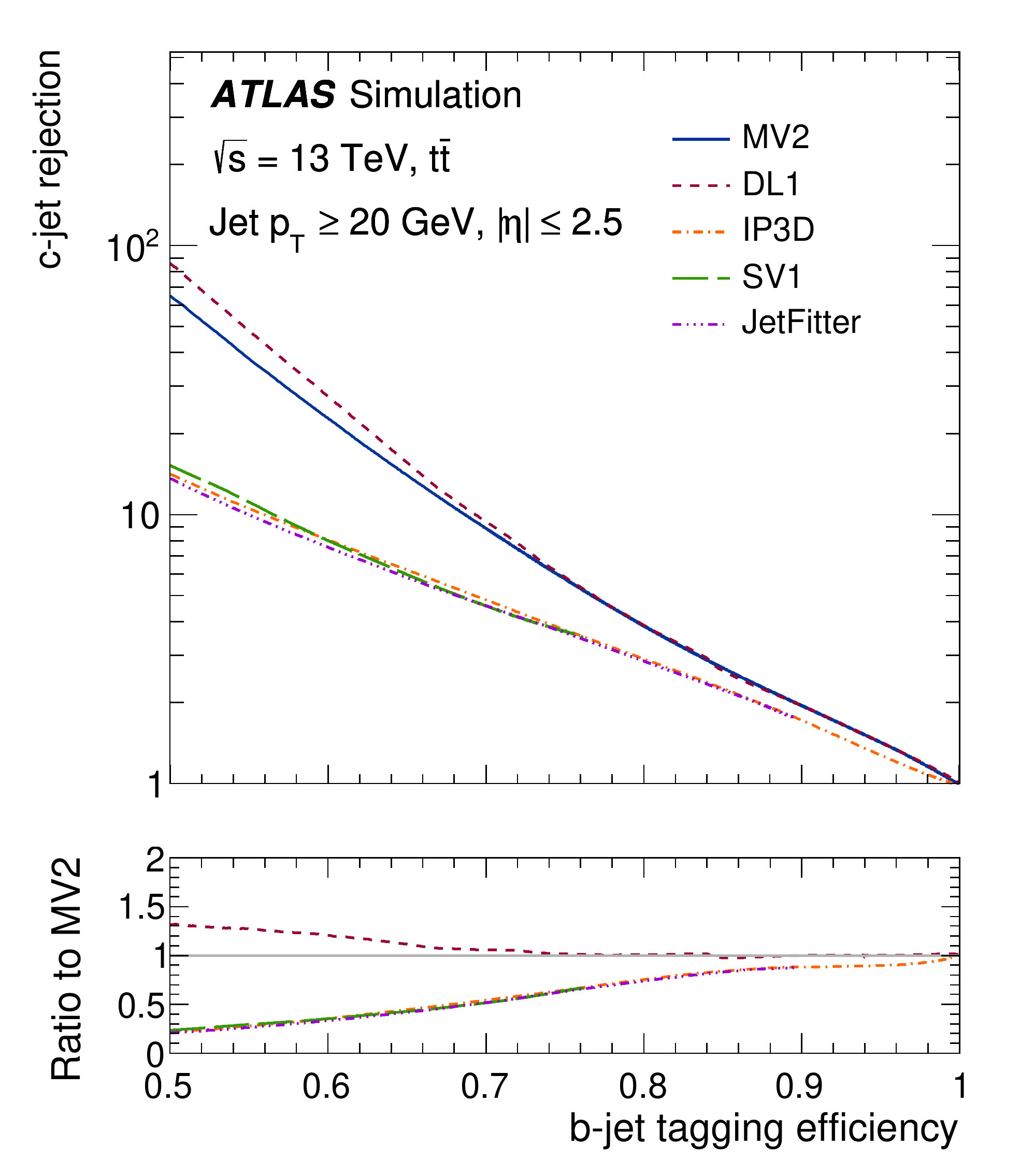}}
\end{center}
\caption{The (a) light-flavour jet and (b) $c$-jet rejections versus the $b$-jet tagging efficiency for the \textsc{IP3D}, \textsc{SV1}, \textsc{JetFitter}, \textsc{MV2} and \textsc{DL1} \btag algorithms evaluated on the baseline \ttbar events.
} \label{fig:ROCsF}
\end{figure}

The light-flavour jet and $c$-jet rejections as a function of the $b$-jet tagging efficiency are shown in Figure~\ref{fig:ROCsF} for the various low- and high-level \btag algorithms.
This demonstrates the advantage of combining the information provided by the low-level taggers, where improvements in the light-flavour jet and $c$-jet rejections by factors of around 10 and 2.5, respectively, are observed at the $\epsilon_b$ = 70\% single-cut OP of the high-level algorithms compared to low-level algorithms. This figure also illustrates the different $b$-jet tagging efficiency range accessible with each low-level algorithm and thereby their complementarity in the multivariate combinations, with the performance of the \textsc{DL1} and \textsc{MV2} discriminants found to be similar. The two algorithms tag a highly correlated sample of $b$-jets, where the relative fraction of jet exclusively tagged by each algorithm is around 3\% at the $\varepsilon_b = 70\%$ single-cut OP. The relative fractions of light-flavour jets exclusively mis-tagged by the \textsc{MV2} or the \textsc{DL1} algorithms at the $\varepsilon_b = 70\%$ single-cut OP reach 0.2\% and 0.1\%, respectively.

However, the additional \textsc{JetFitter} $c$-tagging variables used by \textsc{DL1} bring around 30\% and 10\% improvements in the light-flavour jet and $c$-jet rejections, respectively, at the $\epsilon_b = 70\%$ single-cut OP compared to \textsc{MV2}.

\begin{table}
\caption{Selection and $c$-jet, $\tau$-jet and light-flavour jet rejections corresponding to the different $b$-jet tagging efficiency single-cut operating points for the \textsc{MV2} and the \textsc{DL1} \btag algorithms, evaluated on the baseline \ttbar events.
}
\label{tab:EffCuts}
\begin{center}
\small
\begin{tabular}{|c|c|r|r|r|c|r|r|r|}
\hline
\multirow{2}{*}{$\epsilon_b$} &\multicolumn{4}{c|}{\textsc{MV2}} &\multicolumn{4}{c|}{\textsc{DL1}} \\
\cline{2-9}
& \multirow{2}{*}{Selection} &\multicolumn{3}{c|}{Rejection} & \multirow{2}{*}{Selection} &\multicolumn{3}{c|}{Rejection} \\
\cline{3-5} \cline{7-9}
& & $c$-jet & $\tau$-jet & Light-flavour jet &  & $c$-jet & $\tau$-jet & Light-flavour jet \\
\hline
60\% & $>0.94$ & 23 & 140 & 1200 & $>2.74$ & 27 & 220 & 1300 \\
70\% & $>0.83$ & 8.9 & 36 & 300 & $>2.02$ & 9.4 & 43 & 390 \\
77\% & $>0.64$ & 4.9 & 15 & 110 & $>1.45$ & 4.9 & 14 & 130  \\
85\% & $>0.11$ & 2.7 & 6.1  & 25 & $>0.46$ & 2.6 & 3.9 & 29 \\
\hline
\end{tabular}
\end{center}
\end{table}
 
\section{Data and simulated samples}
\label{sec:samples}
 
In order to use data to evaluate the performance of the high-level $b$-tagging algorithms, a sample of events enriched in $t\bar{t}$ dileptonic decays is selected.
 
The analysis is performed with a $pp$ collision data sample collected at
a centre-of-mass energy of $\sqrt{s}=13~\TeV$ during the years 2015, 2016 and 2017, corresponding to an integrated
luminosity of $80.5$~\ifb{} and a mean number of $pp$ interactions per bunch crossing of 31.9. The uncertainty in the integrated luminosity is 2.0\%~\cite{ATLAS-CONF-2019-021}, obtained using the LUCID-2 detector~\cite{LUCID2} for the primary luminosity measurements. 
All events used were recorded during periods when all relevant ATLAS detector components were functioning normally.
The dataset was collected using triggers requiring the presence of a
single, high-\pt\ electron or muon,
with \pt\ thresholds that yield an approximately constant efficiency for leptons passing an offline selection of $\pt\ge 28~\GeV$.

The baseline \ttbar full simulation sample was produced using
\powhegbox~v2~\cite{Nason:2004rx,Frixione:2007vw,Frixione:2007nw,Alioli:2010xd} where the first-gluon-emission cut-off scale parameter $h_{\textrm{damp}}$ is set to $1.5m_{t}$, with $m_{\textrm{top}}=172.5~\GeV$ used for the top-quark mass.
\powhegbox was interfaced to \textsc{Pythia}~8.230~\cite{Sjostrand:2007gs} with the A14 set of tuned parameters~\cite{ATL-PHYS-PUB-2014-021} and \textsc{NNPDF30NNLO} (\textsc{NNPDF2.3LO})~\cite{Ball:2014uwa,Ball:2012cx} parton distribution functions in the matrix elements (parton shower).
This set-up was found to produce the best modelling of the multiplicity of additional jets and both the individual top-quark and \ttbar system \pt~\cite{ATL-PHYS-PUB-2016-020}.
 
Alternative \ttbar simulation samples were generated using \powhegbox~v2
interfaced to \textsc{Herwig}~7.0.4~\cite{Bellm:2015jjp} with the
H7-UE-MMHT set of tuned parameters. 
The effects of initial- and final-state radiation (ISR, FSR) are explored by reweighting the
baseline \ttbar events in a manner that reduces (reduces and increases) initial (final) parton shower
radiation~\cite{ATL-PHYS-PUB-2016-020} and by using an alternative \powhegbox~v2 + \textsc{Pythia}~8.230 sample
with $h_{\textrm{damp}}$ set to $3m_{\textrm{top}}$ and
parameter variation group \textsc{Var3} (described in Ref.~\cite{ATL-PHYS-PUB-2016-020})
increased, leading to increased ISR.
 
The majority of events with at least one `fake' lepton in the selected sample arise from \ttbar production where only one of the $W$ bosons, which originated from a top-quark decay, decays leptonically.
These fake leptons come from several sources, including
non-prompt leptons produced from bottom or charm hadron decays, electrons arising from a photon conversion, jets
misidentified as electrons, or muons produced from in-flight pion or kaon decays.
This background is also modelled using the \ttbar\ production described above. The rate of events with two fake leptons is found to be negligible.
 
Non-\ttbar processes, which are largely subdominant in this analysis, can be
classified into two types: those with two real prompt leptons from $W$ or $Z$
decays (dominant) and those where at least one of the reconstructed lepton
candidates is `fake' (subdominant). 
Backgrounds containing two real prompt leptons include single top production in
association with a $W$ boson ($Wt$), diboson production ($WW$, $WZ$, $ZZ$) where at least
two leptons are produced in the electroweak boson decays, and $Z$+jets, with
$Z$ decaying into leptons.
The $Wt$ single top production was modelled using \powhegbox~v2
interfaced to \textsc{Pythia}~8.230 using the `diagram removal' scheme~\cite{Frixione:2008yi,Re:2010bp} with the A14 set of tuned parameters and the \textsc{NNPDF30NNLO} (\textsc{NNPDF2.3LO})~\cite{Ball:2014uwa,Ball:2012cx} parton distribution functions in the matrix elements (parton shower).
Diboson production with additional jets was simulated using
\textsc{Sherpa}~\cite{Gleisberg:2008ta,Hoeche:2009rj} v2.2.1 (for events where one boson decays
hadronically) or \textsc{Sherpa} v2.2.2 (for events where no bosons
decay hadronically), using the PDF
set \textsc{NNPDF30NNLO}~\cite{Ball:2014uwa}. This includes the $4\ell$, $\ell\ell\ell\nu$, $\ell\ell\nu\nu$, $\ell\nu\nu\nu$, $\ell\ell qq$ and $\ell\nu qq$ final states, which cover
$WW$, $WZ$ and $ZZ$ production including off-shell $Z$ contributions.
$Z$+jets production (including both $Z\rightarrow\tau\tau$ and
$Z\rightarrow ee/\mu\mu$) was modelled using \textsc{Sherpa} v2.2.1 with
PDF set \textsc{NNPDF30NNLO}.
Processes with one real lepton include $t$-channel and $s$-channel single top production~\cite{Alioli:2009je}. These processes were modelled with the same generator and parton shower combination as the $Wt$ channel.
$W$+jets production, with the $W$ boson decaying into $e\nu$, $\mu\nu$ or $\tau\nu$ with the
$\tau$-lepton decaying leptonically, was modelled in a similar way to the $Z$+jets production described above.
 
Alternative samples of non-\ttbar processes include the $Wt$ single top production
using \powhegbox~v2 interfaced to \textsc{Herwig}~7.0.4. 
The effects of ISR and FSR are evaluated by reweighting the baseline single-top events in a manner that either reduces or increases parton shower radiation. An additional $Wt$ sample using \powhegbox~v2 interfaced to \textsc{Pythia}~8.230 with the alternative `diagram subtraction' scheme~\cite{Frixione:2008yi,Re:2010bp} is used to investigate the impact of the interference between \ttbar\ and $Wt$ production. Uncertainties in diboson and $Z$+jets production are estimated by reweighting the baseline samples, whereas uncertainties in processes with one real lepton are evaluated directly from data, as described later in Section~\ref{sec:uncertainties}.

As described in Section~\ref{sec:algorithms}, the new Run 2 \btag algorithm training strategy is based on the use of a hybrid sample composed of both the baseline \ttbar event sample and a dedicated sample of $Z^\prime$ decaying into hadronic jet pairs. This $Z^\prime$ sample was generated using \textsc{Pythia}~8.2.12 with the \textsc{A14} set of tuned parameters for the underlying event and the leading-order \textsc{NNPDF2.3LO}~\cite{Ball:2012cx} parton distribution functions.
 
The \textsc{Evtgen}~\cite{Lange:2001uf} package was used to handle the decay of heavy-flavour hadrons for all samples except for those generated with the \textsc{Sherpa} generator, for which the default \textsc{Sherpa} configuration recommended by the \textsc{Sherpa} authors was used. All MC events have additional overlaid minimum-bias interactions generated with \textsc{Pythia}~8.160 with the A3 set of tunes parameters~\cite{ATL-PHYS-PUB-2016-017} and \textsc{NNPDF2.3LO} parton distribution functions to simulate 
pile-up background and are weighted to reproduce the observed distribution of the average number of interactions per bunch crossing of the corresponding data sample. The nominal MC samples were processed through the full ATLAS detector simulation~\cite{SOFT-2010-01} based on \textsc{GEANT4}~\cite{Agostinelli:2002hh}, but most samples used for systematic uncertainty evaluation were processed with a faster simulation making use of parameterised showers in the calorimeters~\cite{ATL-PHYS-PUB-2010-013}. The simulated events were reconstructed using the same algorithms as the data.

 
\section{Event selection and classification}
\label{sec:selection}

A sample of events enriched in $t\bar{t}$ dileptonic decays is selected by requiring exactly two well-identified lepton candidates and two jets to be present in each event. Events are further classified on the basis of two topological variables to control processes including non-\bjets. 
The lepton definition, and the event selection and classification are described in this section.
 
\subsection{Lepton object definition}
 
In addition to the objects reconstructed for $b$-tagging, described in Section~\ref{sec:objects}, the event selection for the efficiency measurement requires electron and muon candidates, defined as follows.
 
 
\begin{description}

\item {\textbf{Electron candidates}} are reconstructed from an isolated energy deposit in the electromagnetic calorimeter matched to an ID track~\cite{new_egamma}. Electrons are selected for inclusion in the analysis within the fiducial region of transverse energy $E_{\text{T}} \ge  28$~\GeV{} and $|\eta| < 2.47$. Candidates within the transition region between the barrel and endcap electromagnetic calorimeters, $1.37 \le |\eta| < 1.52 $, are removed in order to avoid
large trigger efficiency uncertainties in the turn-on region of the
lowest \pt trigger. A tight likelihood-based electron identification requirement is used to further suppress the background from multi-jet production.
Isolation criteria are used to reject candidates coming from sources other than prompt decays from massive bosons (hadrons faking an electron signature, heavy-flavour decays or photon conversions). 
Scale factors (SFs), of order unity,  derived in
$Z\rightarrow e^+e^-$ events are applied to simulated events to account for differences in
reconstruction, identification and isolation efficiencies between data
and simulation.  Electron energies are calibrated using the  
$Z$ mass peak.
 
\item{\textbf{Muon candidates}} are reconstructed by combining tracks found in the ID with tracks found in the muon spectrometer~\cite{PERF-2015-10}. Muons are selected for inclusion in the analysis within the fiducial region of transverse momentum $p_{\text{T}} \ge 28$~\GeV{} and $|\eta| < 2.5$.
If the event contains a muon reconstructed
from high hit multiplicities in the muon spectrometer due to very energetic
punch-through jets or from badly measured inner detector tracks in jets wrongly
matched to muon spectrometer track segments, the whole event is vetoed.
A tight muon identification requirement is applied to the muon candidates to further suppress the background.
Isolation selections similar to the ones applied to the electron candidates are imposed to reject candidates coming from sources other than prompt massive boson decays (hadrons faking a muon signature or heavy-flavour decays).
SFs of order unity, similar to those for electrons and derived in
$Z\rightarrow \mu^+\mu^-$ events, are applied to account for differences in
reconstruction, identification and isolation efficiencies between data
and simulated events.
Muon momenta are calibrated using the $Z$ mass peak.

\end{description}
 
If electrons, muons or jets overlap with each other, all but one object
must be removed from the event.
The distance metric used to define overlapping objects is defined as $\Delta R' = \sqrt{(\Delta\phi)^{2}+(\Delta y)^{2}}$ where $\Delta y$ represents the rapidity difference. To prevent double-counting of electron energy deposits as jets, jets within
$\Delta R'$ = 0.2 of a reconstructed electron candidate are removed.
If the nearest remaining jet is within $\Delta R'$ = 0.4 of the electron,
the electron is discarded.
To reduce the background from muons from heavy-flavour decays inside jets,
muons are required to be separated by $\Delta R'\ge0.4$ from the nearest jet.
In cases where a muon and a jet are reconstructed within $\Delta R'<0.4$,
the muon is removed if the jet has at least three associated tracks;
the jet is removed otherwise.
This avoids an inefficiency for high-energy muons undergoing significant energy
loss in the calorimeter.

 
\subsection{Event selection}

To be considered in this analysis, events must have at least one lepton identified in the
trigger system.
This triggered lepton must match
an offline electron or muon candidate. For each applicable trigger,
scale factors are applied to the simulation in order to
correct for known differences in trigger efficiencies between the simulation and
collision data~\cite{TRIG-2016-01}.
 
In order to reject backgrounds with fewer than two prompt leptons, exactly two reconstructed leptons with opposite charges are required.
Contributions from backgrounds with $Z$ bosons are reduced by requiring
that one lepton is an electron and the other is a muon.
The residual contribution from $Z\rightarrow\tau\tau$ events, which populate the low mass region, is further reduced by considering only events with $m_{e\mu}\ge50~\GeV$.
The contribution from \ttbar\ events with light-flavour jets from ISR or FSR
or from $W$ bosons is reduced by requiring exactly two
reconstructed jets.

Since the aim of the study is to measure the \bjet\ tagging efficiency,
it is useful to label simulated events according to the generator-level flavour of the two selected jets, following the definitions introduced in Section~\ref{sec:objects},
instead of the physics process they originate from. Events with two \bjets (non-\bjets)
are labelled $bb$ ($ll$). Events with one selected $b$-jet and one non-\bjet\ are labelled
$bl$ events if the $b$-jet $p_{\text{T}}$ is larger than the non-\bjet\ $p_{\text{T}}$ and $lb$
in the opposite case. According to the simulation, more than 90\% of the non-\bjets are light-flavour jets, the rest being composed of $c$-jets, and more than 95\% of the \bjets\ originate from a top-quark decay. The fraction of $\tau$-jets is predicted to be negligible.

In order to create $bb$, $bl$, $lb$ and $ll$-enriched regions in the selected sample, each of
the two leptons is paired with a jet in an exclusive way to determine whether they
originate from the same top-quark decay.
The pairing is performed such that it  minimises $(m_{j1,\ell_i}^2 + m_{j2,\ell_j}^2)$,
where $j1$ ($j2$) is the highest-\pt\ (second highest-\pt) jet, $\ell_{i,j}$ are the
two leptons and $m_{j1,\ell}$  ($m_{j2,\ell}$) is the invariant mass of the system including the highest (second highest) \pt jet and its associated lepton.
Choosing the pairing that mimimises this quantity relies on the fact that if the pairs of objects are from the same original particles then they are likely to have similar masses. Using the minimum of squared masses penalises asymmetric pairings with one high-mass lepton--jet pair, as well as combinations including two very high invariant masses, which are unlikely for those arising from top-quark decay.
Events are required to have $m_{j1,\ell} \ge 20$~\GeV{} and  $m_{j2,\ell} \ge 20$~\GeV{} in order to avoid configurations in which a soft jet and a soft lepton are close to each other, which are not well described by the simulation. The event classification based on these variables is described in more detail in the next section.

According to the simulation, about 85\% of the
events passing the selection are dileptonic $t\bar{t}$ events, about 65\% of which are $bb$ events.  Single top production in association with
a $W$ boson accounts for 8\% of the events, with about 30\% of these events
containing two selected $\bjets$. Diboson and $Z$+jet production
represent respectively about 5\% and 2\% of the selected
events, 85\% of these events being $ll$ events. Events originating from $W$+jets production are negligible ($<0.1\%$). The main source of non-\bjet{}s\ therefore originates from $t\bar{t}$ $bl$ or $lb$ events,
i.e. dileptonic $t\bar{t}$ events with a high-\pt\ light-flavour
jet originating from ISR or FSR.

Figure~\ref{fig:jet_pteta_control} shows the level of agreement between data and simulation as a
function of the \pt and $\eta$ of the selected jets as well as the expected fraction of $t\bar{t}$ events.
The overall level of agreement between data and simulation is
fairly good, although some mismodelling is present at high jet \pt, possibly related to the modelling of the top-quark \pt~\cite{TOPQ-2017-01}, which motivates the extraction of the \bjet\ tagging efficiency in jet \pt\ bins.
The distribution of the discriminant of the \textsc{MV2} algorithm for
events passing the selection is shown in Figure~\ref{fig:jet_mv2c10_control}.
Generally, good modelling is observed, indicating similar \bjet\ tagging efficiencies in
data and simulation.

\begin{figure}[!htbp]
\centering
\subfigure[][]{
\includegraphics[width=0.45\textwidth]{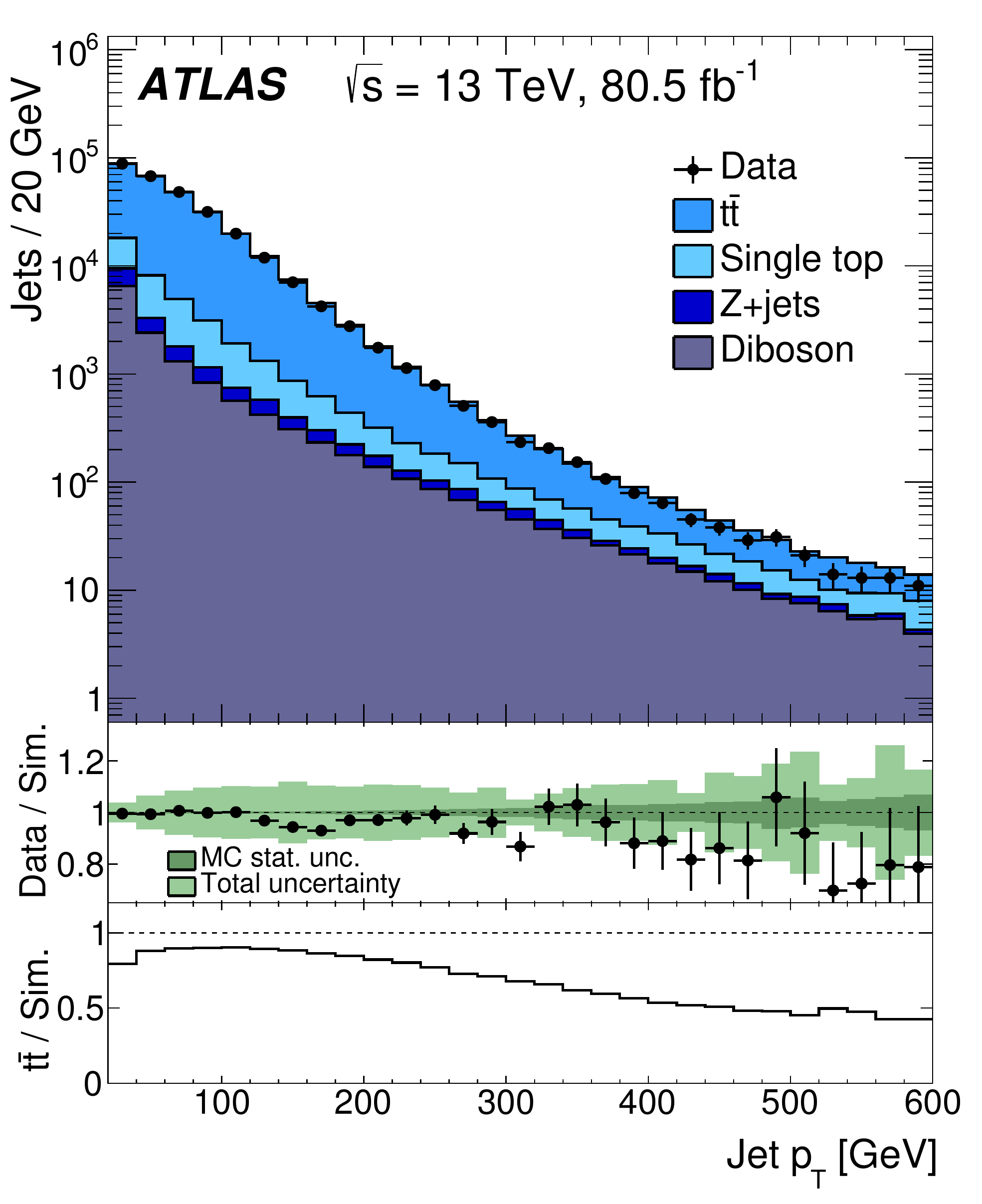}
}
\subfigure[][]{
\includegraphics[width=0.45\textwidth]{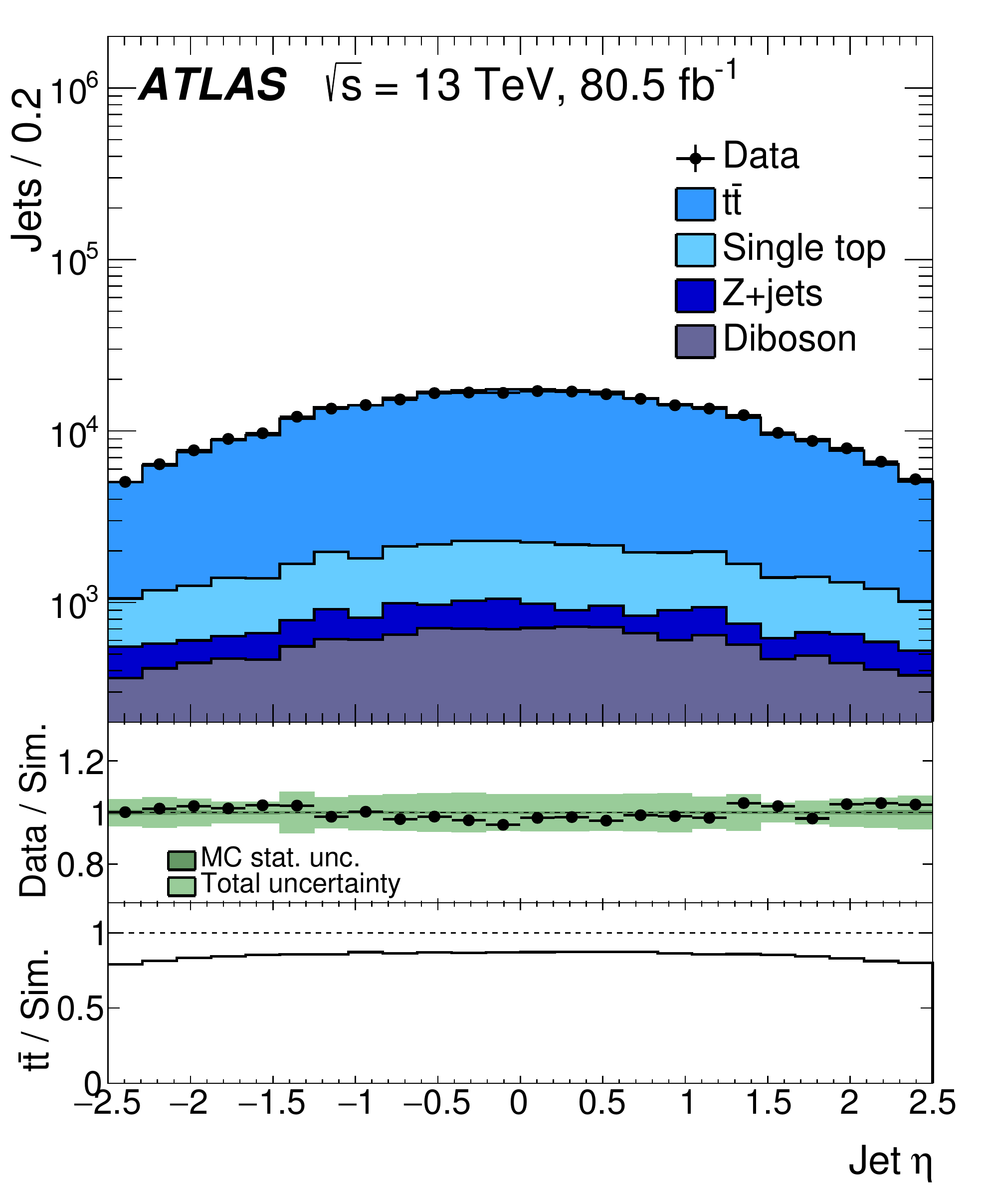}
}
 
\caption{
Distribution of the jet (a) \pt\ and (b) $\eta$ in the events passing the selection. Simulated events are split into physics process. The ratio panels show the data-to-simulation ratio as well as the fraction of \ttbar\ events among the simulated events.
} \label{fig:jet_pteta_control}
\end{figure}
 
\begin{figure}[!htbp]
\centering
\includegraphics[width=0.45\textwidth]{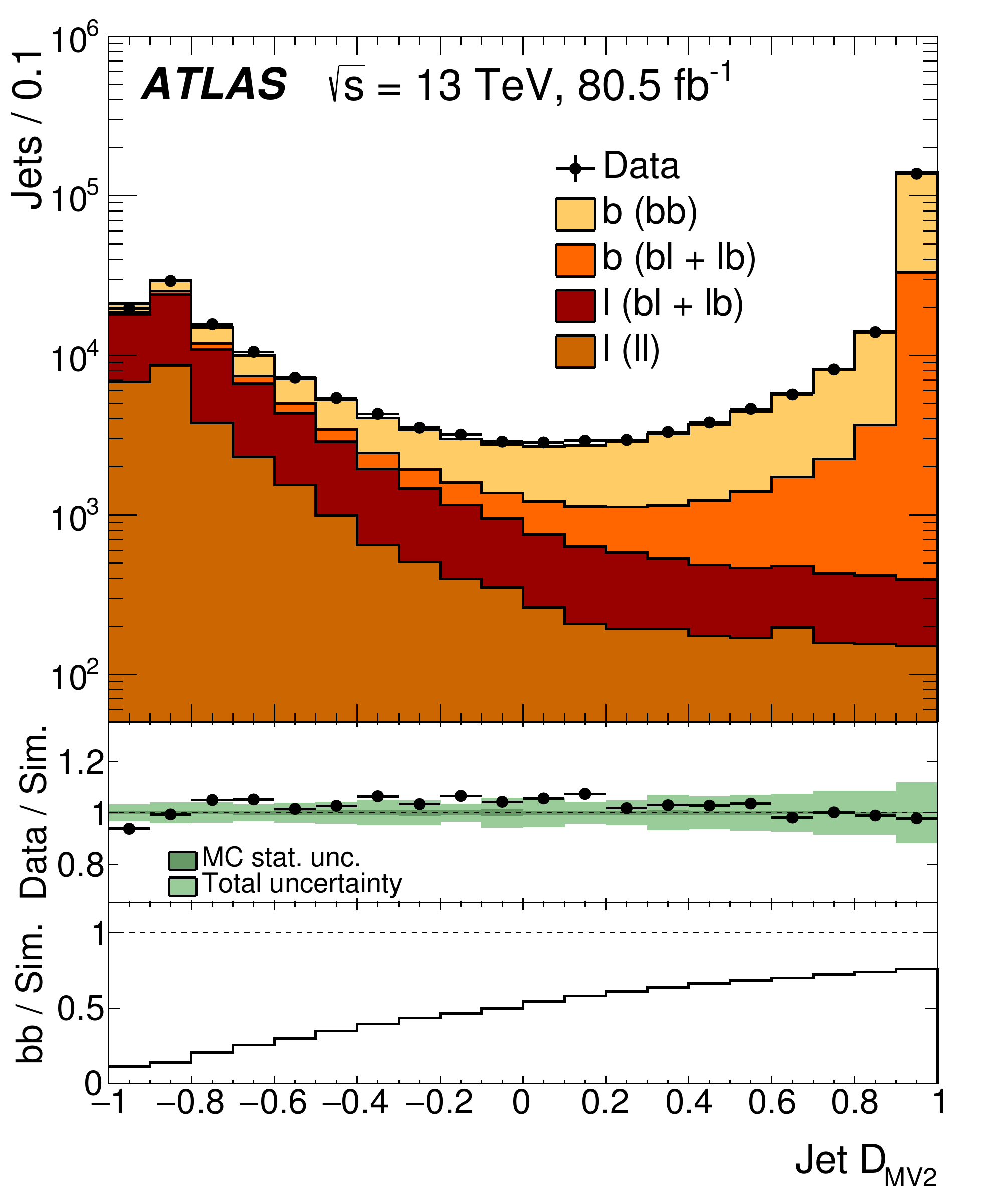}
\caption{
Distribution of the output discriminant of the \textsc{MV2} algorithm for the jets in events passing the selection. Simulated events are classified according to the flavour composition of the two jets, where the first term in each legend entry represents the flavour of the jet which is plotted ($b$ or $l$) and the following term in parenthesis represents the flavour composition of the event ($bb$, $bl+lb$ or $ll$). The ratio panels show the data-to-simulation ratio as well as the fraction of $bb$ events among the simulated events.
} \label{fig:jet_mv2c10_control}
\end{figure}

\subsection{Event classification}

The distributions of the $m_{j1,\ell}$ and $m_{j2,\ell}$ observables are shown in
Figure~\ref{fig:jet_mlj}.
In the case of $t\bar{t}$ events with two \bjets, both $m_{j1,\ell}$ and $m_{j2,\ell}$ have an
upper limit around $m_{t} = 172.5~\GeV$ and are usually significantly smaller due
to the undetected neutrino. This is generally not the case for $bl$, $lb$ and $ll$ events, which result in high
$m_{j1,\ell}$ and/or $m_{j2,\ell}$ values more often at high jet \pt.
Therefore, the $m_{j1,\ell}$ and $m_{j2,\ell}$ observables discriminate
between $bb$, $bl$, $lb$ and $ll$ events while being
uncorrelated with the $b$-tagging discriminants, which do not make use of leptons outside jets.

The selected events are classified into 45 different bins according
to the \pt of the two jets, allowing the $b$-jet tagging efficiency to be measured as a function
of the jet \pt.
In addition, in each leading jet \pt, subleading jet \pt bin (\ptone, \pttwo),
the events are further classified into four bins according to the $m_{j1,\ell}$ and $m_{j2,\ell}$ values:
 
\begin{itemize}
 
\item $m_{j1,\ell}, m_{j2,\ell}<175~\GeV$, signal region (SR): high $bb$ purity
region used to measure the \bjet tagging efficiency,
 
\item $m_{j1,\ell}, m_{j2,\ell}\ge 175~\GeV$, $ll$ control region (CR$_{\textrm{LL}}$): high $ll$
purity control region used to constrain the $bb$, $bl$, $lb$ and $ll$ fractions
in the SR,
 
\item $m_{j1,\ell}<175~\GeV, m_{j2,\ell} \ge 175~\GeV$, $bl$ control region (CR$_{\textrm{BL}}$):
high $bl$ purity control region used to constrain the $bb$, $bl$, $lb$ and $ll$
fractions in the SR,
 
\item $m_{j1,\ell}\ge 175~\GeV, m_{j2,\ell}<175~\GeV$, $lb$ control region (CR$_{\textrm{LB}}$):
high $lb$ purity control region used to constrain the $bb$, $bl$, $lb$ and $ll$
fractions in the SR.
 
\end{itemize}
 
Finally, the events in the SR are further classified as
a function of the pseudo-continuous binned
$b$-tagging discriminant of the two jets, denoted $w_1$ and $w_2$,
as defined in Section~\ref{sec:TaggerPerformance}.
 
These classifications result in a total of 1260 orthogonal categories.
A schematic diagram  illustrating the event categorisation is shown in
Figure~\ref{fig:event_class}. The $bb$ event purity in the signal
regions for the different \ptone, \pttwo\ bins is shown in
Figure~\ref{fig:bb_fraction}. The lowest purity ($19\%$) occurs
when both jets have very low $p_{\text{T}}$; however, the majority of
bins have a $bb$ event purity greater than $70\%$ and the highest purity (where
both jets have high $p_{\text{T}}$) reaches $93\%$.
The CR$_{\textrm{LL}}$, CR$_{\textrm{BL}}$ and CR$_{\textrm{LB}}$ control regions are enriched in their targeted
backgrounds relative to the corresponding SR. Their purity in $ll$, $bl$ and $lb$
events varies across the \ptone, \pttwo\ plane and ranges in the simulation from 30\% to 90\% (CR$_{\textrm{LL}}$), 32\% to 79\% (CR$_{\textrm{BL}}$) and 20\% to 74\% (CR$_{\textrm{LB}}$), respectively. The dominant background in each SR always benefits from
a high-purity (i.e. $\ge 50\%$) control region.
 
\begin{figure}[!htbp]
\centering
\subfigure[][]{
\includegraphics[width=0.45\textwidth]{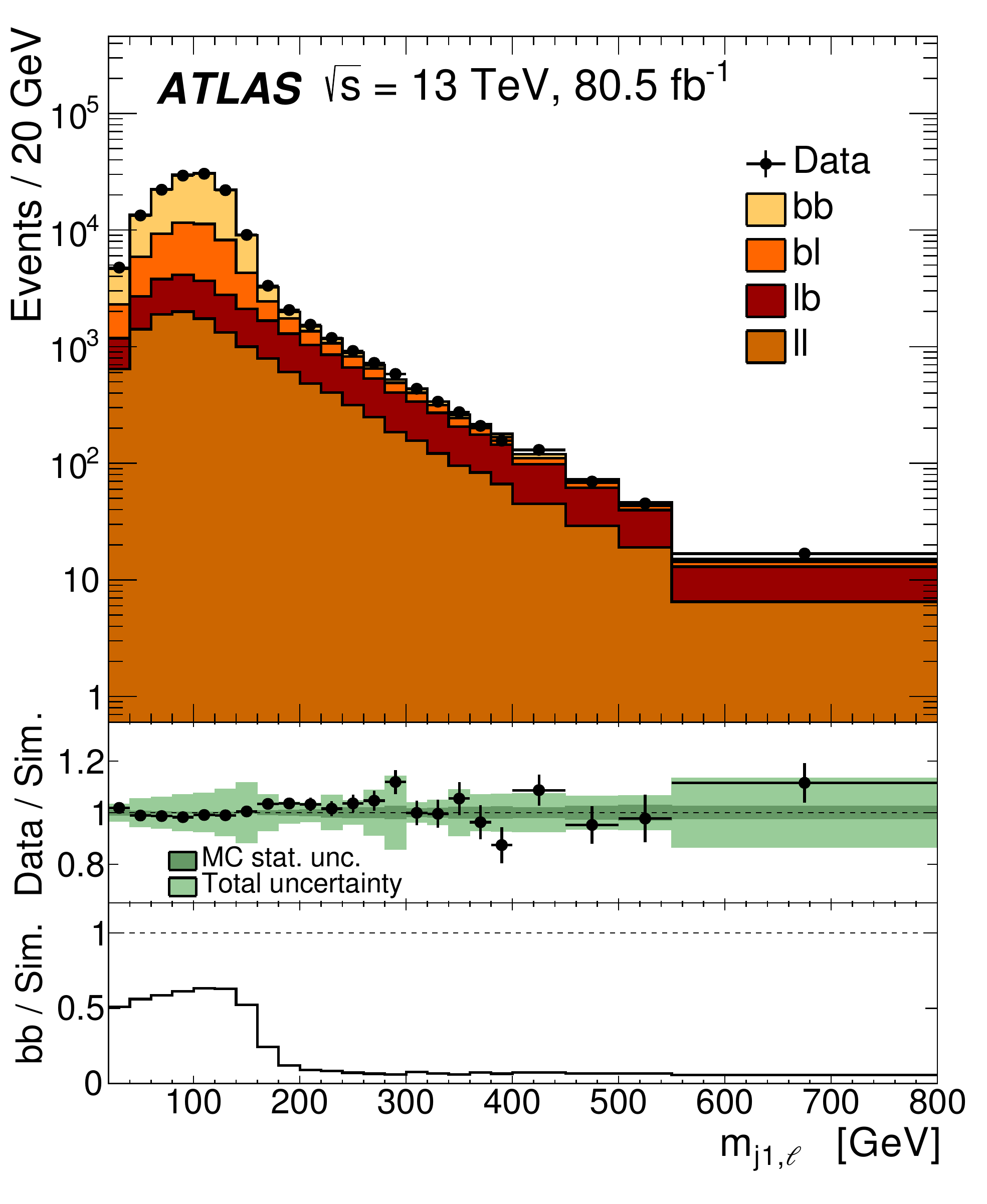}
}
\subfigure[][]{
\includegraphics[width=0.45\textwidth]{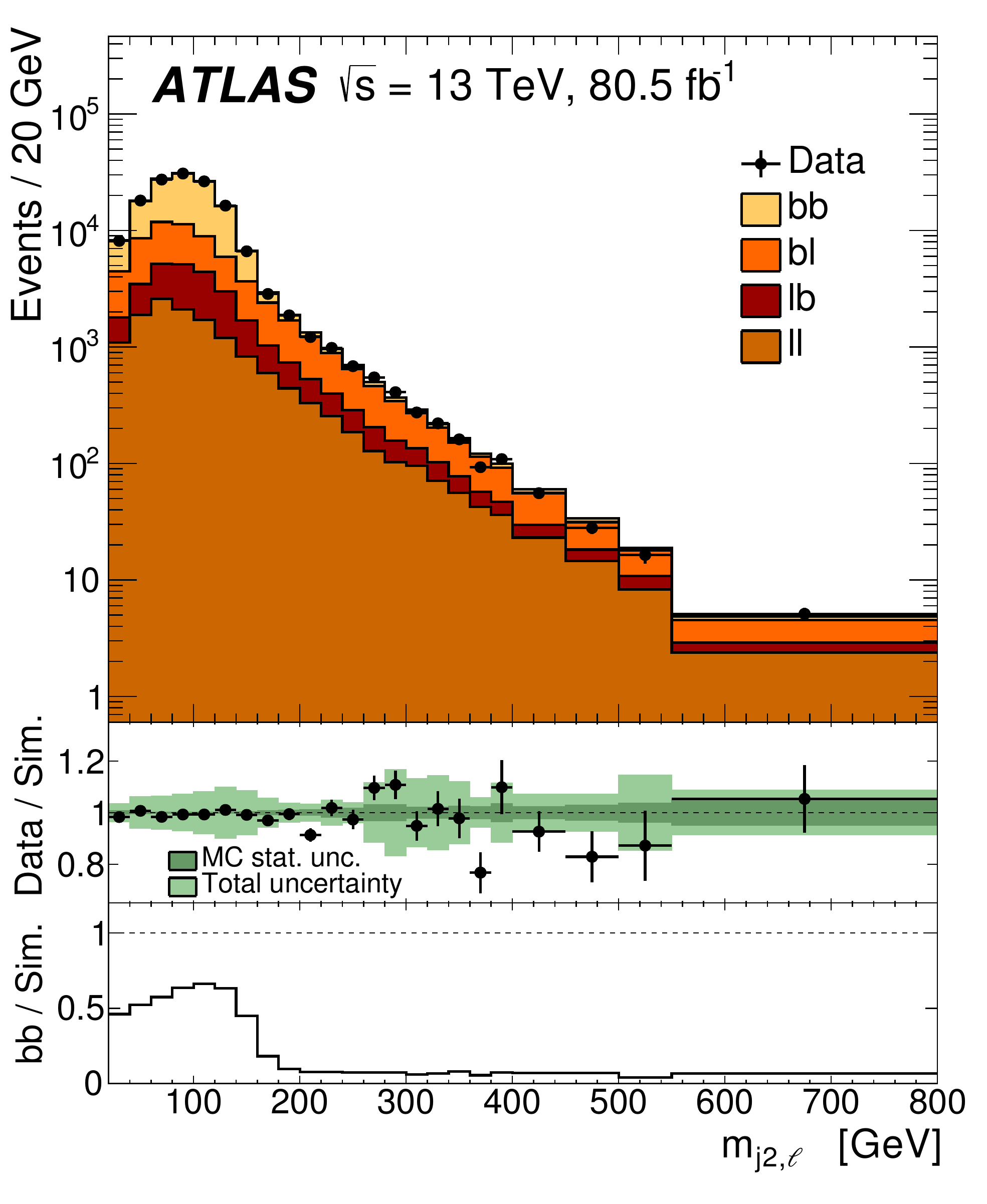}
}
 
\caption{
Distribution of the (a) \mljone\ and (b) \mljtwo\ observables in events
passing the selection. Simulated events are classified in terms of
the flavour composition of the two jets. The
ratio panels show the data-to-simulation ratio as well as the fraction of $bb$ events among the simulated events.
} \label{fig:jet_mlj}
\end{figure}
 
\begin{figure}[!htbp]
\centering
\includegraphics[width=0.95\textwidth]{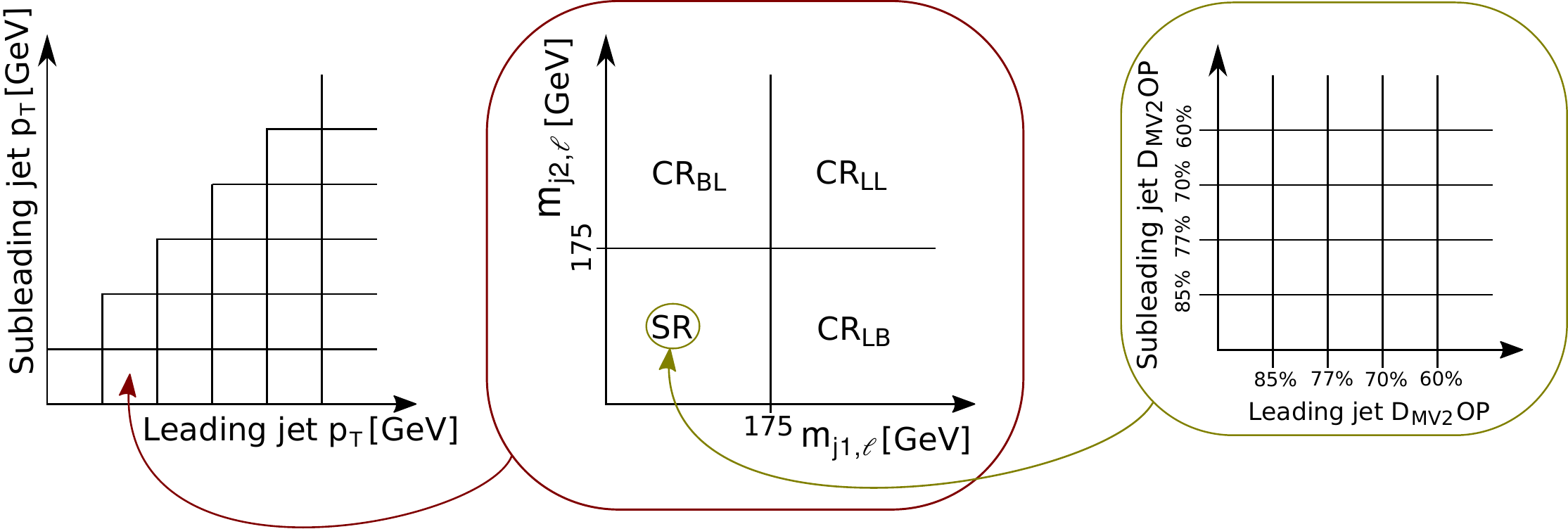}
\caption{
Schematic illustration of the event categorisation. Events are binned according to the \pt\ of both the leading and subleading jet in the event. In each of these bins, events are further assigned to control (CR) and signal (SR) regions. The events in the SR are then classified according to the \btag\ discriminant of the two jets.
} \label{fig:event_class}
\end{figure}

\begin{figure}[!htbp]
\centering
\includegraphics[width=0.7\textwidth]{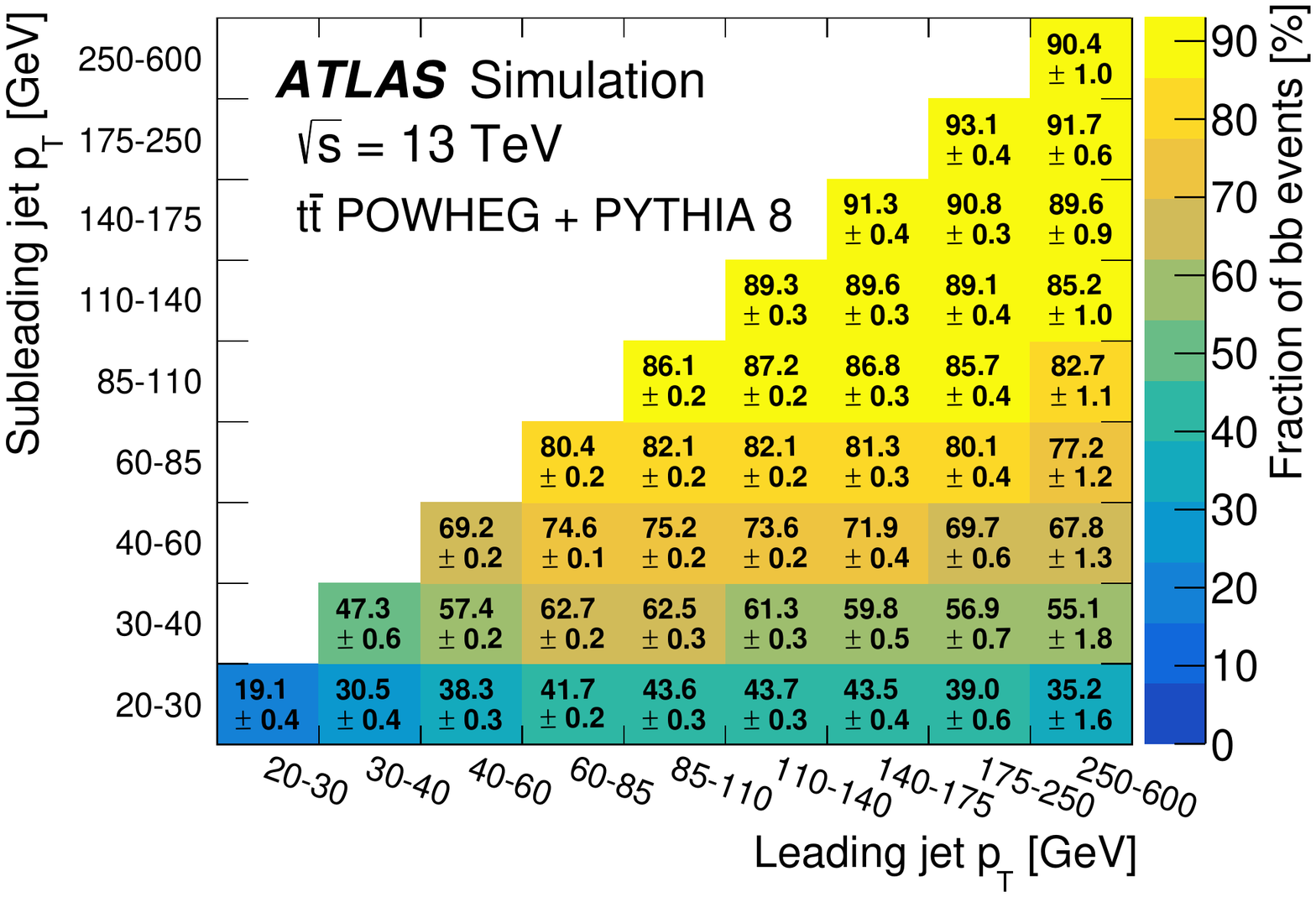}
\caption{
Fraction of $bb$ events in the signal region as predicted by the simulation as a function of the leading and subleading jet \pt. Uncertainties are statistical only.
} \label{fig:bb_fraction}
\end{figure}

 
\section{Extraction of $b$-jet tagging efficiency}
\label{sec:efficiency}
 
Once events have been selected and classified, the measurement of the $b$-jet tagging probabilities is performed. The precision of the previous ATLAS measurement~\cite{PERF-2016-05} was limited by the uncertainty in the fractions of $bb$, $bl$, $lb$ and $ll$ events in the selected sample, which is driven by the modelling of top-quark pair production. The main novelty of this work in comparison to Ref.~\cite{PERF-2016-05} lies in the measurement method, which uses both signal and control region data to define a joint log-likelihood function allowing the simultaneous estimate of the $b$-jet tagging probabilities and flavour compositions. This new technique leads to a reduction in the total uncertainties by up to a factor of two, as discussed in Section~\ref{sec:uncertainties}.
 
The general form of an extended binned log-likelihood function, after dropping terms that do not depend on the parameters to be estimated, is provided in Eq.~(\ref{eqn:general_likelihood}):
 
\begin{equation}
\label{eqn:general_likelihood}
\log \mathcal{L} \left( \nu_{\textrm{tot}}, \hat \Theta \right) = - \nu_{\textrm{tot}} + \sum_{i}^{N} n_i \log \nu_i \left( \nu_{\textrm{tot}},\hat \Theta \right),
\end{equation}
 
where $\nu_{\textrm{tot}}$ is the total expected number of events, $\hat \Theta=(\Theta_1,...,\Theta_m)$ is the list of parameters to be estimated, including the parameters of interest (POI) and nuisance parameters, $\nu_i$ ($n_i$) is the number of expected (observed) events in the bin $i$ and $N$ bins are considered in total. In this work, the POIs are the $b$-jet tagging probabilities, $\Pro_b$, introduced in Section~\ref{sec:TaggerPerformance}. They are defined in this measurement per \pt\ bin, i.e.\ as the conditional probabilities for a \bjet\ with a transverse momentum falling in one of the nine \pt\ bins $(T^m)_{m=1..9}$ of the measurement to have a $b$-tagging discriminant $w$ falling in one of the five pseudo-continuous bins $(O^k)_{k=1..5}$. The $b$-jet tagging efficiency of the single-cut OP $X$ in that jet \pt bin, \beff, relates to the POIs as outlined below:
 
\begin{equation*}
\beff(X|T^m) = \sum_{O^k \subset~X} \Pro_b(O^k|T^m).
\end{equation*}
 
In each control region, the number of events in a given \ptone, \pttwo\  bin $(T^m, T^n)$ is written as the sum of the $bb$, $bl$, $lb$ and $ll$ yields expected in that bin ($\nu_{bb}^{m,n}$, $\nu_{bl}^{m,n}$, $\nu_{lb}^{m,n}$, $\nu_{ll}^{m,n}$), corrected by \ptone, \pttwo\ dependent correction factors ($c_{bb}^{m,n}$, $c_{bl}^{m,n}$, $c_{lb}^{m,n}$, $c_{ll}^{m,n}$), forming the nuisance parameters:
 
\begin{equation*}
\nu_{CR}(T^m, T^n) = c_{bb}^{m,n}\nu_{CR,bb}^{m,n} + c_{bl}^{m,n}\nu_{CR,bl}^{m,n} + c_{lb}^{m,n}\nu_{CR,lb}^{m,n} + c_{ll}^{m,n}\nu_{CR,ll}^{m,n}.
\end{equation*}
 
In each signal region, the events are further binned according to the $b$-tagging discriminants of the two jets, $w_1$, $w_2$. The number of events expected in a given \ptone, \pttwo, $w_1$, $w_2$ bin ($T^m$, $T^n$, $O^k$, $O^p$) is thus written:
 
\begin{align*}
\nu_{SR}(T^m, T^n, O^k, O^p)  &= c_{bb}^{m,n}\nu_{SR,bb}^{m,n}\cdot\Pro_b(O^k|T^m)\cdot\Pro_b(O^p|T^n) \nonumber\\
&+ c_{bl}^{m,n}\nu_{SR,bl}^{m,n}\cdot\Pro_b(O^k|T^m)\cdot\Pro_l(O^p|T^n) \nonumber\\
&+ c_{lb}^{m,n}\nu_{SR,lb}^{m,n}\cdot\Pro_l(O^k|T^m)\cdot\Pro_b(O^p|T^n) \nonumber\\
&+ c_{ll}^{m,n}\nu_{SR,ll}^{m,n}\cdot\Pro_l(O^k|T^m)\cdot\Pro_l(O^p|T^n),
\end{align*}
 
where $\Pro_l$ is the effective $b$-jet tagging probability of the mix of $c$-jets and light-flavour jets predicted by the simulation in each \ptone, \pttwo\ bin. 
The POIs and correction factors are estimated by minimising the negative log-likelihood function defined above with the \textrm{MINUIT} algorithm~\cite{James:2004xla}. Both the POIs and correction factors are free parameters during the minimisation procedure. Signal and control region data are provided as input as well as the $\Pro_l$ conditional probabilities, which are estimated from the MC simulation corrected to match data (see Section~\ref{sec:intro}). The simulation is also used to determine the $bb$, $bl$, $lb$ and $ll$ yield fractions according to the type of region (SR, CRs), as the correction factors are defined as a function of \ptone\ and \pttwo\ only. 
 
The extraction method is validated using pseudo-data generated with a known flavour composition. This is created by combining events from either nominal or alternative MC simulation fluctuated according to the statistical uncertainty expected from the actual dataset. The input parameters of the minimisation procedure are taken from the nominal MC simulation in all cases. The size of the non-closure effects observed when using pseudo-data based on the nominal and alternative MC simulation are compared respectively with the expected data statistical uncertainty (0.6-3.7\%), and to the sum in quadrature of the expected data statistical uncertainty, the MC statistical uncertainty and the physics modelling uncertainties quoted for the final measurement (0.9--5.4\%). The non-closure effects are found to be within uncertainties in each jet \pt\ bin such that no additional uncertainty related to the signal extraction method is considered.

 
\section{Uncertainties}
\label{sec:uncertainties}
 
Uncertainties affecting the measurement which originate from statistical sources are considered together with systematic uncertainties related to the detector calibration and physics modelling.
 
The data statistical uncertainty in the $b$-jet tagging probabilities, and their bin-to-bin correlations, are obtained from the error matrix
returned by \textrm{MINUIT}
~\cite{James:2004xla} and propagated to the $b$-jet tagging efficiencies via a basis transformation. 
The data statistical uncertainty reaches about 4\% (2\%) for jets within $20\le\pt<30~\GeV$ ($30\le\pt<40~\GeV$), ranges from 1\% to 3\% for jet $\pt\ge 140~\GeV$ and is below 1\% elsewhere.

The bootstrap resampling technique~\cite{Bohm:389738} is used to assess
the MC statistical uncertainty by creating an ensemble of statistically equivalent
measurements in which the weight of each simulated event used in the nominal measurement is multiplied by
an additional term, randomly chosen for each event from a Poisson distribution with a mean of one. The standard deviation of the distribution of these measurements is taken as the MC statistical uncertainty. This method allows all correlations to be preserved and the uncertainty in the value of any parameter to be extracted. One hundred bootstrap replicas of each simulated sample are used for this evaluation. The MC statistical uncertainty in the $b$-jet tagging efficiencies is found to be non-negligible only for jet $\pt\le 40~\GeV$, where it reaches about 2\% and 1\% for jets within $20~\GeV\le \pt<30~\GeV$ and $30~\GeV\le \pt<40~\GeV$, respectively.
 
The systematic uncertainties are derived by varying a parameter in the simulated events, repeating the complete analysis with this varied parameter and taking the difference between the updated measurement of the $b$-jet tagging efficiency or probability and the nominal measurement as the (bin-wise correlated) uncertainty. For $b$-jet tagging efficiencies, the bootstrap replicas of simulated events are then used to evaluate the MC statistical uncertainty in each systematic variation. Variations of the $b$-jet tagging efficiency that are not statistically significant undergo a bin-merging procedure over an increasing number of \pt\ bins  to improve their significance. Following this procedure, only statistically significant variations are considered as systematic uncertainties.
 
Uncertainty sources related to the energy scale and resolution of hadronic jets~\cite{PERF-2016-04} encompass both the modelling of the detector response and the analysis techniques used to derive the calibration. The impact of the jet energy scale uncertainty reaches from 4\%--5\% for jet ${\pt \le 30~\GeV}$, 1\% for ${30~\GeV \le \pt<40~\GeV}$, and is negligible elsewhere. It is dominated by the prediction of the quark/gluon origin of the light-flavour jets and by the difference in their energy response, as well as the difference in the calorimeter energy response as a function of $\eta$. The uncertainty originating from the jet energy resolution is negligible. Uncertainty sources related to the performance of the JVT algorithm~\cite{ATLAS-CONF-2014-018}, the $b$-tagging performance for light-flavour jets~\cite{ATLAS-CONF-2018-006} and $c$-jets~\cite{ATLAS-CONF-2018-045} as well as the modelling of pile-up interactions were investigated and found to be negligible, as were lepton-related uncertainties, including energy/momentum scale and resolution, identification, isolation, trigger and track--vertex association efficiency.
 
The uncertainty in the physics modelling of top-quark events is evaluated by changing the parton shower and hadronisation model from \textsc{Pythia}~8 to \textsc{Herwig}~7 and increasing or decreasing the amount of ISR and FSR within \textsc{Pythia}~8~\cite{ATL-PHYS-PUB-2016-020}. The uncertainties originating from parton distribution functions (PDF) are quantified following the PDF4LHC recommendations~\cite{Butterworth:2015oua}. An additional source of uncertainty originates from the mismodelling of the interference between single top $Wt$ and \ttbar\ production. It is evaluated by switching the nominal single-top simulation sample, based on the `diagram removal' scheme, to the one based on the `diagram subtraction' scheme~\cite{Frixione:2008yi}. The final \ttbar\ modelling uncertainty reaches 3\% (2\%) for jet $\pt<30~\GeV$ ($30\le\pt<40~\GeV$) and about 1\% for $\pt\ge 40~\GeV$. It is dominated at low \pt\ by PDF and ISR/FSR variations whereas at higher \pt\ the choice of parton shower and hadronisation model is the dominant contribution. The single-top uncertainty reaches about 3\% for jet $\pt<30~\GeV$ due to the parton shower and hadronisation model variation, and 1\% for jet $\pt\ge 250~\GeV$, where the uncertainty in the interference with $t\bar{t}$ events is the dominant contribution. It is below $1\%$ elsewhere. The uncertainties associated with the modelling of top-quark events are reduced by up to a factor of two relative to the previous ATLAS analysis~\cite{PERF-2016-05} due to the new \bjet\ tagging efficiency extraction method, which allows the $bb$ event yield to be determined at a precision of a few percent in each \ptone, \pttwo\  bin. 
 
The uncertainty in the modelling of diboson and $Z$+jet production~\cite{ATL-PHYS-PUB-2017-005,ATL-PHYS-PUB-2017-006} is evaluated by varying the total cross-section and the factorisation and renormalisation scales for these processes, as well as propagating the uncertainty from the PDF. The total cross-section is kept constant when performing the scale and PDF variations such that only the shapes of the kinematic distributions are impacted. The total cross-section is varied by $\pm6\%$ ($\pm5\%$) for $Z$+jets (diboson) production. The scale uncertainties are estimated simultaneously by varying the nominal values by a factor of two up and down and taking the largest deviations from the nominal predictions in each direction as uncertainties. PDF uncertainties are evaluated using the 100 bootstrap replicas provided by the NNPDF30NNLO~\cite{Ball:2014uwa} using the same method as outlined for the MC statistical uncertainty earlier in this section. The final diboson and $Z$+jets uncertainties are found to be negligible in the entire range covered by the analysis.
 
The number of events with a selected muon not originating from a $Z$- or $W$-boson decay is predicted by the simulation to be negligible after the event selection. This is due to the tight muon identification and isolation criteria applied. The number of events with a selected electron not originating from a $Z$- or $W$-boson decay  passing the event selection (1NPel, for 1 non-prompt electron) is also predicted by the simulation to be very small but one order of magnitude higher, reaching about 0.3\% of the total event yield after selection. An uncertainty in this yield is derived by comparing the number of data and MC events in an alternative region defined by requiring two same-sign (SS) leptons instead of opposite-sign (OS). The SS region is predicted by the simulation to have a composition that is 12\% 1NPel events, with the remaining 88\% of the sample coming from non-1NPel events, which is dominated by diboson production. This is estimated from simulation and subtracted from the data. The remaining data events are then compared with MC predictions in bins of electron \pt. The data-to-simulation ratio ranges from values close to 3 for $\pt<120~\GeV$ to values close to 1 for $\pt\ge 300~\GeV$. These values are used as simulation-to-data scale factors to correct the yield of simulated 1NPel events in the OS region in order to estimate an uncertainty in the fake-lepton modelling. The $b$-jet tagging efficiency measurement is then repeated with these scale factors applied and compared with the nominal measurement. Differences of about 1\% to 2\% for jet $\pt<40~\GeV$ and negligible elsewhere are observed and accounted for as an additional systematic uncertainty.
 
 
\section{Results}
\label{sec:result}
 
The goodness-of-fit is evaluated by computing a Pearson's $\chi^2$ and comparing it with the number of degrees of freedom ($ndf$) of the fit~\cite{chi2}. This procedure tests the hypothesis that the remaining differences between observed and expected yields post-fit originate only from the limited size of the dataset. The $\chi^2/ndf$ value obtained for the nominal measurement is 0.98, corresponding to a $p$-value of about 0.65. This result illustrates the high goodness-of-fit already observed before accounting for the other sources of uncertainty discussed in Section~\ref{sec:uncertainties}.
 
The $bb$, $bl$, $lb$ and $ll$ yield post-fit correction factors are of order unity, compatible with unity within uncertainties, and typically constrained within 2\%--5\% for $bb$, 5\%--10\% for $bl$, and 7\%--20\% for $lb$ and $ll$. The central values of the $bb$ yield correction factors tend to be a few percent below unity, pointing to a slight underestimate of the number of light-flavour jets in the nominal simulation. The yield correction factors deviate more strongly from unity when running on the alternative simulated samples.

The $b$-jet tagging efficiency measurement for the $\varepsilon_b = 70\%$ single-cut OP of the \textsc{MV2} algorithm is presented in Figure~\ref{fig:results}(a) as a function of jet \pt\ together with the efficiency derived from \ttbar\ simulated events passing the signal region selection. The corresponding $b$-jet tagging efficiency simulation-to-data scale factors, defined as the ratio between the measured $b$-jet tagging efficiency to the $b$-jet tagging efficiency derived from the simulation, are shown in Figure~\ref{fig:results}(b). Scale factors are derived for all single-cut OPs and for the \textsc{DL1} tagger using the same technique, resulting in similar results, as illustrated in Figure~\ref{fig:results_summary}. The scale factors have values very close to one and are approximately constant throughout the entire \pt\ range, illustrating the good modelling of the $b$-jet tagging performance.
The $b$-jet tagging efficiency measurement for the $\varepsilon_b = 70\%$ single-cut OP of the \textsc{MV2} algorithm as a function of jet $|\eta|$ and the corresponding simulation-to-data scale factors are presented in Figure~\ref{fig:results_eta}. The $b$-jet tagging probability and efficiency measurement was also repeated considering only data and simulated events with either less or more than 28 additional $pp$ interactions per bunch crossing, and separately for 2015--2016 and 2017 data. In all cases, consistent results were observed.

\begin{figure}[!htbp]
\centering
 
\subfigure[][]{
\includegraphics[width=0.45\textwidth]{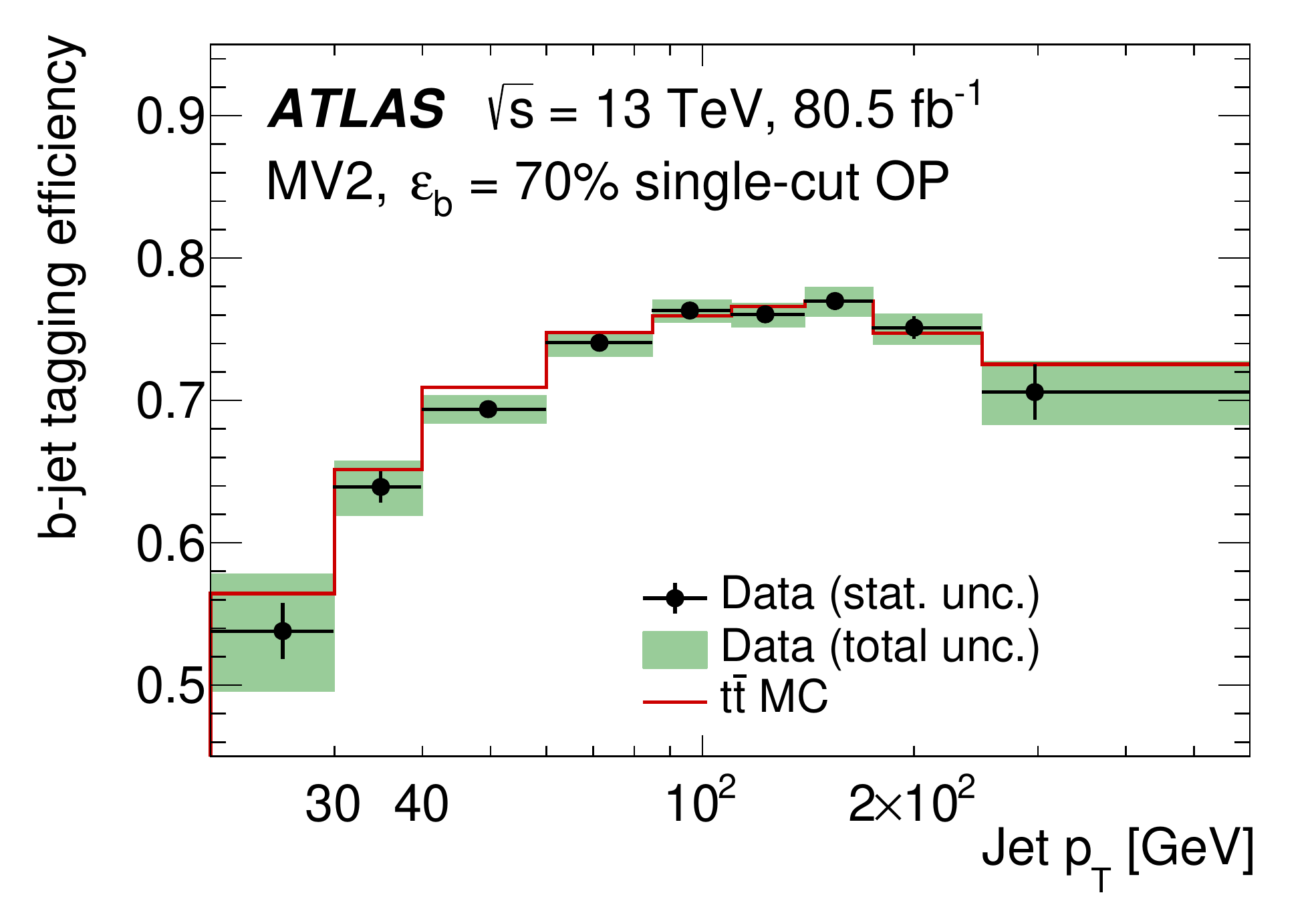}
}
\subfigure[][]{
\includegraphics[width=0.45\textwidth]{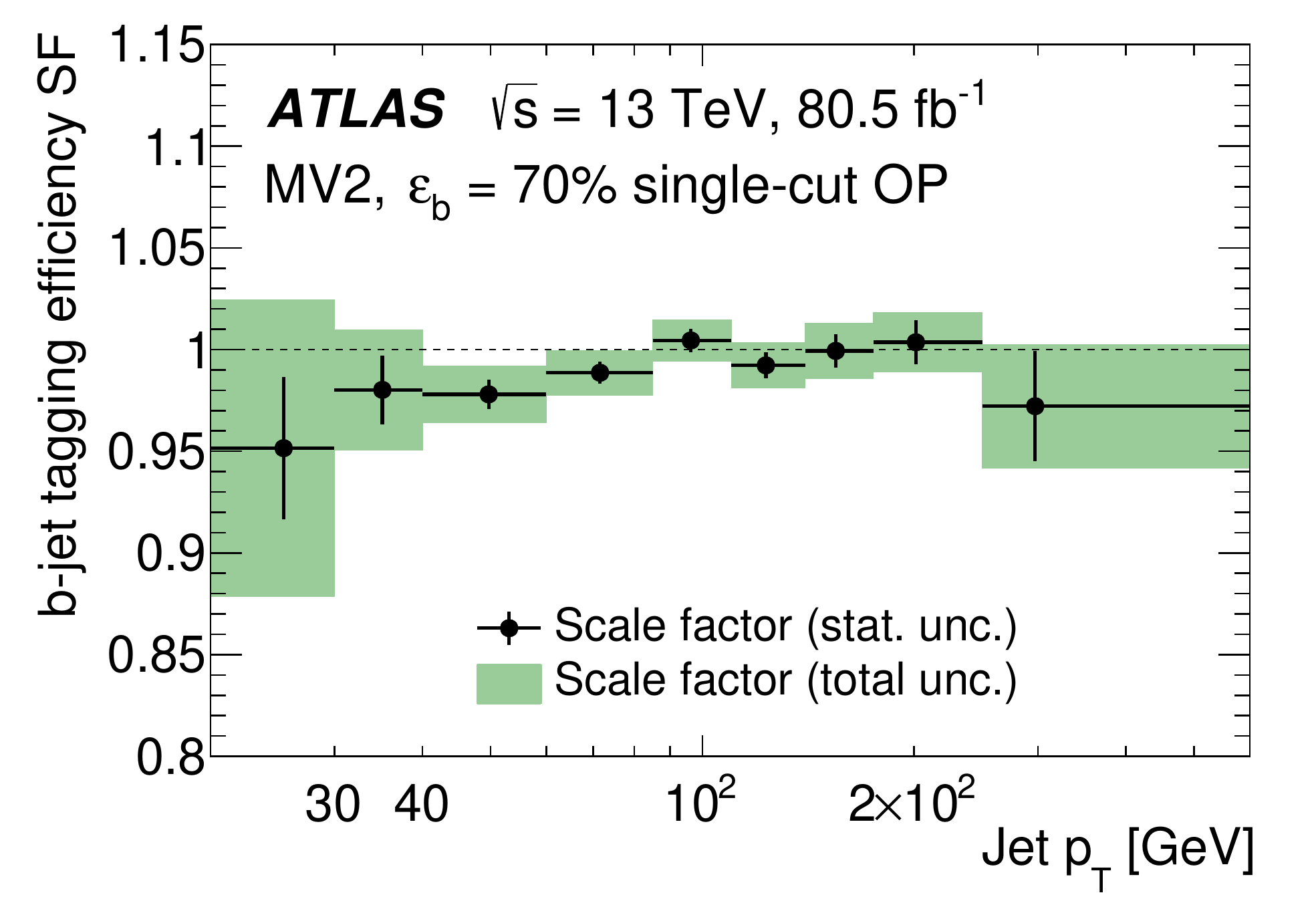}
}
\caption{
The (a) $b$-jet tagging efficiency and (b) $b$-jet tagging efficiency simulation-to-data scale factors for the $\varepsilon_b = 70\%$ single-cut OP of the \textsc{MV2} tagger as a function of jet \pt. The efficiency measurement is shown together with the efficiency derived from \ttbar\ simulated events passing the signal region selection. Vertical error bars include data statistical uncertainties only while the green bands correspond to the sum in quadrature of statistical and systematic uncertainties. The dots are located at the mean of the $b$-jet \pt\ distribution in each \pt\ bin.
} \label{fig:results}
\end{figure}
 
\begin{figure}[!htbp]
\centering
 
\subfigure[][]{
\includegraphics[width=0.45\textwidth]{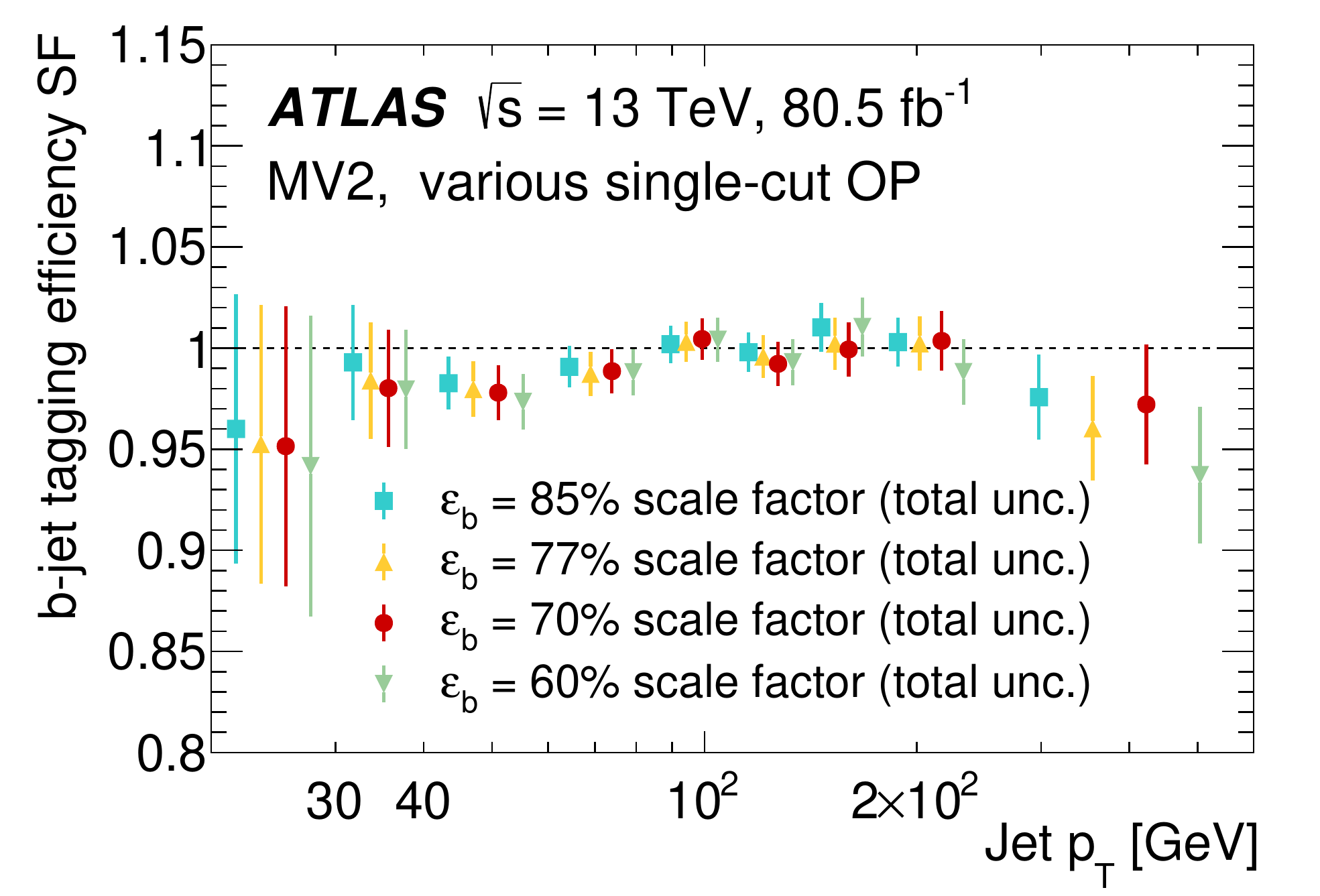}
}
\subfigure[][]{
\includegraphics[width=0.45\textwidth]{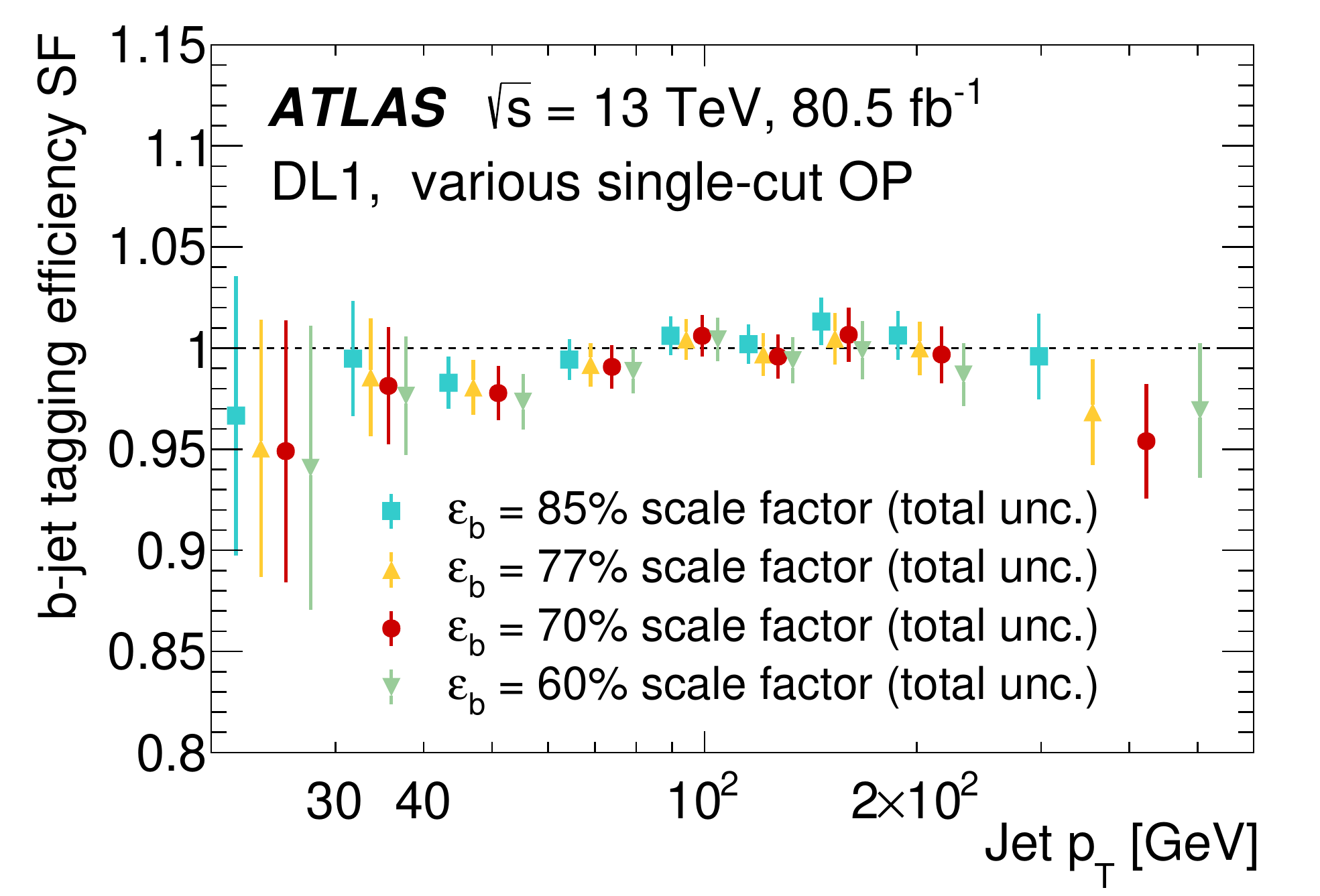}
}
\caption{
$b$-jet tagging efficiency simulation-to-data scale factors as a function of jet \pt. The 60\%, 70\%, 77\% and 85\% single-cut OP of the (a) \textsc{MV2} and (b) \textsc{DL1} taggers are shown. The various groups of points are offset for visual effects but computed in the same jet \pt\ range. Vertical error bars represent the total uncertainty.
} \label{fig:results_summary}
\end{figure}
 
\begin{figure}[!htbp]
\centering
\subfigure[][]{
\includegraphics[width=0.45\textwidth]{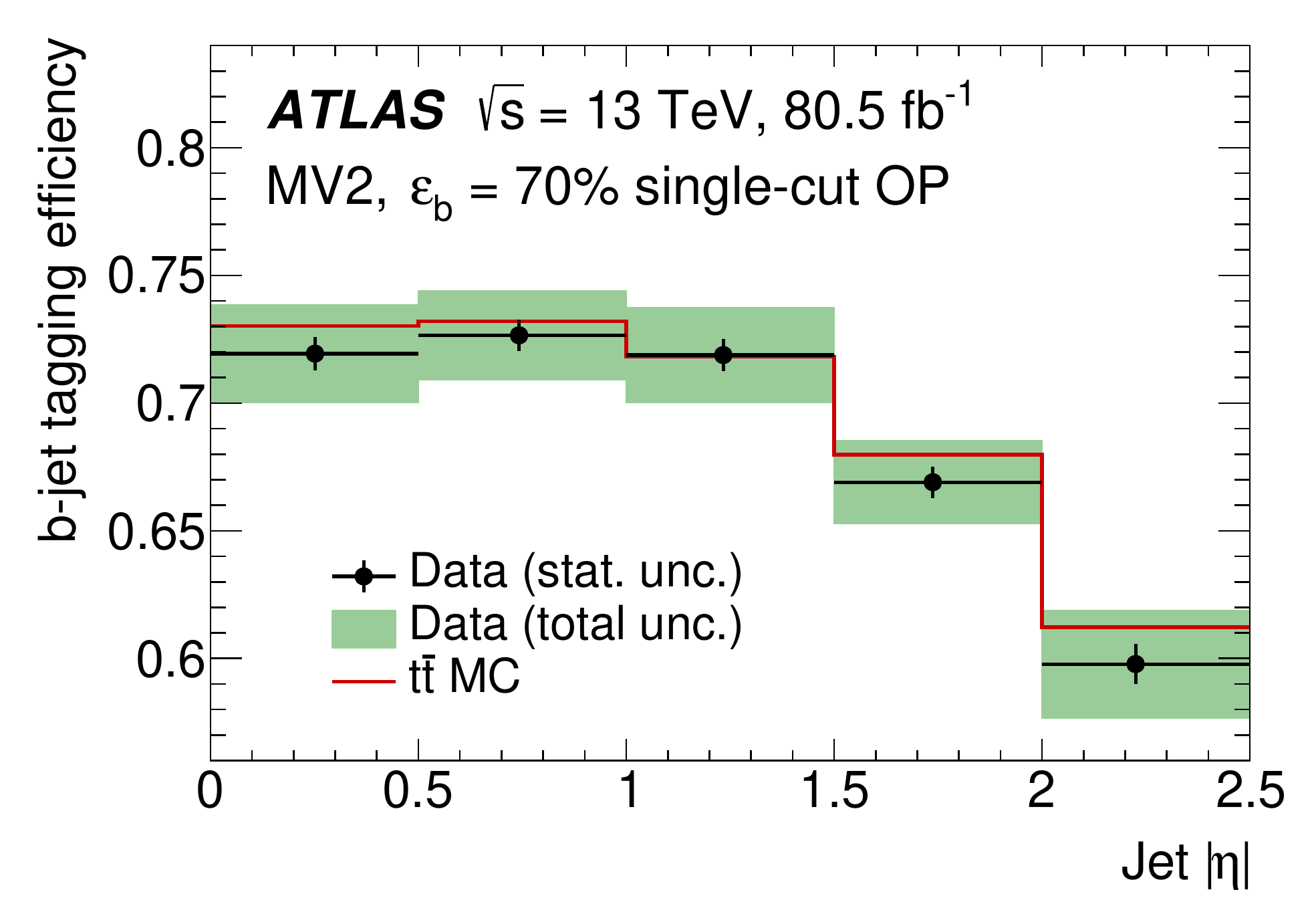}
}
\subfigure[][]{
\includegraphics[width=0.45\textwidth]{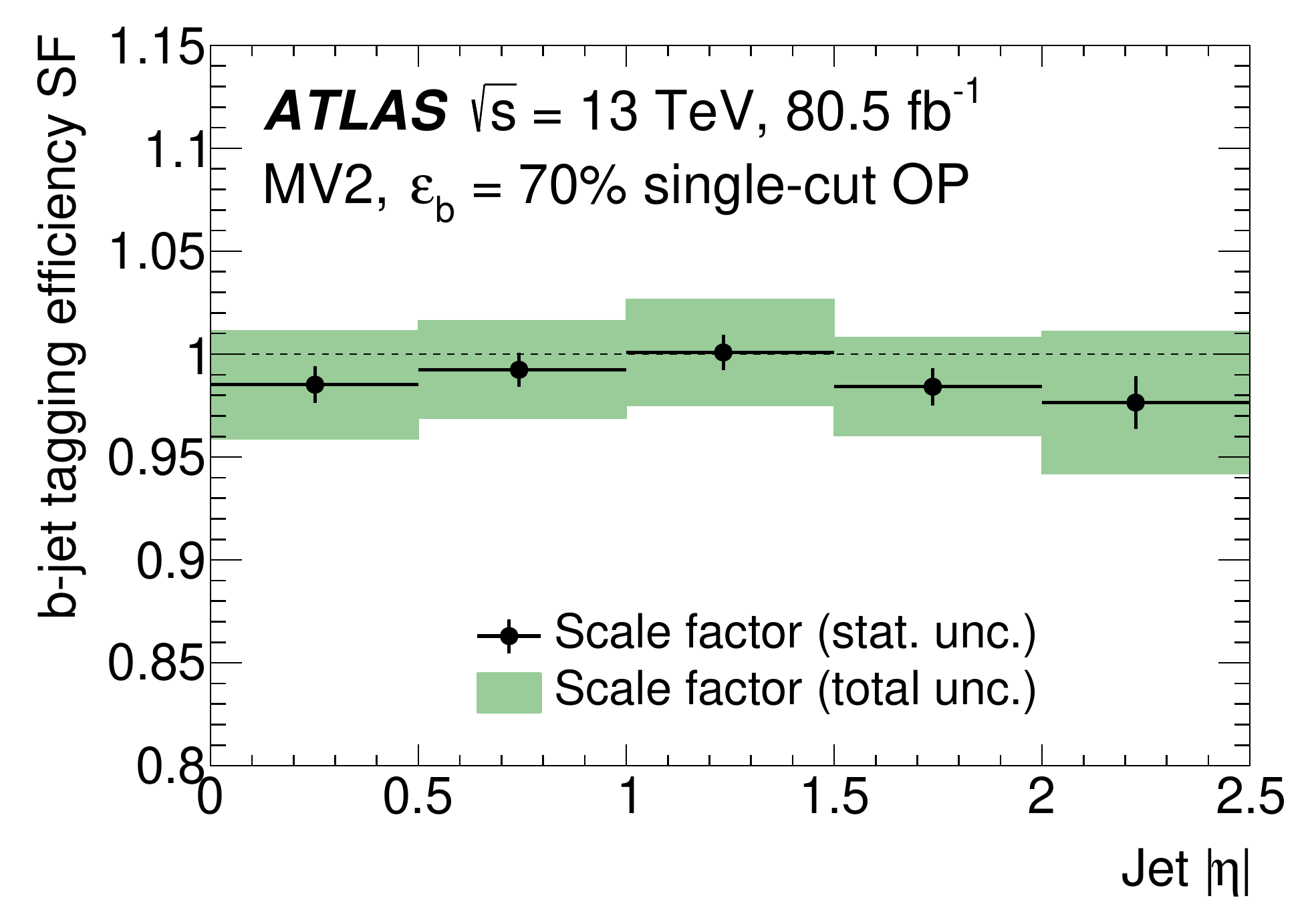}
}
\caption{
The (a) $b$-jet tagging efficiency and (b) $b$-jet tagging efficiency simulation-to-data scale factor for ${\varepsilon_b=70\%}$ single-cut OP of the \textsc{MV2} tagger as a function of jet $|\eta|$. Vertical error bars include data statistical uncertainties only while the green bands correspond to the sum in quadrature of statistical and systematic uncertainties. The dots are located at the mean of the $b$-jet $|\eta|$ distribution in each $|\eta|$ bin.
} \label{fig:results_eta}
\end{figure}

The uncertainty in the efficiency measurement for the $\varepsilon_b = 70\%$ single-cut OP of the \textsc{MV2} tagger is summarised in Table~\ref{tab:results_table}. The total uncertainty reaches about 1\% for ${40~\GeV\le\pt<250~\GeV}$, where it is dominated by the uncertainty in the physics modelling of \ttbar\ events and the data statistical uncertainty. At lower \pt\ values ($20~\GeV\le\pt<40~\GeV$), the total uncertainty increases to 8\% due to higher uncertainties in the jet energy scale, the modelling of \ttbar\ and single-top-quark events, the limited number of data and MC events and the modelling of fake leptons. For jet $\pt\ge250~\GeV$, the uncertainty increases to about 3\% due to the limited number of data events. These observations are consistent across single-cut OPs and taggers.

\begin{table}[htbp]
\caption{Breakdown of the systematic uncertainties in the $b$-jet tagging efficiency measurement for the 70\% single-cut OP of the \textsc{MV2} tagger as a function of the jet \pt\ bin. The `\ttbar\ modelling' and the `Single top modelling' uncertainties correspond to the sum in quadrature of the uncertainty in the parton shower, hadronisation model, initial-state and final-state radiation and PDF for \ttbar\ and single top-quark production, respectively. The `Single top modelling' uncertainties include an additional source originating from the interference between single top and \ttbar\ production. `Other sources' corresponds to the sum in quadrature of the uncertainties related to jet energy resolution, electron and muon performance, $b$-tagging performance for light-flavour jets and $c$-jets,  JVT performance, diboson and $Z$+jet modelling (including normalisation and shape uncertainties) and pile-up modelling. All systematic uncertainties are fully correlated bin-by-bin whereas the statistical uncertainty correlations are evaluated following the procedures described in Section~\ref{sec:uncertainties}. In the case of correlated systematic uncertainties, the relative sign of the uncertainty in each bin is taken into account, even if not shown here.}
\label{tab:results_table}
\setlength\extrarowheight{3pt}
\centering
\small
\begin{tabular}{|l|lllllllll|}
\hline
\multirow{2}{*}{Source of uncertainty} & \multicolumn{9}{c|}{Relative uncertainty on \beff\ [\%] per jet \pt\ bin [\GeV]} \\
& 20--30    & 30--40      & 40--60 & 60--85 & 85--110 & 110--140 & 140--175 & 175--250 & 250--600 \\
\hline
Data statistics       & 3.7 & 1.7 & 0.7 & 0.6 & 0.6 & 0.6 & 0.8 & 1.1 & 2.8 \\
MC statistics          & 2.2 & 1.0 & 0.4 & 0.2 & 0.2 & 0.2 & 0.2 & 0.2 & 0.5 \\
Jet energy scale       & 4.5 & 0.8 & 0.3 & 0.1 & 0.1 & 0.1 & 0.1 & 0.2 & 0.4 \\
\ttbar\ modelling      & 3.2 & 1.5 & 1.0 & 0.7 & 0.7 & 0.8 & 1.0 & 0.8 & 0.5 \\
Single top modelling    & 2.5 & 0.5 & 0.6 & 0.6 & 0.4 & 0.3 & 0.3 & 0.4 & 1.1 \\
Fake leptons modelling  & 1.8 & 1.1 & 0.1 & 0.2 & $<0.1$ & $<0.1$ & 0.2 & $<0.1$ & 0.2 \\
Other sources          &      1.4      &   0.9    &   0.2    &  0.3     &   0.2    &     0.1  &   0.1    &  0.1     & 0.3 \\[3pt]
\hline
Total          & 7.7 & 3.0 & 1.4 & 1.1 & 1.0 & 1.1 & 1.3 & 1.5 & 3.1 \\[3.5pt]
\hline
\end{tabular}
\end{table}

The measurement of the \bjet\ tagging probabilities in the \textsc{MV2} and \textsc{DL1} algorithm output bins is presented in Figures~\ref{fig:results_2}(a) and~\ref{fig:results_2}(c) together with the \bjet\ tagging probabilities derived from \ttbar\ simulated events passing the signal region selection. The probabilities are shown for jets with $110~\GeV\le\pt<140~\GeV$, which is located close to the \bjet\ tagging efficiency maximum.
The corresponding $b$-jet tagging probability scale factors are shown in Figures~\ref{fig:results_2}(b) and~\ref{fig:results_2}(d). Scale factors are derived for all single-cut OPs using the same technique, resulting in similar results. The scale factors have values close to one and are about constant throughout the pseudo-continuous bins.
 
\begin{figure}[!htbp]
\centering
 
\subfigure[][]{
\includegraphics[width=0.45\textwidth]{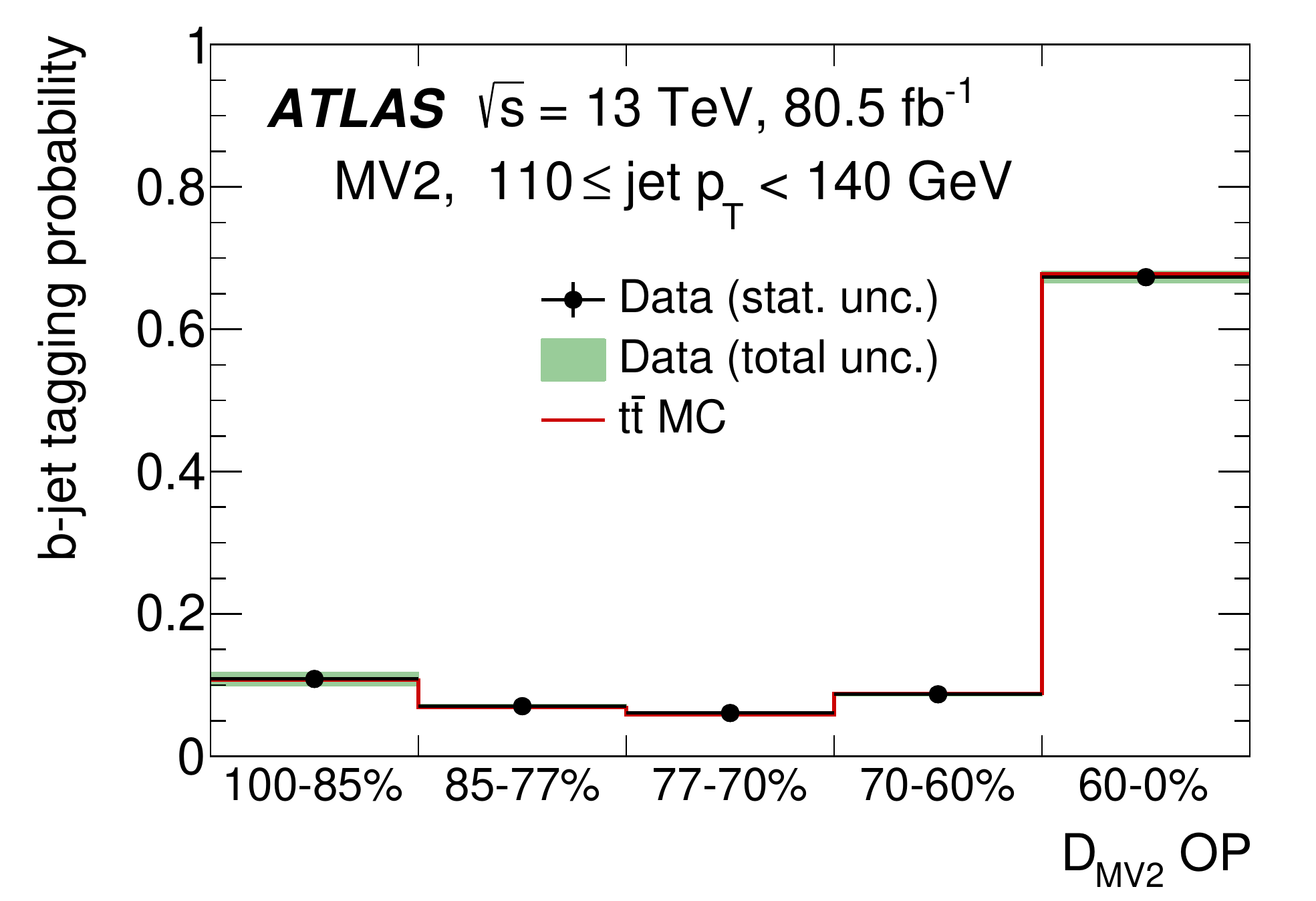}
}
\subfigure[][]{
\includegraphics[width=0.45\textwidth]{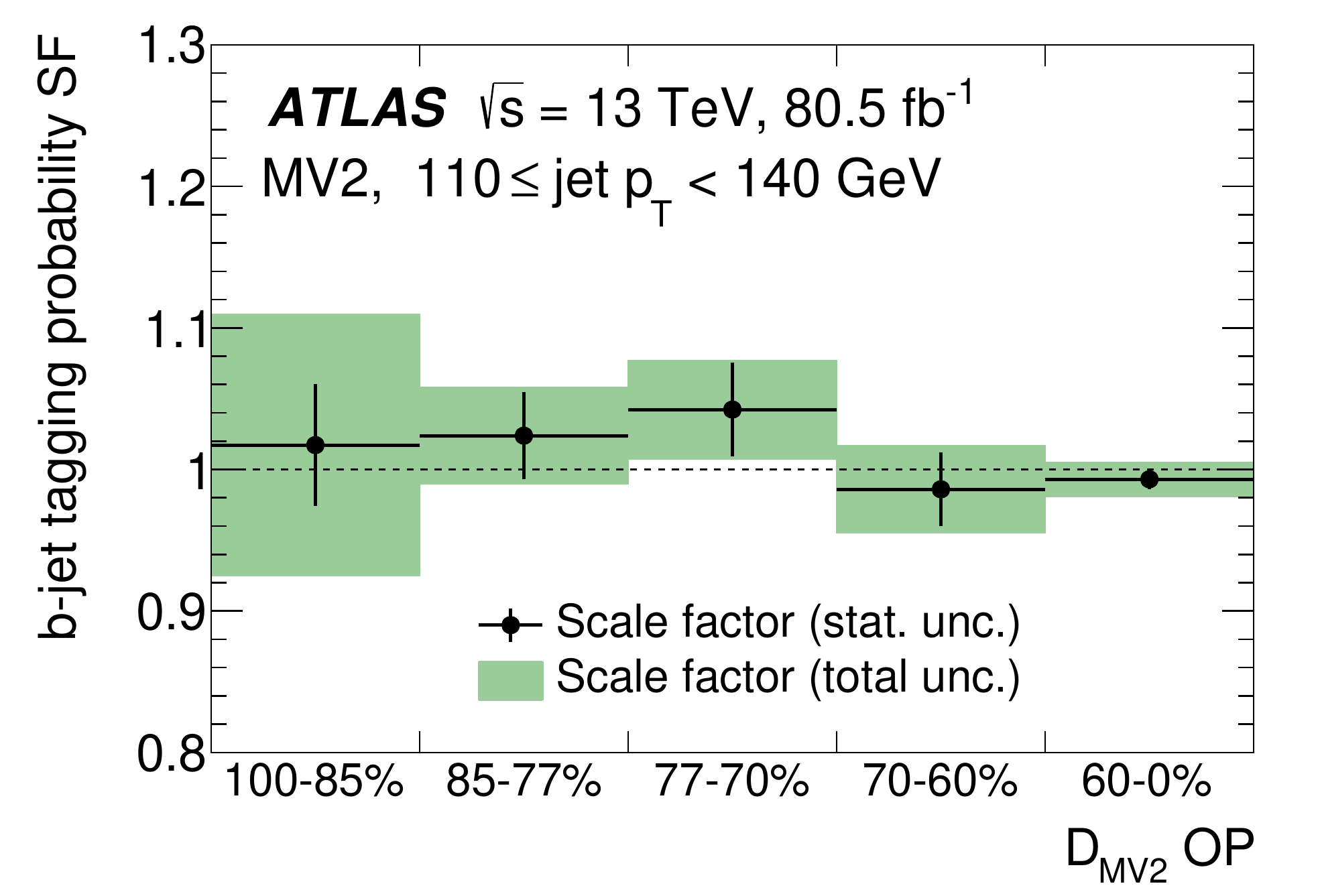}
}
\subfigure[][]{
\includegraphics[width=0.45\textwidth]{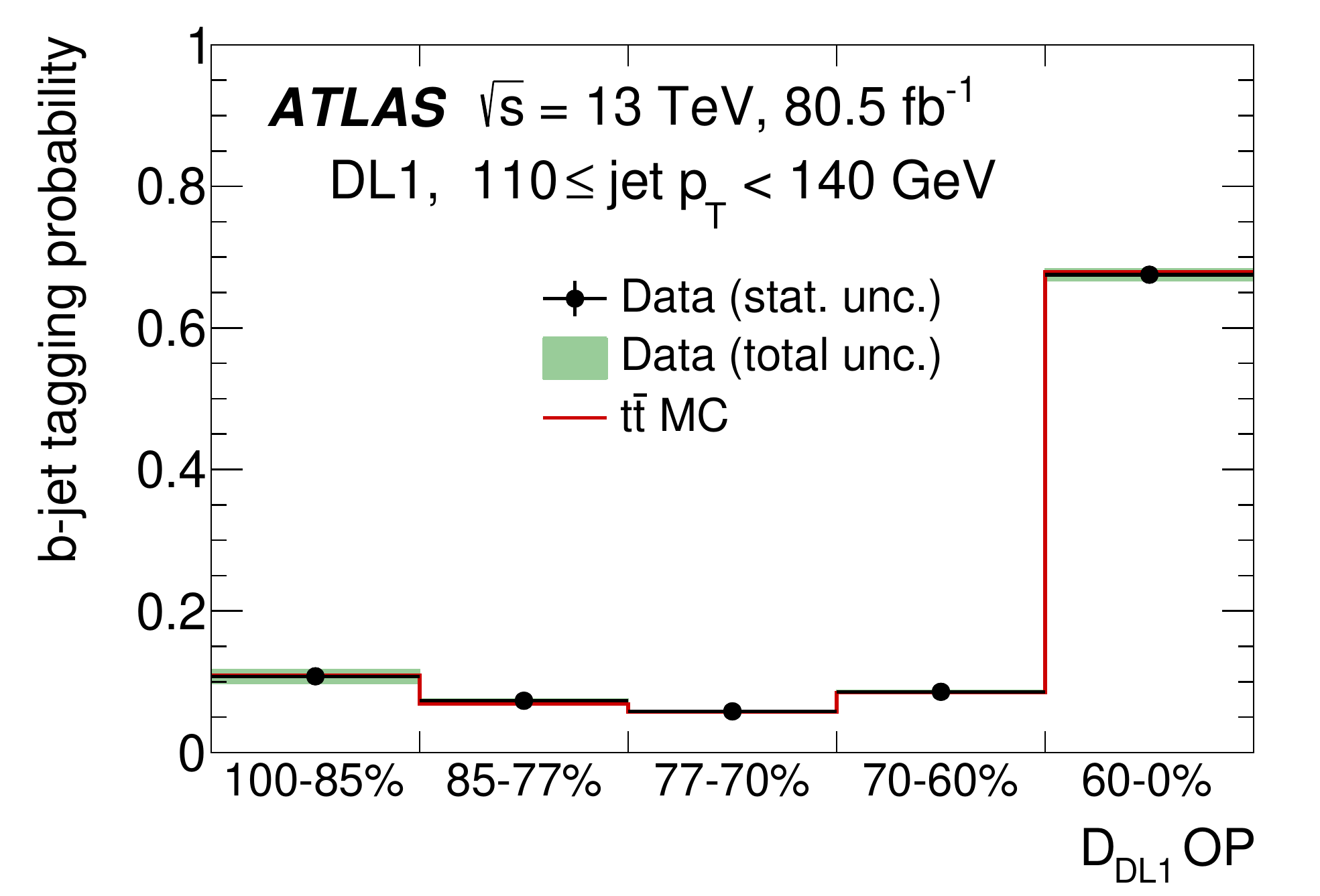}
}
\subfigure[][]{
\includegraphics[width=0.45\textwidth]{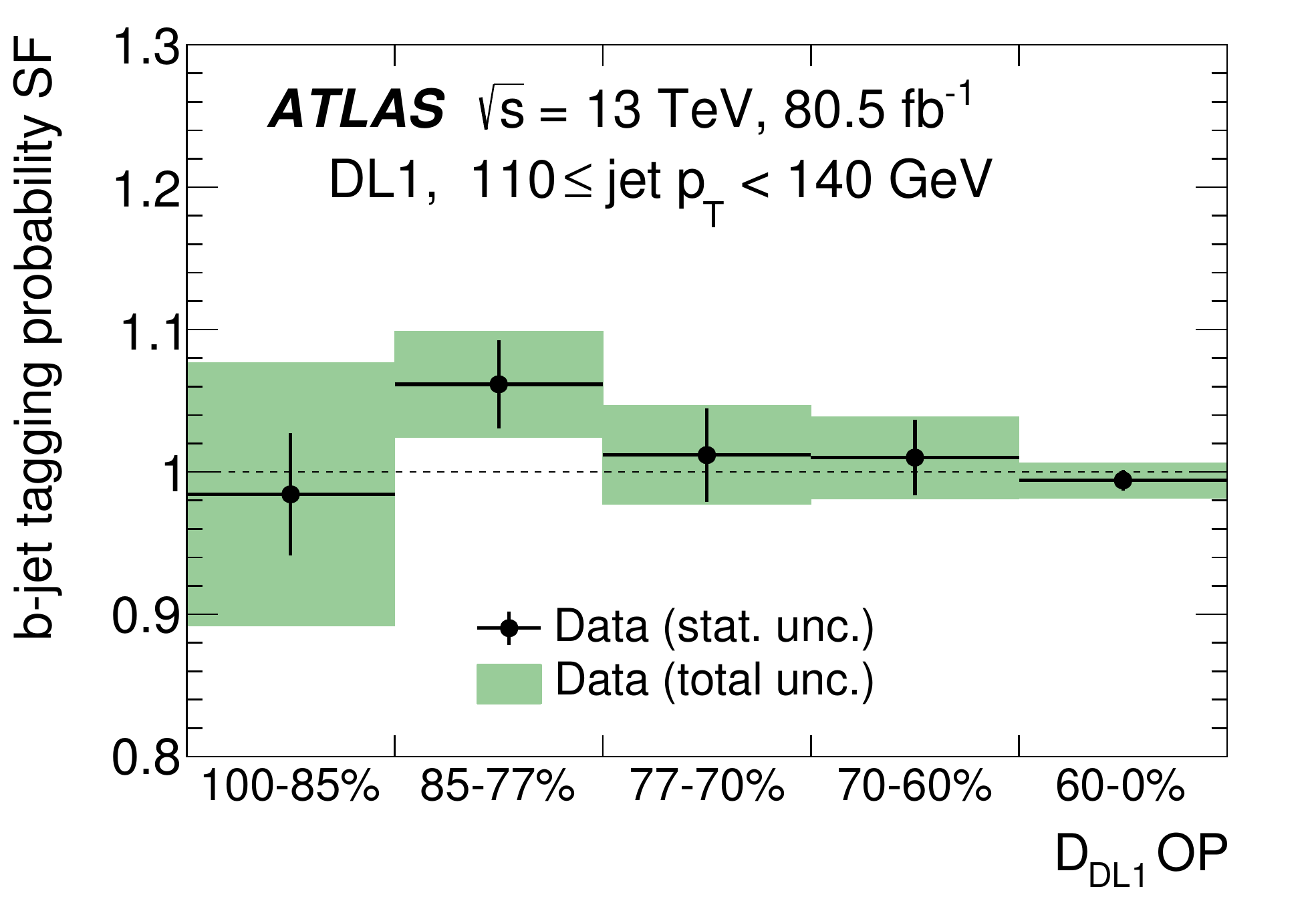}
}
\caption{
The (a, c) $b$-jet tagging probability and (b, d) $b$-jet tagging probability simulation-to-data scale factors for the (a, b) \textsc{MV2} and (c, d) \textsc{DL1} tagger for jets with $110\le\pt<140~\GeV$ in the various pseudo-continuous bins. The probability measurement is shown together with the probabilities derived from \ttbar\ simulated events passing the signal region selection. Vertical error bars include data statistical uncertainties only while the green bands correspond to the sum in quadrature of statistical and systematic uncertainties. The dots are located at the bin centres.
} \label{fig:results_2}
\end{figure}
 
The uncertainty in this measurement is summarised in Table~\ref{tab:results_table_2}. The total uncertainty varies from about 9\% in the 100\%--85\% bin to about 1\% in the 60\%--0\% bin. It is driven by the $t\bar{t}$ modelling uncertainties and data statistics, which is consistent with the result reported for the 70\% single-cut OP in this \pt\ range.

\begin{table}[htbp]
\caption {
Breakdown of the systematic uncertainties in the $b$-jet tagging probability measurement of the \textsc{MV2} tagger as a function of the `pseudo-continuous' bins for jets satisfying $110\le\pt<140~\GeV$. The `\ttbar\ modelling' and the `Single top modelling' uncertainties correspond to the sum in quadrature of the uncertainty in the parton shower, hadronisation model, initial-state and final-state radiation and PDF for \ttbar\ and single-top-quark production, respectively. The `Single top modelling' uncertainties include an additional source originating from the interference between single top and \ttbar\ production. `Other sources' corresponds to the sum in quadrature of the uncertainties related to jet energy resolution, electron and muon performance, $b$-tagging performance for light-flavour jets and $c$-jets,  JVT performance, diboson and $Z$+jet modelling (including normalisation and shape uncertainties) and pile-up modelling. All systematic uncertainties are fully correlated bin-by-bin whereas the statistical uncertainty correlations are evaluated following the procedures described in Section~\ref{sec:uncertainties}. In the case of correlated systematic uncertainties, the relative sign of the uncertainty in each bin is taken into account, even if not shown here.}
\label{tab:results_table_2}
\setlength\extrarowheight{3pt}
\centering
\small
\begin{tabular}{|l|lllll|}
\hline
\multirow{2}{*}{Source of uncertainty} & \multicolumn{5}{c|}{Impact on $\Pro_b$\ [\%] per pseudo-continuous OP} \\
& 100\%--85\%  & 85\%--77\%  & 77\%--70\% & 70\%--60\% & 60\%--0\% \\
\hline
Data statistics        &      4.2      &      3.0   &   3.2    &    2.6   &    0.7    \\
MC statistics          &       1.4     &      0.5   &    0.3   &    0.3   &     0.2   \\
Jet energy scale       &       1.0     &      0.2   &    0.2   &    0.2   &    0.1    \\
\ttbar\ modelling       &        7.2    &      0.9   &     0.8  &   0.9   &    0.9    \\
Single top modelling    &      2.4      &   0.5      &    0.3   &   0.2    &    0.3    \\
Fake leptons modelling  &     0.1       &    0.1     &   $<0.1$  &  $<0.1$    &   $<0.1$    \\
Other sources combined &      2.1      &    0.5     &    0.5   &  2.2     &  0.1      \\[3pt]
\hline
Total                  &      9.1      &     3.3    &     3.3  &   3.2    &  1.2     \\[3.5pt]
\hline
\end{tabular}
\end{table}

 
\section{Usage in ATLAS analysis}
\label{sec:usage}
 
This section details how the simulation-to-data scale factors are incorporated into ATLAS physics analyses.
Scale factors are smoothed, extrapolated beyond the jet~\pt range of the data measurement and corrected taking into account the generator dependence in the simulation.
The number of systematic uncertainties is reduced while preserving the bin-by-bin correlations.
The scale factors are then applied to ATLAS physics analyses by correcting the $b$-jet tagging response in simulation and by applying related uncertainties to the correction.
 
\subsection{Smoothing}
 
The simulation-to-data scale factors for single cut OPs are smoothed in jet~\pt using a local polynomial kernel estimator with a bandwith parameter of 0.2 following the procedure described in Ref.~\cite{PERF-2016-05}.
This procedure prevents distortions in the variables of interest induced by the application of the scale factors.
 
\subsection{Extrapolation to high-\pt jets}
 
The analysis described in this paper provides a precise measurement of the $b$-jet tagging efficiency in data and compares it with the one obtained from MC simulation.
Since there are currently not many $b$-jets in data for jet~\pt\ above $400~\GeV$ in di-lepton \ttbar\ events, an alternative assessment of the uncertainty in the $b$-jet tagging efficiency for jet~\pt\ in this range is developed to extend the single cut OP calibration to the entire jet~\pt range inspected by physics analyses in ATLAS.
Underlying quantities that are known to affect the $b$-tagging performance are varied in the simulation one by one and the $b$-jet tagging efficiency is recomputed in each case. The difference from the $b$-jet tagging efficiency obtained in the nominal simulation is then taken as an additional systematic uncertainty.
 
Four distinct sets of variables, related to the reconstruction of tracks, of jets, the modelling of the $b$-hadrons and the interaction of long-lived $b$-hadrons with the detector material, are considered.
Among the uncertainties related to the reconstruction of tracks, the ones that are found to most affect the $b$-tagging performance are those related to the track impact-parameter resolution, the fraction of fake tracks, the description of the detector material, and the track multiplicity per jet.
The uncertainty in the impact-parameter resolution includes the effects of alignment, dead modules and additional material not accurately modelled in the simulation.
The uncertainty is derived from several event topologies, including dijet events where effects due to tracking in dense environments, such as in the cores of high-energy jets, are included~\cite{PERF-2015-08}.
No dedicated studies of samples enriched in high-energy $b$-jets, where collimated tracks from displaced decay vertices conspire to create a challenging environment for the track reconstruction algorithm, are included at this stage.
The effect of the parton shower simulation and $b$-quark fragmentation function is evaluated by comparing the $b$-jet tagging efficiency with the one obtained from the alternative \ttbar~event simulations described in Section~\ref{sec:samples}.
In standard ATLAS MC simulations, interactions with detector material are simulated only for the decay products of the $b$-hadron and not for the $b$-hadron itself. 
Given that about $5$\% of the $b$-hadrons within $b$-jets with jet~$\pt=150~\GeV$ decay after the innermost pixel detector layer, differences in the $b$-jet tagging efficiency at high \pt\ are expected.
In order to evaluate the size of the effect, the $Z'$ sample described in Section~\ref{sec:samples} was enhanced to include the interaction of $b$-hadrons with the detector material, and the $b$-jet tagging efficiency derived from this sample is compared with the one obtained from the nominal $Z'$ sample.
 
These sources of uncertainties are found to have a similar impact on the $b$-jet tagging efficiency of the \textsc{MV2} and \textsc{DL1} taggers in the jet~\pt\ range $400~\GeV$ to $1~\TeV$.
In this jet \pt\ regime, the modelling uncertainties are dominant, reaching $~2\%$ for $\pt\sim400~\GeV$ and growing linearly to $\sim4\%$ at the~\TeV\ scale.
The uncertainty due to the interaction with the detector material is also important and found to be $\sim1\%$ at $\pt\sim700~\GeV$, growing to $\sim2\%$ at $\sim1~\TeV$.
Other leading uncertainties include the jet energy scale and track impact-parameter resolution uncertainties, reaching about 2.5\% and 1\% at $\sim1~\TeV$, respectively.
At the~\TeV\ scale, the impact of the extrapolation uncertainty is different for \textsc{MV2} and \textsc{DL1}, due to the differing efficiency profile of the two $b$-taggers. The $b$-jet tagging efficiency of \textsc{DL1} falls more steeply at high \pt compared to that of \textsc{MV2}, which is approximately constant. This results in the jet energy scale uncertainty having a much larger impact for the \textsc{DL1} tagger, due to the increased impact of the migration of jets between the \pt\ bins.
 
The simulation-to-data scale factor measured in the highest \pt\ bin considered in the collision data analysis is extrapolated for $\pt\ge400~\GeV$.
The mean value and uncertainties after smoothing at $\pt=400~\GeV$ are assumed to stay valid for higher jet \pt.
An extrapolation uncertainty is then constructed as the sum in quadrature of all the uncertainties described above, rescaled in proportion to their respective values in the highest \pt bin of the data measurement, and added in quadrature to the pre-existing uncertainties.
The result of the procedures of smoothing and extrapolating the single cut OP scale factors are shown in Figure~\ref{fig:calib_extrap}, where both the $b$-jet tagging efficiency as directly measured in data and its extrapolation are shown for the $\varepsilon_b = 70\%$ single cut OP of the \textsc{MV2} and \textsc{DL1} taggers.
 
\begin{figure}[!htbp]
\centering
\subfigure[][]{
\includegraphics[width=0.45\textwidth]{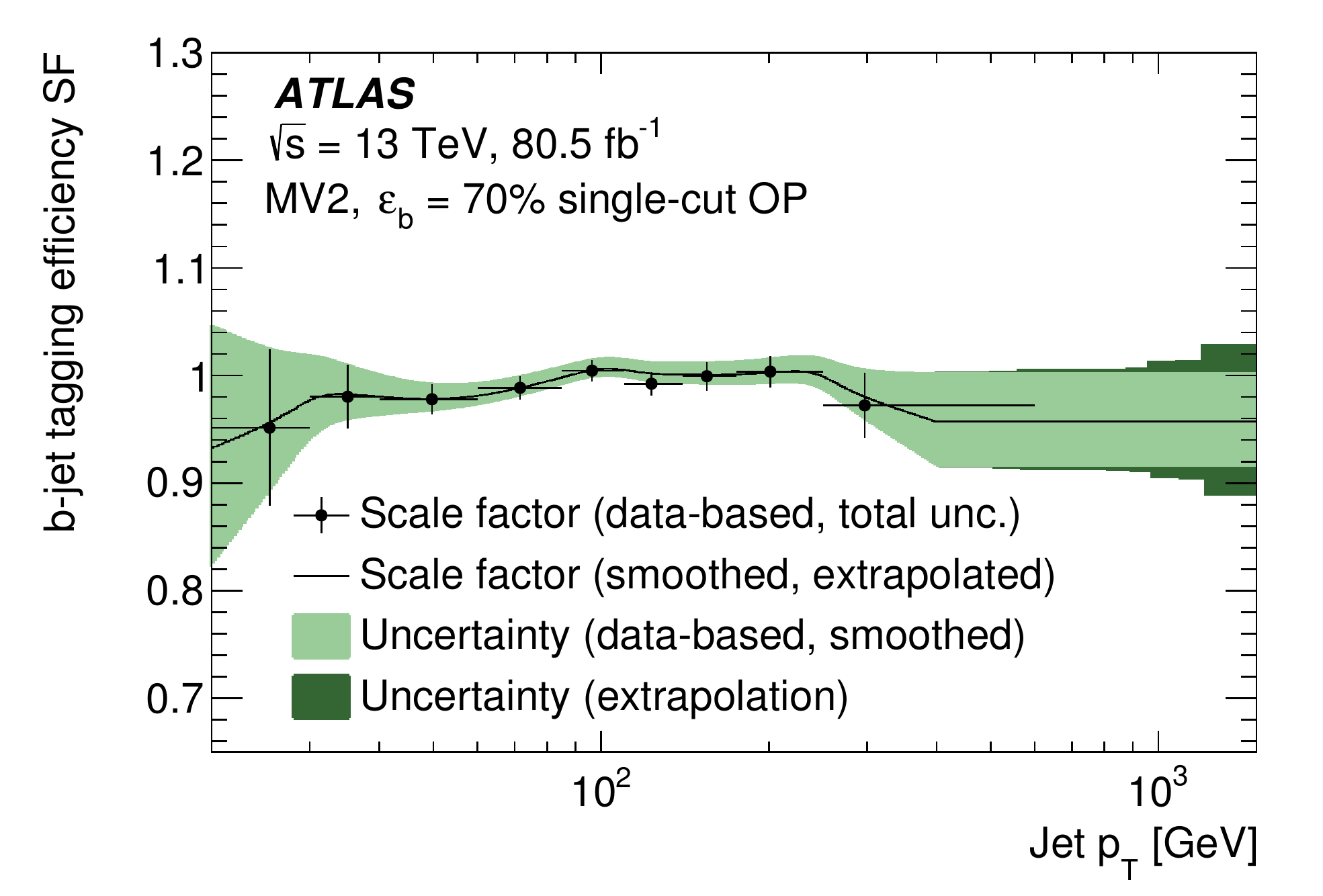}
}
\subfigure[][]{
\includegraphics[width=0.45\textwidth]{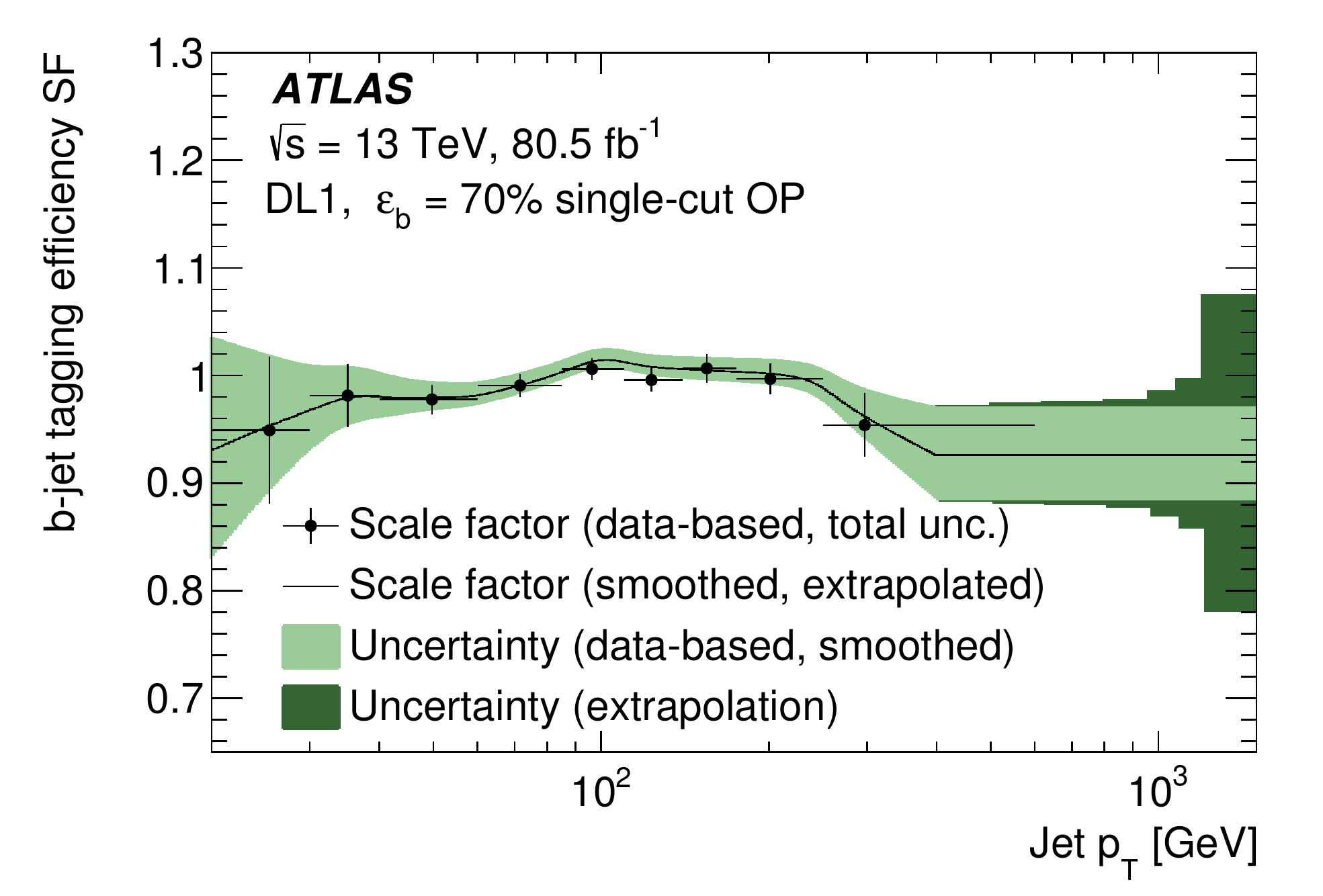}
}
\caption{$b$-jet tagging efficiency simulation-to-data scale factors for the $\varepsilon_b = 70\%$ single cut OP of the (a) \textsc{MV2} and (b) \textsc{DL1} taggers, including the smoothed and extrapolated results.
The bin centres are used for the smoothing whereas the dots are located at the mean of the $b$-jet \pt\ distribution in each \pt\ bin.}
\label{fig:calib_extrap}
\end{figure}
 
\subsection{Generator dependence}
 
The $b$-jet tagging efficiency in the simulation depends on several properties, such as the production fractions of the different $b$-hadron species, the fragmentation function and the number of additional charged particles near the $b$-hadron, which are not necessarily identical among the different MC event generators.
Simulation-to-simulation scale factors are therefore derived to take into account differences in the $b$-jet tagging efficiency due to the usage of a different fragmentation model to that used to derive the simulation-to-data scale factors.
The simulation-to-simulation scale factors are computed as ratios of $b$-jet tagging efficiencies in the same jet~\pt\ bins of the alternative and nominal \ttbar~samples. For $b$-jets, they range from 1\% to 3\% as a function of jet~\pt.
These scale factors are used when the $b$-jet tagging efficiency simulation-to-data scale factors are applied to a sample produced with a showering generator different from the one used for the nominal \ttbar\ sample used in the denominator of the scale factor calculation.
 
\subsection{Reduction of systematic uncertainties}
 
The individual application in a physics analysis of each independent systematic uncertainty included in Figure~\ref{fig:calib_extrap} would lead to a large number of variations.
A method for reducing the total number of uncertainties while preserving the bin-by-bin correlations is provided for use in ATLAS  physics analyses and is described in Ref.~\cite{PERF-2016-05}.
This is achieved by constructing the covariance matrix for each source of uncertainty and by summing these matrices together.
Bin-by-bin correlations are kept as non-zero off-diagonal elements.
As this equates to the total covariance matrix, which is symmetric and positive-definite, an eigenvector decomposition is performed.
The resulting number of variations equals the number of jet \pt bins and is further reduced where eigenvalue variations are shown to have a negligible impact on a result. 
 
\subsection{Application to physics analyses}
 
For each jet where $b$-jet tagging is applied in ATLAS physics analyses, a weight is applied in simulation to match the tagging rate as measured in data by the calibration analyses.
The weight is jet-flavour dependent.
The calibration analysis described in this paper is the baseline correction for jets labelled as $b$-jets.
If the jet is tagged using a single cut OP in MC simulation the weight is simply the smoothed simulation-to-data scale factor itself:
 
\begin{equation}
\label{eqn:weight_tagged}
w_{\text{jet}} = SF(\pt) \,,
\end{equation}
 
where $SF(\pt)$ is the smoothed $b$-jet tagging efficiency scale factor evaluated at a given \pt. If the jet is not tagged the weight becomes:
 
\begin{equation}
\label{eqn:weight_not_tagged}
w_{\text{jet}} = \frac{1-\epsilon_{b}^{\textrm{data}}(\pt)}{1-\epsilon_{b}^{\textrm{MC}}(\pt)}
= \frac{1-SF(\pt)\epsilon_{b}^{\textrm{MC}}(\pt)}{1-\epsilon_{b}^{\textrm{MC}}(\pt)} \,.
\end{equation}
 
The latter form of Eq.~(\ref{eqn:weight_not_tagged}) is adopted because, in this way, and by constructing high-granularity efficiency distributions, possible differences in the tagging rate induced by event topologies are minimised. These weights are necessary to ensure that the number of events remains the same after corrections.
The final event weight is then computed as the product of all jet weights.
In cases where the physics analysis does not rely on the single cut OP but on the pseudo-continuous bins of the discriminant distribution, Eq.~(\ref{eqn:weight_tagged}) becomes dependent on the \pt\ and pseudo-continuous bins $T^m,O^k$, and $SF(T^m,O^k)$ becomes the $b$-jet tagging probability SF measured in the pseudo-continuous bin $O^k$ and \pt\ bin $T^m$ while Eq.~(\ref{eqn:weight_not_tagged}) becomes unnecessary.
 
 
\FloatBarrier
 
\section{Conclusion}
 
Several $b$-tagging algorithms are used to analyse data recorded by the ATLAS experiment during Run 2 of the LHC. Their performance is evaluated in simulation, and the $b$-jet tagging efficiencies of the \textsc{MV2} and \textsc{DL1} algorithms are measured in $pp$ collision data.

The $b$-jet identification strategy combines
the results of low-level algorithms (\textsc{IP2D}, \textsc{IP3D}, \textsc{SV1}, \textsc{JetFitter})
into high-level algorithms based on multivariate classifiers (\textsc{MV2}, \textsc{DL1}).
The low-level algorithms either exploit the large impact parameters of the tracks originating from the $b$-hadron decay products or attempt to directly reconstruct heavy-flavour
hadron vertices. Large increases in light-flavour jet and $c$-jet rejection are obtained by the
\textsc{MV2} and \textsc{DL1} algorithms compared to each individual low-level algorithm,
illustrating the high complementarity of the latter and validating the overall strategy followed by the ATLAS Collaboration.
 
The $b$-jet tagging efficiency of the \textsc{MV2} and \textsc{DL1} algorithms are measured
in $80.5$~\ifb{} of proton--proton collision data collected by the ATLAS detector over the period
2015--2017 at a centre-of-mass energy $\sqrt{s}=13~\TeV$. A high-purity sample of
dileptonic $t\bar{t}$ events is obtained by retaining events with exactly one muon, one electron and
two hadronic jets. Events are classified according to the transverse momentum of each of the two jets
as well as the jet--lepton invariant masses obtained when pairing exclusively each jet with the lepton
most likely to originate from the same top-quark decay. A combinatorial likelihood approach is then used
to simultaneously extract the jet flavour composition of the sample and the $b$-jet tagging probabilities for jets in a transverse momentum range from $20$ to $600~\GeV$, from which the
$b$-jet tagging efficiencies for single cut OPs are obtained. Simulation-to-data scale factors are computed by comparing the efficiency measured in collision data with that observed
in the simulation. The measured simulation-to-data scale factors are close to unity with a total
uncertainty ranging from 1\% to 8\% for single cut OPs. The precision of the measurement is limited at low \pt\ by the uncertainty in the jet energy scale, which is detector-related, and at high \pt\ by the size of the dataset, which will grow in the future. It was previously limited by the modelling of top-quark pair production. Further procedures relating to the simulation-to-data scale factors are undertaken to correct for generator dependences, smooth the shape, extrapolate beyond the range of the data measurement, reduce the number of nuisance parameters and apply them in ATLAS analyses. 
 
This result demonstrates a significant improvement in the precision of the $b$-jet tagging efficiency
measurement for the ATLAS experiment, which is typically limited by systematic uncertainties. The total uncertainty
is reduced by up to a factor of two relative to the previous ATLAS results. 
This is achieved by the improvement in the measurement method, which simultaneously extracts the $b$-jet tagging efficiency and jet flavour composition.
 

\section*{Acknowledgements}
 
 
We thank CERN for the very successful operation of the LHC, as well as the
support staff from our institutions without whom ATLAS could not be
operated efficiently.
 
We acknowledge the support of ANPCyT, Argentina; YerPhI, Armenia; ARC, Australia; BMWFW and FWF, Austria; ANAS, Azerbaijan; SSTC, Belarus; CNPq and FAPESP, Brazil; NSERC, NRC and CFI, Canada; CERN; CONICYT, Chile; CAS, MOST and NSFC, China; COLCIENCIAS, Colombia; MSMT CR, MPO CR and VSC CR, Czech Republic; DNRF and DNSRC, Denmark; IN2P3-CNRS, CEA-DRF/IRFU, France; SRNSFG, Georgia; BMBF, HGF, and MPG, Germany; GSRT, Greece; RGC, Hong Kong SAR, China; ISF and Benoziyo Center, Israel; INFN, Italy; MEXT and JSPS, Japan; CNRST, Morocco; NWO, Netherlands; RCN, Norway; MNiSW and NCN, Poland; FCT, Portugal; MNE/IFA, Romania; MES of Russia and NRC KI, Russian Federation; JINR; MESTD, Serbia; MSSR, Slovakia; ARRS and MIZ\v{S}, Slovenia; DST/NRF, South Africa; MINECO, Spain; SRC and Wallenberg Foundation, Sweden; SERI, SNSF and Cantons of Bern and Geneva, Switzerland; MOST, Taiwan; TAEK, Turkey; STFC, United Kingdom; DOE and NSF, United States of America. In addition, individual groups and members have received support from BCKDF, CANARIE, CRC and Compute Canada, Canada; COST, ERC, ERDF, Horizon 2020, and Marie Sk{\l}odowska-Curie Actions, European Union; Investissements d' Avenir Labex and Idex, ANR, France; DFG and AvH Foundation, Germany; Herakleitos, Thales and Aristeia programmes co-financed by EU-ESF and the Greek NSRF, Greece; BSF-NSF and GIF, Israel; CERCA Programme Generalitat de Catalunya, Spain; The Royal Society and Leverhulme Trust, United Kingdom.
 
The crucial computing support from all WLCG partners is acknowledged gratefully, in particular from CERN, the ATLAS Tier-1 facilities at TRIUMF (Canada), NDGF (Denmark, Norway, Sweden), CC-IN2P3 (France), KIT/GridKA (Germany), INFN-CNAF (Italy), NL-T1 (Netherlands), PIC (Spain), ASGC (Taiwan), RAL (UK) and BNL (USA), the Tier-2 facilities worldwide and large non-WLCG resource providers. Major contributors of computing resources are listed in Ref.~\cite{ATL-GEN-PUB-2016-002}.
 

\clearpage
\printbibliography

@article{Sjostrand:2007gs,
      author         = "Sj{\"o}strand, Torbjorn and Mrenna, Stephen and Skands, Peter
                        Z.",
      title          = "{A brief introduction to PYTHIA 8.1}",
      journal        = "Comput. Phys. Commun.",
      volume         = "178",
      year           = "2008",
      pages          = "852-867",
      doi            = "10.1016/j.cpc.2008.01.036",
      eprint         = "0710.3820",
      archivePrefix  = "arXiv",
      primaryClass   = "hep-ph",
      reportNumber   = "CERN-LCGAPP-2007-04, LU-TP-07-28,
                        FERMILAB-PUB-07-512-CD-T",
      SLACcitation   = "%%CITATION = ARXIV:0710.3820;%%"
}

@article{Nason:2004rx,
      author         = "Nason, Paolo",
      title          = "{A New method for combining NLO QCD with shower Monte
                        Carlo algorithms}",
      journal        = "JHEP",
      volume         = "11",
      year           = "2004",
      pages          = "040",
      doi            = "10.1088/1126-6708/2004/11/040",
      eprint         = "hep-ph/0409146",
      archivePrefix  = "arXiv",
      OPTprimaryClass   = "hep-ph",
      reportNumber   = "BICOCCA-FT-04-11",
      SLACcitation   = "%%CITATION = HEP-PH/0409146;%%"
}

@article{Frixione:2007vw,
      author         = "Frixione, Stefano and Nason, Paolo and Oleari, Carlo",
      title          = "{Matching NLO QCD computations with parton shower
                        simulations: the POWHEG method}",
      journal        = "JHEP",
      volume         = "11",
      year           = "2007",
      pages          = "070",
      doi            = "10.1088/1126-6708/2007/11/070",
      eprint         = "0709.2092",
      archivePrefix  = "arXiv",
      primaryClass   = "hep-ph",
      reportNumber   = "BICOCCA-FT-07-9, GEF-TH-21-2007",
      SLACcitation   = "%%CITATION = ARXIV:0709.2092;%%"
}

@article{Alioli:2010xd,
      author         = "Alioli, Simone and Nason, Paolo and Oleari, Carlo and Re,
                        Emanuele",
      title          = "{A general framework for implementing NLO calculations in
                        shower Monte Carlo programs: the POWHEG BOX}",
      journal        = "JHEP",
      volume         = "06",
      year           = "2010",
      pages          = "043",
      doi            = "10.1007/JHEP06(2010)043",
      eprint         = "1002.2581",
      archivePrefix  = "arXiv",
      primaryClass   = "hep-ph",
      reportNumber   = "DESY-10-018, SFB-CPP-10-22, IPPP-10-11, DCPT-10-22",
      SLACcitation   = "%%CITATION = ARXIV:1002.2581;%%"
}

@article{Frixione:2007nw,
      author         = "Frixione, Stefano and Nason, Paolo and Ridolfi, Giovanni",
      title          = "{A Positive-weight next-to-leading-order Monte Carlo for
                        heavy flavour hadroproduction}",
      journal        = "JHEP",
      volume         = "09",
      pages          = "126",
      doi            = "10.1088/1126-6708/2007/09/126",
      year           = "2007",
      eprint         = "0707.3088",
      archivePrefix  = "arXiv",
      primaryClass   = "hep-ph",
      reportNumber   = "BICOCCA-FT-07-12, GEF-TH-19-2007",
      SLACcitation   = "%%CITATION = ARXIV:0707.3088;%%",
}

@article{Gleisberg:2008ta,
      author         = "Gleisberg, T. and Hoeche, Stefan. and Krauss, F. and
                        Schonherr, M. and Schumann, S. and Siegert, F. and Winter,
                        J.",
      title          = "{Event generation with SHERPA 1.1}",
      journal        = "JHEP",
      volume         = "02",
      year           = "2009",
      pages          = "007",
      doi            = "10.1088/1126-6708/2009/02/007",
      eprint         = "0811.4622",
      archivePrefix  = "arXiv",
      primaryClass   = "hep-ph",
      reportNumber   = "FERMILAB-PUB-08-477-T, SLAC-PUB-13420, ZU-TH-17-08,
                        DCPT-08-138, IPPP-08-69, EDINBURGH-2008-30, MCNET-08-14",
      SLACcitation   = "%%CITATION = ARXIV:0811.4622;%%"
}

@article{Cacciari:2008gp,
      author         = "Cacciari, Matteo and Salam, Gavin P. and Soyez, Gregory",
      title          = "{The anti-$k_{t}$ jet clustering algorithm}",
      journal        = "JHEP",
      volume         = "04",
      pages          = "063",
      doi            = "10.1088/1126-6708/2008/04/063",
      year           = "2008",
      eprint         = "0802.1189",
      archivePrefix  = "arXiv",
      PrimaryClass   = "hep-ph",
      OPTprimaryClass   = "hep-ph",
      reportNumber   = "LPTHE-07-03",
      SLACcitation   = "%%CITATION = ARXIV:0802.1189;%%",
}

@article{Cacciari:2008gn,
      author         = "Cacciari, Matteo and Salam, Gavin P. and Soyez, Gregory",
      title          = "{The catchment area of jets}",
      journal        = "JHEP",
      volume         = "04",
      year           = "2008",
      pages          = "005",
      doi            = "10.1088/1126-6708/2008/04/005",
      eprint         = "0802.1188",
      archivePrefix  = "arXiv",
      primaryClass   = "hep-ph",
      reportNumber   = "LPTHE-07-02",
      SLACcitation   = "%%CITATION = ARXIV:0802.1188;%%"
}

@article{Ball:2014uwa,
      author         = "Ball, Richard D. and others",
      title          = "{Parton distributions for the LHC Run II}",
      collaboration  = "NNPDF",
      journal        = "JHEP",
      volume         = "04",
      year           = "2015",
      pages          = "040",
      doi            = "10.1007/JHEP04(2015)040",
      eprint         = "1410.8849",
      archivePrefix  = "arXiv",
      primaryClass   = "hep-ph",
      reportNumber   = "EDINBURGH-2014-15, IFUM-1034-FT, CERN-PH-TH-2013-253,
                        OUTP-14-11P, CAVENDISH-HEP-14-11",
      SLACcitation   = "%%CITATION = ARXIV:1410.8849;%%"
}

@article{Ball:2012cx,
      author         = "Ball, Richard D. and others",
      title          = "{Parton distributions with LHC data}",
      journal        = "Nucl. Phys. B",
      volume         = "867",
      year           = "2013",
      pages          = "244-289",
      doi            = "10.1016/j.nuclphysb.2012.10.003",
      eprint         = "1207.1303",
      archivePrefix  = "arXiv",
      primaryClass   = "hep-ph",
      reportNumber   = "EDINBURGH-2012-08, IFUM-FT-997, FR-PHENO-2012-014,
                        RWTH-TTK-12-25, CERN-PH-TH-2012-037, SFB-CPP-12-47",
      SLACcitation   = "%%CITATION = ARXIV:1207.1303;%%"
}

@Article{Agostinelli:2002hh,
     author    = "Agostinelli, S. and others",
OPTcollaboration = "GEANT4",
     title     = "{GEANT4 - a simulation toolkit}",
     journal   = "Nucl. Instrum. Meth. A",
     volume    = "506",
     year      = "2003",
     pages     = "250-303",
     doi       = "10.1016/S0168-9002(03)01368-8",
     SLACcitation  = "%%CITATION = NUIMA,A506,250;%%"
}

@article{Lange:2001uf,
      author         = "Lange, D. J.",
      title          = "{The EvtGen particle decay simulation package}",
      booktitle      = "{Proceedings, 7th International Conference on B physics
                        at hadron machines (BEAUTY 2000)}",
      journal        = "Nucl. Instrum. Meth. A",
      volume         = "462",
      year           = "2001",
      pages          = "152-155",
      doi            = "10.1016/S0168-9002(01)00089-4",
      SLACcitation   = "%%CITATION = NUIMA,A462,152;%%"
}

@article{Butterworth:2015oua,
      author         = "Butterworth, Jon and others",
      title          = "{PDF4LHC recommendations for LHC Run II}",
      journal        = "J. Phys. G",
      volume         = "43",
      year           = "2016",
      pages          = "023001",
      doi            = "10.1088/0954-3899/43/2/023001",
      eprint         = "1510.03865",
      archivePrefix  = "arXiv",
      primaryClass   = "hep-ph",
      reportNumber   = "OUTP-15-17P, SMU-HEP-15-12, TIF-UNIMI-2015-14,
                        LCTS-2015-27, CERN-PH-TH-2015-249",
      SLACcitation   = "%%CITATION = ARXIV:1510.03865;%%"
}

@article{Re:2010bp,
      author         = "Re, Emanuele",
      title          = "{Single-top Wt-channel production matched with parton
                        showers using the POWHEG method}",
      journal        = "Eur. Phys. J. C",
      volume         = "71",
      year           = "2011",
      pages          = "1547",
      doi            = "10.1140/epjc/s10052-011-1547-z",
      eprint         = "1009.2450",
      archivePrefix  = "arXiv",
      primaryClass   = "hep-ph",
      reportNumber   = "IPPP-10-74, DCPT-10-148",
      SLACcitation   = "%%CITATION = ARXIV:1009.2450;%%"
}

@article{Alioli:2009je,
      author         = "Alioli, Simone and Nason, Paolo and Oleari, Carlo and Re,
                        Emanuele",
      title          = "{NLO single-top production matched with shower in POWHEG:
                        s- and t-channel contributions}",
      journal        = "JHEP",
      volume         = "09",
      year           = "2009",
      pages          = "111",
      doi            = "10.1007/JHEP02(2010)011, 10.1088/1126-6708/2009/09/111",
      note           = "[Erratum: JHEP02,011(2010)]",
      eprint         = "0907.4076",
      archivePrefix  = "arXiv",
      primaryClass   = "hep-ph",
      SLACcitation   = "%%CITATION = ARXIV:0907.4076;%%"
}

@article{Bellm:2015jjp,
      author         = "Bellm, Johannes and others",
      title          = "{Herwig 7.0/Herwig++ 3.0 release note}",
      journal        = "Eur. Phys. J. C",
      volume         = "76",
      year           = "2016",
      number         = "4",
      pages          = "196",
      doi            = "10.1140/epjc/s10052-016-4018-8",
}

@article{James:2004xla,
      author         = "James, Fred and Winkler, Matthias",
      title          = "{MINUIT User's Guide}",
      year           = "2004",
      url          = {http://seal.web.cern.ch/seal/documents/minuit/mnusersguide.pdf},
      SLACcitation   = "%%CITATION = INSPIRE-1258345;%%"
}

@BOOK{Bohm:389738,
      author       = {Bohm, Gerhard and Zech, Günter},
      title        = {{I}ntroduction to {S}tatistics and {D}ata {A}nalysis for
                      {P}hysicists; 3rd revised},
      address      = {Hamburg},
      publisher    = {Verlag Deutsches Elektronen-Synchrotron},
      reportid     = {PUBDB-2017-08987},
      isbn         = {978-3-945931-13-4},
      pages        = {488},
      year         = {2017},
      cin          = {L},
      cid          = {I:(DE-H253)L-20120731},
      pnm          = {899 - ohne Topic (POF3-899)},
      pid          = {G:(DE-HGF)POF3-899},
      experiment   = {EXP:(DE-MLZ)NOSPEC-20140101},
      typ          = {PUB:(DE-HGF)3},
      doi          = {10.3204/PUBDB-2017-08987},
      url          = {http://bib-pubdb1.desy.de/record/389738},
}

@article{Fruhwirth:1987fm,
      author         = "{Fr\"uhwirth, R.}",
      title          = "{Application of Kalman filtering to track and vertex
                        fitting}",
      journal        = "Nucl. Instrum. Meth. A",
      volume         = "262",
      year           = "1987",
      pages          = "444-450",
      doi            = "10.1016/0168-9002(87)90887-4",
      reportNumber   = "HEPHY-PUB-87-503",
      SLACcitation   = "%%CITATION = NUIMA,A262,444;%%"
}

@Article{TMVA,
    author         = "H{\"o}cker, Andreas and others",
    title          = "{TMVA - Toolkit for Multivariate Data Analysis}",
    year           = "2007",
    eprint         = "physics/0703039",
    archivePrefix  = "arXiv",
    primaryClass   = "physics.data-an",
}

@Article{2016arXiv160502688short,
    author = {{Theano Development Team}},
    title = "{Theano: A {Python} framework for fast computation of mathematical expressions}",
    journal = {arXiv - 1605.02688 },
    year = "2016",
    eprint         = "1605.02688",
    archivePrefix  = "arXiv",
    primaryClass   = "cs.SC",
}

@Article{Kingma:2014aa,
    author= {{D.P. Kingma, J. Ba}},
    title = "{Adam: A Method for Stochastic Optimization}",
    journal = {arXiv - 1412.6980},
    year = "2014",
    eprint         = "1412.6980",
    archivePrefix  = "arXiv",
    primaryClass   = "cs.SC",
}

@Article{Goodfellow:2013aa,
    author= {{I.J. Goodfellow, D. Warde-Farley, M. Mirza, A. Courville, Y. Bengio}},
    title = "{Maxout Networks}",
    journal = {arXiv - 1302.4389},
    year = "2013",
    eprint         = "1302.4389",
    archivePrefix  = "arXiv",
    primaryClass   = "stat.ML",
}

@Article{Ioffe:2015aa,
    author= {{S. Ioffe; C. Szegedy}},
    title = "{Batch Normalization: Accelerating Deep Network Training by Reducing
    Internal Covariate Shift}",
    journal = {arXiv -1502.03167},
    year = "2015",
    url = {http://arxiv.org/abs/1502.03167},
    eprint         = "1502.03167",
    archivePrefix  = "arXiv",
    primaryClass   = "cs.LG",
}

@misc{chollet2015keras,
    title={Keras},
    author={Chollet, Fran\c{c}ois and others},
    year={2015},
    howpublished={\url{https://keras.io}},
}

@article{chi2,
author = {K. Pearson},
title = {On the criterion that a given system of deviations from the probable in the case of a correlated system of variables is such that it can be reasonably supposed to have arisen from random sampling},
journal = {The London, Edinburgh, and Dublin Philosophical Magazine and Journal of Science},
volume = {50},
number = {302},
pages = {157-175},
year  = {1900},
publisher = {Taylor \& Francis},
doi = {10.1080/14786440009463897},
}

@article{Billoir:245288,
    author        = "Billoir, P and Qian, S",
    title         = "Fast vertex fitting with a local parametrization of tracks",
    journal       = "Nucl. Instrum. Methods Phys. Res., A",
    volume        = "311",
    pages         = "139-150",
    year          = "1992",
    doi           = "10.1016/0168-9002(92)90859-3",
}

@article{LUCID2,
  author={G. Avoni and others},
  title={The new LUCID-2 detector for luminosity measurement and monitoring in ATLAS},
  journal={JINST},
  volume={13},
  number={07},
  pages={P07017},
  doi="10.1088/1748-0221/13/07/P07017",
  year={2018},
}

@Booklet{ATLAS-CONF-2019-021,
    author         = "{ATLAS Collaboration}",
    title          = "{Luminosity Determination in \(pp\) collisions at \(\sqrt{s} = 13\,\text{TeV}\) using the ATLAS Detector at the LHC}",
    howpublished   = "{ATLAS-CONF-2019-021}",
    url            = "https://cds.cern.ch/record/2677054",
    year           = "2019",
}

@Article{new_egamma,
    author         = "{ATLAS Collaboration}",
    title          = "{Electron and photon performance measurements with the ATLAS detector using the 2015--2017 LHC proton--proton collision data}",
    reportNumber   = "CERN-EP-2019-145",
    year           = "2019",
    eprint         = "1908.00005",
    archivePrefix  = "arXiv",
    primaryClass   = "hep-ex",
}

@Article{Evans:2008zzb,
      author         = "Evans, Lyndon and Bryant, Philip",
      title          = "{LHC Machine}",
      journal        = "JINST",
      volume         = "3",
      pages          = "S08001",
      doi            = "10.1088/1748-0221/3/08/S08001",
      year           = "2008",
      SLACcitation   = "%%CITATION = JINST,3,S08001;%%",
}

@Article{Frixione:2008yi,
     author    = "Frixione, Stefano and Laenen, Eric and Motylinski, Patrick
                  and Webber, Bryan R. and White, Chris D.",
     title     = "{Single-top hadroproduction in association with a W
                  boson}",
     journal   = "JHEP",
     volume    = "07",
     year      = "2008",
     pages     = "029",
     eprint    = "0805.3067",
     archivePrefix = "arXiv",
     primaryClass  =  "hep-ph",
     doi       = "10.1088/1126-6708/2008/07/029"
}

@Article{Hoeche:2009rj,
      author         = "H{\"o}che, Stefan and Krauss, Frank and Schumann, Steffen and Siegert, Frank",
      title          = "{QCD matrix elements and truncated showers}",
      journal        = "JHEP",
      volume         = "05",
      pages          = "053",
      doi            = "10.1088/1126-6708/2009/05/053",
      year           = "2009",
      eprint         = "0903.1219",
      archivePrefix  = "arXiv",
      primaryClass   = "hep-ph",
}

@Article{PERF-2007-01,
    author         = "{ATLAS Collaboration}",
    title          = "{The ATLAS Experiment at the CERN Large Hadron Collider}",
    journal        = "JINST",
    volume         = "3",
    year           = "2008",
    pages          = "S08003",
    doi            = "10.1088/1748-0221/3/08/S08003",
    primaryClass   = "hep-ex",
}

@Article{SOFT-2010-01,
    author         = "{ATLAS Collaboration}",
    title          = "{The ATLAS Simulation Infrastructure}",
    journal        = "Eur. Phys. J. C",
    volume         = "70",
    year           = "2010",
    pages          = "823",
    doi            = "10.1140/epjc/s10052-010-1429-9",
    eprint         = "1005.4568",
    archivePrefix  = "arXiv",
    primaryClass   = "physics.ins-det",
}

@Article{PERF-2012-04,
    author         = "{ATLAS Collaboration}",
    title          = "{Performance of \(b\)-jet identification in the ATLAS experiment}",
    journal        = "JINST",
    volume         = "11",
    year           = "2016",
    pages          = "P04008",
    doi            = "10.1088/1748-0221/11/04/P04008",
    reportNumber   = "CERN-PH-EP-2015-216",
    eprint         = "1512.01094",
    archivePrefix  = "arXiv",
    primaryClass   = "hep-ex",
}

@Article{PERF-2014-07,
    author         = "{ATLAS Collaboration}",
    title          = "{Topological cell clustering in the ATLAS calorimeters and its performance in LHC Run~1}",
    journal        = "Eur. Phys. J. C",
    volume         = "77",
    year           = "2017",
    pages          = "490",
    doi            = "10.1140/epjc/s10052-017-5004-5",
    reportNumber   = "CERN-PH-EP-2015-304",
    eprint         = "1603.02934",
    archivePrefix  = "arXiv",
    primaryClass   = "hep-ex",
}

@Article{PERF-2015-01,
    author         = "{ATLAS Collaboration}",
    title          = "{Reconstruction of primary vertices at the ATLAS experiment in Run~1 proton--proton collisions at the LHC}",
    journal        = "Eur. Phys. J. C",
    volume         = "77",
    year           = "2017",
    pages          = "332",
    doi            = "10.1140/epjc/s10052-017-4887-5",
    reportNumber   = "CERN-EP-2016-150",
    eprint         = "1611.10235",
    archivePrefix  = "arXiv",
    primaryClass   = "hep-ex",
}

@Article{PERF-2015-08,
    author         = "{ATLAS Collaboration}",
    title          = "{Performance of the ATLAS track reconstruction algorithms in dense environments in LHC Run~2}",
    journal        = "Eur. Phys. J. C",
    volume         = "77",
    year           = "2017",
    pages          = "673",
    doi            = "10.1140/epjc/s10052-017-5225-7",
    reportNumber   = "CERN-EP-2017-045",
    eprint         = "1704.07983",
    archivePrefix  = "arXiv",
    primaryClass   = "hep-ex",
}

@Article{PERF-2015-10,
    author         = "{ATLAS Collaboration}",
    title          = "{Muon reconstruction performance of the ATLAS detector in proton--proton collision data at \(\sqrt{s} = 13\,\text{TeV}\)}",
    journal        = "Eur. Phys. J. C",
    volume         = "76",
    year           = "2016",
    pages          = "292",
    doi            = "10.1140/epjc/s10052-016-4120-y",
    reportNumber   = "CERN-EP-2016-033",
    eprint         = "1603.05598",
    archivePrefix  = "arXiv",
    primaryClass   = "hep-ex",
}

@Article{PERF-2016-04,
    author         = "{ATLAS Collaboration}",
    title          = "{Jet energy scale measurements and their systematic uncertainties in proton--proton collisions at \(\sqrt{s} = 13\,\text{TeV}\) with the ATLAS detector}",
    journal        = "Phys. Rev. D",
    volume         = "96",
    year           = "2017",
    pages          = "072002",
    doi            = "10.1103/PhysRevD.96.072002",
    reportNumber   = "CERN-EP-2017-038",
    eprint         = "1703.09665",
    archivePrefix  = "arXiv",
    primaryClass   = "hep-ex",
}

@Article{PERF-2016-05,
    author         = "{ATLAS Collaboration}",
    title          = "{Measurements of \(b\)-jet tagging efficiency with the ATLAS detector using \(t\bar{t}\) events at \(\sqrt{s} = 13\,\text{TeV}\)}",
    journal        = "JHEP",
    volume         = "08",
    year           = "2018",
    pages          = "089",
    doi            = "10.1007/JHEP08(2018)089",
    reportNumber   = "CERN-EP-2018-047",
    eprint         = "1805.01845",
    archivePrefix  = "arXiv",
    primaryClass   = "hep-ex",
}

@Article{TRIG-2016-01,
    author         = "{ATLAS Collaboration}",
    title          = "{Performance of the ATLAS trigger system in 2015}",
    journal        = "Eur. Phys. J. C",
    volume         = "77",
    year           = "2017",
    pages          = "317",
    doi            = "10.1140/epjc/s10052-017-4852-3",
    reportNumber   = "CERN-EP-2016-241",
    eprint         = "1611.09661",
    archivePrefix  = "arXiv",
    primaryClass   = "hep-ex",
}

@Article{TOPQ-2017-01,
    author         = "{ATLAS Collaboration}",
    title          = "{Measurements of differential cross sections of top quark pair production in association with jets in \(pp\) collisions at \(\sqrt{s} = 13\,\text{TeV}\) using the ATLAS detector}",
    journal        = "JHEP",
    volume         = "10",
    year           = "2018",
    pages          = "159",
    doi            = "10.1007/JHEP10(2018)159",
    reportNumber   = "CERN-EP-2017-227",
    eprint         = "1802.06572",
    archivePrefix  = "arXiv",
    primaryClass   = "hep-ex",
}

@Article{FTAG-2018-01,
    author         = "{ATLAS Collaboration}",
    title          = "{ATLAS \(b\)-jet identification performance and efficiency measurement with \(t\bar{t}\) events in \(pp\) collisions at \(\sqrt{s} = 13\,\text{TeV}\)}",
    year           = "2019",
    reportNumber   = "CERN-EP-2019-132",
    eprint         = "1907.05120",
    archivePrefix  = "arXiv",
    primaryClass   = "hep-ex",
}

@Article{HIGG-2018-04,
    author         = "{ATLAS Collaboration}",
    title          = "{Observation of \(H \rightarrow b\bar{b}\) decays and \(VH\) production with the ATLAS detector}",
    journal        = "Phys. Lett. B",
    volume         = "786",
    year           = "2018",
    pages          = "59",
    doi            = "10.1016/j.physletb.2018.09.013",
    reportNumber   = "CERN-EP-2018-215",
    eprint         = "1808.08238",
    archivePrefix  = "arXiv",
    primaryClass   = "hep-ex",
}

@Article{HIGG-2018-13,
    author         = "{ATLAS Collaboration}",
    title          = "{Observation of Higgs boson production in association with a top quark pair at the LHC with the ATLAS detector}",
    journal        = "Phys. Lett. B",
    volume         = "784",
    year           = "2018",
    pages          = "173",
    doi            = "10.1016/j.physletb.2018.07.035",
    reportNumber   = "CERN-EP-2018-138",
    eprint         = "1806.00425",
    archivePrefix  = "arXiv",
    primaryClass   = "hep-ex",
}

@Article{PIX-2018-001,
    author         = "Abbott, B. and others",
    title          = "{Production and integration of the ATLAS Insertable B-Layer}",
    journal        = "JINST",
    volume         = "13",
    year           = "2018",
    pages          = "T05008",
    doi            = "10.1088/1748-0221/13/05/T05008",
    eprint         = "1803.00844",
    archivePrefix  = "arXiv",
    primaryClass   = "physics.ins-det",
}

@Booklet{ATL-GEN-PUB-2016-002,
    author         = "{ATLAS Collaboration}",
    title          = "{ATLAS Computing Acknowledgements}",
    howpublished   = "{ATL-GEN-PUB-2016-002}",
    url            = "https://cds.cern.ch/record/2202407",
}

@Booklet{ATLAS-TDR-2010-19,
    author         = "{ATLAS Collaboration}",
    title          = "{ATLAS Insertable B-Layer Technical Design Report}",
    howpublished   = "ATLAS-TDR-19",
    url            = "https://cds.cern.ch/record/1291633",
    year           = "2010",
    related        = "ATLAS-TDR-2010-19-add",
    relatedstring  = "Addendum:",
}

@Booklet{ATLAS-CONF-2014-018,
    author         = "{ATLAS Collaboration}",
    title          = "{Tagging and suppression of pileup jets with the ATLAS detector}",
    howpublished   = "{ATLAS-CONF-2014-018}",
    url            = "https://cds.cern.ch/record/1700870",
    year           = "2014",
}

@Booklet{ATLAS-CONF-2015-029,
    author         = "{ATLAS Collaboration}",
    title          = "{Selection of jets produced in \(13~\text{TeV}\) proton--proton collisions with the ATLAS detector}",
    howpublished   = "{ATLAS-CONF-2015-029}",
    url            = "https://cds.cern.ch/record/2037702",
    year           = "2015",
}

@Booklet{ATLAS-CONF-2018-006,
    author         = "{ATLAS Collaboration}",
    title          = "{Calibration of light-flavour \(b\)-jet mistagging rates using ATLAS proton--proton collision data at \(\sqrt{s} = 13~\text{TeV}\)}",
    howpublished   = "{ATLAS-CONF-2018-006}",
    url            = "https://cds.cern.ch/record/2314418",
    year           = "2018",
}

@Booklet{ATLAS-CONF-2018-045,
    author         = "{ATLAS Collaboration}",
    title          = "{Calibration of the ATLAS \(b\)-tagging algorithm in \(t\bar{t}\) semileptonic events}",
    howpublished   = "{ATLAS-CONF-2018-045}",
    url            = "https://cds.cern.ch/record/2638455",
    year           = "2018",
}

@Booklet{ATL-PHYS-PUB-2010-013,
    author         = "{ATLAS Collaboration}",
    title          = "{The simulation principle and performance of the ATLAS fast calorimeter simulation FastCaloSim}",
    howpublished   = "{ATL-PHYS-PUB-2010-013}",
    url            = "https://cds.cern.ch/record/1300517",
    year           = "2010",
}

@Booklet{ATL-PHYS-PUB-2014-021,
    author         = "{ATLAS Collaboration}",
    title          = "{ATLAS Pythia~8 tunes to \(7~\text{TeV}\) data}",
    howpublished   = "{ATL-PHYS-PUB-2014-021}",
    url            = "https://cds.cern.ch/record/1966419",
    year           = "2014",
}

@Booklet{ATL-PHYS-PUB-2015-026,
    author         = "{ATLAS Collaboration}",
    title          = "{Vertex Reconstruction Performance of the ATLAS Detector at \(\sqrt{s} = 13~\text{TeV}\)}",
    howpublished   = "{ATL-PHYS-PUB-2015-026}",
    url            = "https://cds.cern.ch/record/2037717",
    year           = "2015",
}

@Booklet{ATL-PHYS-PUB-2015-051,
    author         = "{ATLAS Collaboration}",
    title          = "{Early Inner Detector Tracking Performance in the 2015 Data at \(\sqrt{s} = 13~\text{TeV}\)}",
    howpublished   = "{ATL-PHYS-PUB-2015-051}",
    url            = "https://cds.cern.ch/record/2110140",
    year           = "2015",
}

@Booklet{ATL-PHYS-PUB-2016-017,
    author         = "{ATLAS Collaboration}",
    title          = "{The Pythia~8 A3 tune description of ATLAS minimum bias and inelastic measurements incorporating the Donnachie--Landshoff diffractive model}",
    howpublished   = "{ATL-PHYS-PUB-2016-017}",
    url            = "https://cds.cern.ch/record/2206965",
    year           = "2016",
}

@Booklet{ATL-PHYS-PUB-2016-020,
    author         = "{ATLAS Collaboration}",
    title          = "{Studies on top-quark Monte Carlo modelling for Top2016}",
    howpublished   = "{ATL-PHYS-PUB-2016-020}",
    url            = "https://cds.cern.ch/record/2216168",
    year           = "2016",
}

@Booklet{ATL-PHYS-PUB-2017-005,
    author         = "{ATLAS Collaboration}",
    title          = "{Multi-Boson Simulation for \(13~\text{TeV}\) ATLAS Analyses}",
    howpublished   = "{ATL-PHYS-PUB-2017-005}",
    url            = "https://cds.cern.ch/record/2261933",
    year           = "2017",
}

@Booklet{ATL-PHYS-PUB-2017-006,
    author         = "{ATLAS Collaboration}",
    title          = "{ATLAS simulation of boson plus jets processes in Run~2}",
    howpublished   = "{ATL-PHYS-PUB-2017-006}",
    url            = "https://cds.cern.ch/record/2261937",
    year           = "2017",
}

@Booklet{ATL-PHYS-PUB-2017-011,
    author         = "{ATLAS Collaboration}",
    title          = "{Secondary vertex finding for jet flavour identification with the ATLAS detector}",
    howpublished   = "{ATL-PHYS-PUB-2017-011}",
    url            = "https://cds.cern.ch/record/2270366",
    year           = "2017",
}

@Booklet{ATL-PHYS-PUB-2017-013,
    author         = "{ATLAS Collaboration}",
    title          = "{Optimisation and performance studies of the ATLAS \(b\)-tagging algorithms for the 2017-18 LHC run}",
    howpublished   = "{ATL-PHYS-PUB-2017-013}",
    url            = "https://cds.cern.ch/record/2273281",
    year           = "2017",
}

@Booklet{ATL-PHYS-PUB-2018-025,
    author         = "{ATLAS Collaboration}",
    title          = "{Topological \(b\)-hadron decay reconstruction and identification of \(b\)-jets with the JetFitter package in the ATLAS experiment at the LHC}",
    howpublished   = "{ATL-PHYS-PUB-2018-025}",
    url            = "https://cds.cern.ch/record/2645405",
    year           = "2018",
}

\clearpage 
 
\begin{flushleft}
{\Large The ATLAS Collaboration}

\bigskip

G.~Aad$^\textrm{\scriptsize 102}$,    
B.~Abbott$^\textrm{\scriptsize 129}$,    
D.C.~Abbott$^\textrm{\scriptsize 103}$,    
A.~Abed~Abud$^\textrm{\scriptsize 71a,71b}$,    
K.~Abeling$^\textrm{\scriptsize 53}$,    
D.K.~Abhayasinghe$^\textrm{\scriptsize 94}$,    
S.H.~Abidi$^\textrm{\scriptsize 167}$,    
O.S.~AbouZeid$^\textrm{\scriptsize 40}$,    
N.L.~Abraham$^\textrm{\scriptsize 156}$,    
H.~Abramowicz$^\textrm{\scriptsize 161}$
% ...
\bigskip
\\

% ...
$^{40}$Niels Bohr Institute, University of Copenhagen, Copenhagen; Denmark.\\
% ...
$^{53}$II. Physikalisches Institut, Georg-August-Universit\"{a}t G\"ottingen, G\"ottingen; Germany.\\
% ...
$^{71}$$^{(a)}$INFN Sezione di Pavia;$^{(b)}$Dipartimento di Fisica, Universit\`a di Pavia, Pavia; Italy.\\
% ...
$^{94}$Department of Physics, Royal Holloway University of London, Egham; United Kingdom.\\
% ...
$^{102}$CPPM, Aix-Marseille Universit\'e, CNRS/IN2P3, Marseille; France.\\
$^{103}$Department of Physics, University of Massachusetts, Amherst MA; United States of America.\\
% ...
$^{129}$Homer L. Dodge Department of Physics and Astronomy, University of Oklahoma, Norman OK; United States of America.\\
% ...
$^{156}$Department of Physics and Astronomy, University of Sussex, Brighton; United Kingdom.\\
% ...
$^{161}$Raymond and Beverly Sackler School of Physics and Astronomy, Tel Aviv University, Tel Aviv; Israel.\\
% ...
$^{167}$Department of Physics, University of Toronto, Toronto ON; Canada.\\
% ...
$^{a}$ Also at Borough of Manhattan Community College, City University of New York, New York NY; United States of America.\\
$^{b}$ Also at CERN, Geneva; Switzerland.\\
% ...

\end{flushleft}


\end{document}